\documentclass[twocolumn,preprintnumbers,amsmath,amssymb,superscriptaddress]{revtex4}

\usepackage{graphicx}
\usepackage{dcolumn}
\usepackage{bm}

\makeatletter
\def\@seccntformat#1{}
\makeatother
\renewcommand{\numberline}[1]{}

\begin{document}

\title{A new experimental approach for the exploration of topological quantum phenomena}


\author{M. Zahid Hasan}
\affiliation{Joseph Henry Laboratories, Department of Physics, Princeton
University, Princeton, NJ 08544, USA}
\affiliation{Princeton Institute for the Science and Technology of Materials, School of Engineering and Applied Science, Princeton University, Princeton NJ 08544,
USA}
\affiliation{Advanced Light Source, Lawrence Berkeley National Laboratory, Berkeley, California 94305, USA}

\author{David Hsieh}
\affiliation{Joseph Henry Laboratories, Department of Physics, Princeton
University, Princeton, NJ 08544, USA}
\affiliation{Department of Physics, Massachusetts Institute of Technology, Cambridge, MA 02139, USA}
\author{Yuqi Xia}
\affiliation{Joseph Henry Laboratories, Department of Physics, Princeton
University, Princeton, NJ 08544, USA}
\author{L. Andrew Wray}
\affiliation{Joseph Henry Laboratories, Department of Physics, Princeton
University, Princeton, NJ 08544, USA}
\author{Su-Yang Xu}
\affiliation{Joseph Henry Laboratories, Department of Physics, Princeton University, Princeton, NJ 08544, USA}

\author{Charles L. Kane}
\affiliation{Department of Physics and Astronomy, University of
Pennsylvania, Philadelphia, PA 19104, USA}

\begin{abstract}
The three-dimensional topological insulator (originally called "topological insulators") is the first example in nature of a topologically ordered electronic phase existing in three dimensions that cannot be reduced to multiple copies of quantum-Hall-like states. \textbf{Their topological order can be realized at room temperatures without magnetic fields and they can be turned into magnets and exotic superconductors leading to world-wide interest and activity in topological insulators}. These experimentally demonstrated unique properties of 3D topological insulators have turned the topic into a field of intense experimental activity. It is also only the third topologically ordered phase in weakly interacting systems to be discovered in nature, with the first two belonging to the quantum Hall-like topological insulator class consisting of the 2D integer quantum Hall state (IQH) and the 2D quantum spin Hall state (QSH). The 2D quantum spin Hall state (QSH) can be thought of as two copies of IQH put together leading to a time-reversal invariant version of IQH state. All of the 2D topological insulator examples (IQH, QSH) including the fractional one (FQH) involving Coulomb interaction are understood in the standard picture of quantized electron orbits in a spin-independent or spin-dependent magnetic field, \textbf{the 3D topological insulator defies such description and is a novel type of topological order which cannot be reduced to multiple copies of quantum-Hall-like states}. In fact, the 3D topological insulator exists not only in zero magnetic field, and differs from the 2D variety in three very important aspects: \textbf{1}) they possess topologically protected 2D metallic surfaces (a new type of 2DEG) rather than the 1D edges, \textbf{2}) they can work at room temperature (300K and beyond, large-gap topological insulators) rather than cryogenic (mK) temperatures required for the QSH effects and \textbf{3}) they occur in standard bulk semiconductors rather than at buried interfaces of ultraclean semiconductor heterostructures thus tolerate stronger disorder than the IQH-like states. One of the major challenges in going from quantum Hall-like 2D states to 3D topological insulators is to \textbf{develop new experimental approaches/methods to precisely probe this novel form of topological-order} since the standard tools and settings that work for IQH-state also work for QSH states. The method to probe 2D topological-order is exclusively with charge transport (pioneered by Von Klitzing in the 1980s), which either measures quantized transverse conductance plateaus in IQH systems or longitudinal conductance in quantum spin Hall (QSH) systems. In a 3D topological insulator, the boundary itself supports a two dimensional electron gas (2DEG) and transport is not (Z$_2$) topologically quantized hence \textit{cannot} directly probe the topological invariants {$\nu_o$} or the topological quantum numbers analogous to the Chern numbers of the IQH systems. This is \textit{unrelated} to the fact that the present materials have some extrinsic or residual/impurity conductivity in their naturally grown bulk. In this paper, we review the birth of momentum- and spin-resolved spectroscopy as a new experimental approach and as a directly boundary sensitive method to study and prove topological-order in three-dimensions via the direct measurements of the topological invariants {$\nu_o$} that are associated with the Z$_2$ topology of the spin-orbit band structure and opposite parity band inversions, which led to the experimental discovery of the first 3D topological insulator in Bi-Sb semiconductors (KITP Proceeding \textbf{2007} http://online.itp.ucsb.edu/online/motterials07/hasan/ (2007), Nature 452, 970 (2008), Submitted \textbf{2007}) which further led to the discovery of the Bi$_{2}$Se$_3$ class - the most widely researched topological insulator to this date. We discuss the fundamental properties of the novel topologically spin-momentum locked half Dirac metal on the surfaces of the 3D topological insulators and how they emerge from topological phase transitions due to increasing spin-orbit coupling in the bulk. These electronic and spin properties of \textbf{topological surface states} discovered via the methods reviewed here are already guiding the interpretation of surface transport measurements as they are beginning to be possible further advancing the field potentially towards device applications. These methods and their derivatives are also being applied by many others world-wide for further investigations of topological-order and for discovering new topological insulator states as well as exotic topological quantum phenomena (the list is too long to review here). We also review how spectroscopic methods are leading to the identification of spin-orbit superconductors that may work as Majorana platforms and can be used to identify topological superconductors - yet another class of new state of matter.

\end{abstract}


\maketitle

\tableofcontents

\section{The birth of momentum-resolved spectroscopy as a direct experimental probe of Topological-Order}

Ordered phases of matter such as a superfluid or a ferromagnet are
usually associated with the breaking of a symmetry and are
characterized by a local order parameter \cite{1}. The typical
experimental probes of these systems are sensitive to order
parameters. In the 1980s, two new phases of matter were realized by subjecting 2D electron gases at buried interfaces of semiconductor heterostructures to large magnetic fields. These new phases of matter, the 2D integer and 2D fractional quantum Hall states, exhibited a new and rare type of order that is derived from an
organized collective quantum entangled motion of electrons \cite{2,3,4,Tsui}. These
so-called ``2D topologically ordered insulators" do not exhibit any
symmetry breaking and are characterized by a topological number \cite{5}
as opposed to a local order parameter. The most striking manifestation of this 2D topological order is the existence of one-way propagating 1D metallic states confined to their edges, which lead to remarkable quantized charge transport phenomena. To date the experimental
probe of their topological quantum numbers is based on charge transport, where measurements of the quantization of transverse magneto-conductivity $\sigma_{xy} =
ne^{2}/h$ (where $e$ is the electric charge and $h$ is Planck's
constant) reveals the value of the topological number $n$ that
characterizes these quantum Hall states \cite{6}.

\begin{figure*}
\includegraphics[scale=0.55,clip=true, viewport=0.0in 0.0in 11.5in 8.5in]{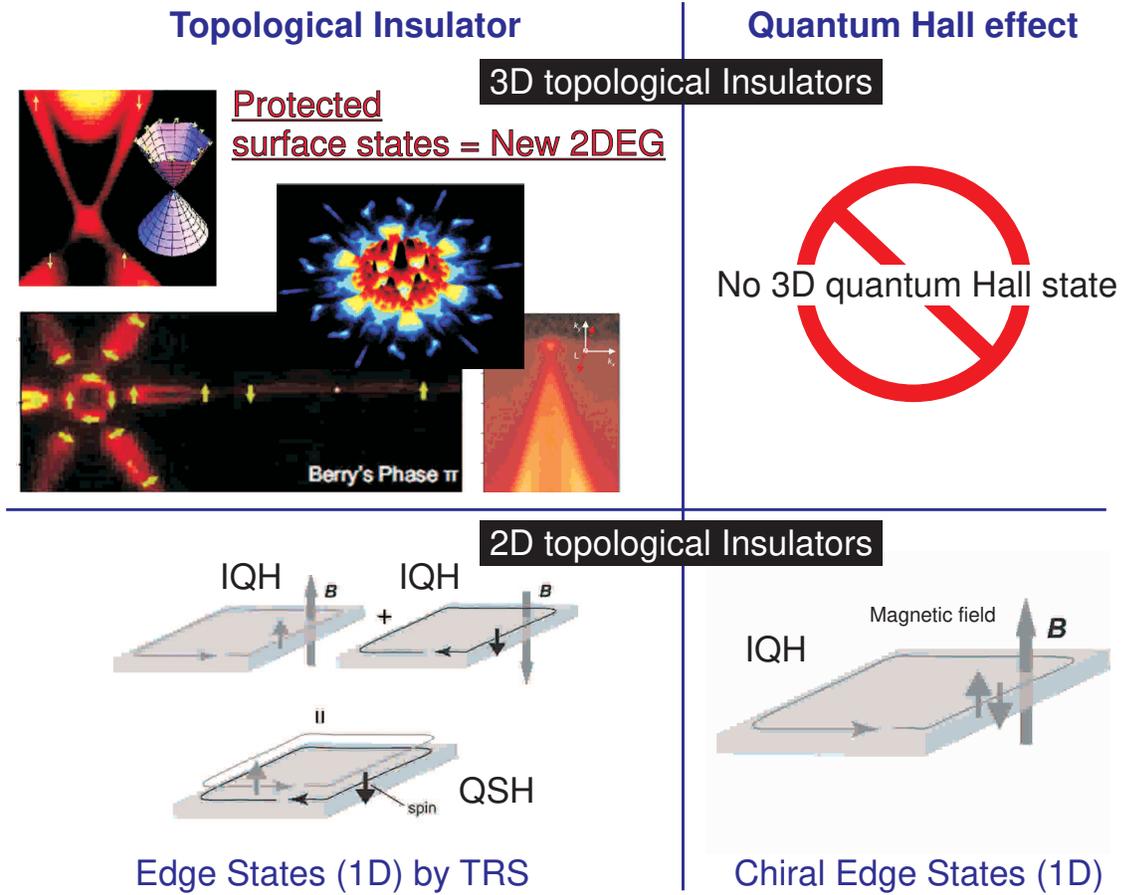}
\caption{\label{Intro} \textbf{2D and 3D topological insulators in nature.} The 2D topological insulator class consists of the 2D quantum Hall states (IQH) and 2D quantum spin Hall state (QSH). The latter is constructed from two copies of the former. On the other hand, a 3D quantum Hall state is forbidden in nature, so the 3D topological insulator represents a new type of topologically ordered phase. The protected surface states form a novel type of topological metal (half Dirac metal) where electron's spin is locked to its momentum but exhibit no spin quantum Hall effect.}
\end{figure*}

Recently, a third type of 2D topological insulator, the spin quantum Hall insulator, was theoretically predicted \cite{14,8} and then experimentally discovered \cite{7}. This class of quantum Hall-like topological phases can exist in spin-orbit materials without external magnetic fields, and can be described as an ordinary quantum Hall state in a spin-dependent magnetic field. Their topological order gives rise to counter-propagating 1D edge states that carry opposite spin polarization, often described as a superposition of a spin up and spin down quantum Hall edge state (Figure 1). Like conventional quantum Hall systems, the 2D spin quantum Hall insulator (QSH) is realized at a buried solid interface. The first, and to date only, realization of this phase was made in (Hg,Cd)Te quantum wells using \textit{charge transport} by measuring a longitudinal conductance of about $2e^2/h$ at mK temperatures \cite{7}. The quantum spin Hall state (QSH) can be thought of as two copies of integer quantum Hall states (IQH) and protected by a Z2 invariant.

It was also realized that a fundamentally new type of genuinely three-dimensional topological order might be realized in bulk crystals without need for an external magnetic field \cite{Fu:STI2,11,15}. Such a 3D topological insulator cannot be reduced to multiple copies of the IQH and such phases would be only the fourth type of topologically ordered phase to be discovered in nature, and the first type to fall outside the quantum Hall-like 2D topological states (IQH, FQH, QSH). Instead of having quantum-Hall type 1D edge state, these so-called 3D topological insulators would possess unconventional metallic 2D topological surface states called spin-textured helical metals, a type of 2D electron gas long thought to be impossible to realize. However, it was recognized that 3D topological insulators would NOT necessarily exhibit a topologically (Z2) quantized charge transport by themselves as carried out in a conventional transport settings of all quantum-Hall-like measurements. Therefore, their 3D topological quantum numbers (Z2), the analogues of $n$ (Chern numbers), could not be measured via the charge transport based methods even if a complete isolation of surface charge transport becomes routinely possible. Owing to the 2D nature of the two surface conduction channels that contribute together in a 3D topological insulator case, it was theoretically recognized that it would not be possible to measure the topological invariants due to the lack of a quantized transport response of the 2D surface that measures the Z2 topological invariants  \cite{Fu:STI2}.

Here we review the development of spin- and angle-resolved photoemission spectroscopy (spin-ARPES) as the new approach/method to probe 3D topological order \cite{10,Science}, which today constitutes the experimental standard for identifying topological order in bulk solids also used by many others world-wide. We will review the procedures for i) separating intrinsic bulk bands from surface electronic structures using incident energy modulated ARPES, ii) mapping the surface electronic structure across the Kramers momenta to establish the topologically non-trivial nature of the surface states, iii) using spin-ARPES to map the spin texture of the surface states to reveal topological quantum numbers and Berry's phases and iv) measuring the topological parent compounds to establish the microscopic origins of 3D topological order. These will be discussed in the context of Bi$_{1-x}$Sb$_x$, which was the first 3D topological insulator to be experimentally discovered in nature and a textbook example of how this method is applied. The confluence of three factors, having a detailed spectroscopic procedure to measure 3D topological order, their discovery in fairly simple bulk semiconductors and being able to work at room temperatures, has led to worldwide efforts to study 3D topological physics and led to over 100 compounds being identified as 3D topological insulators to date.

\begin{figure*}
\includegraphics[scale=0.62,clip=true, viewport=0.0in 0in 10.8in 7.0in]{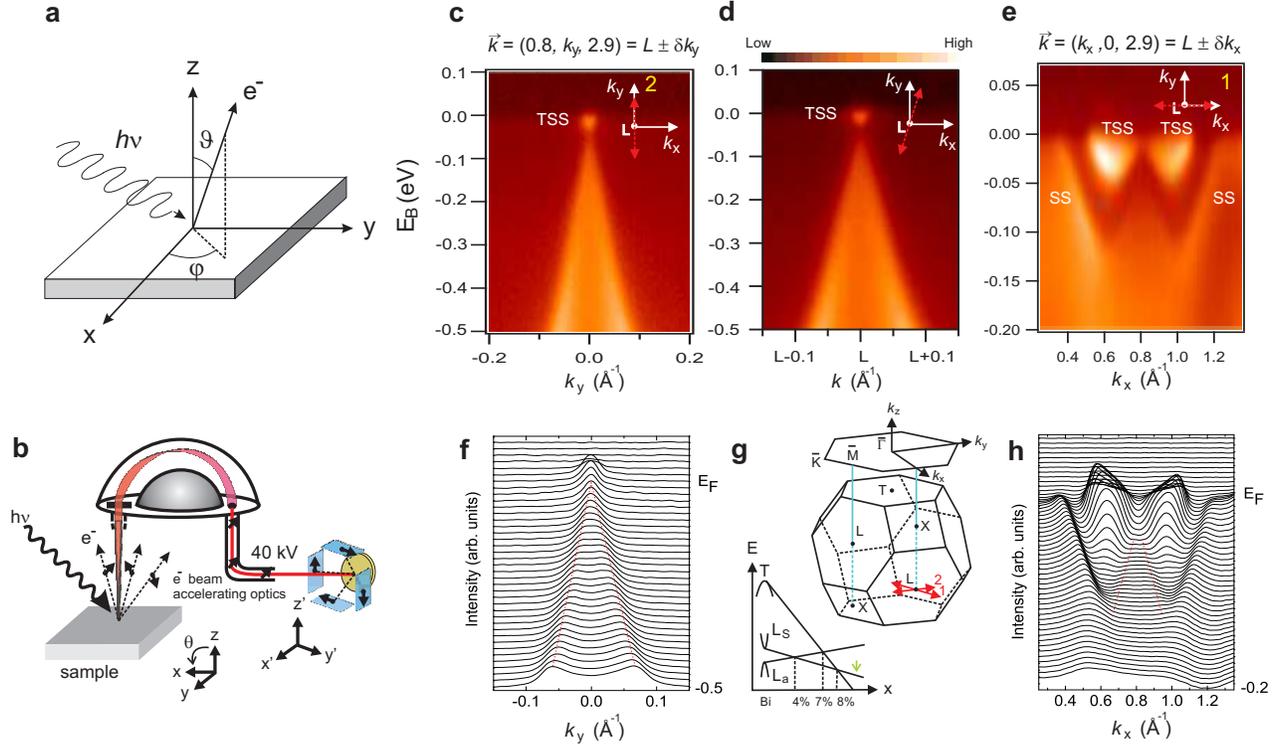}
\caption{\label{fig:BiSb_Fig1} \textbf{The first 3D topological insulator (\textbf{2007}): Dirac-like dispersion signalling band inversion via spin-orbit interaction} \textbf{a}, Schematic of an ARPES experimental geometry. The kinetic energy of photoelectrons and their angles of emission ($\theta$,$\phi$) determine its electronic structure. \textbf{b}, Energy and momentum analysis take place through a hemispherical analyzer and the spin analysis is performed using a Mott detector. Selected ARPES intensity
maps of Bi$_{0.9}$Sb$_{0.1}$ are shown along three $\vec{k}$-space
cuts through the L point of the bulk 3D Brillouin zone (BZ). The
presented data are taken in the third BZ with L$_z$ = 2.9 \AA$^{-1}$
with a photon energy of 29 eV. The cuts are along \textbf{c}, the
$k_y$ direction, \textbf{d}, a direction rotated by approximately
$10^{\circ}$ from the $k_y$ direction, and \textbf{e}, the $k_x$
direction. Each cut shows a $\Lambda$-shaped bulk band whose tip
lies below the Fermi level signalling a bulk gap. The (topological) surface states
are denoted (T)SS and are all identified in Fig.\ref{fig:BiSb_Fig2} [for further
identification via theoretical calculations see Supplementary
Materials (SM)]. \textbf{f}, Momentum distribution curves (MDCs)
corresponding to the intensity map in \textbf{c}. \textbf{h}, Log
scale plot of the MDCs corresponding to the intensity map in
\textbf{e}. The red lines are guides to the eye for the bulk
features in the MDCs. \textbf{g}, Schematic of the bulk 3D BZ of
Bi$_{1-x}$Sb$_x$ and the 2D BZ of the projected (111) surface. The
high symmetry points $\bar{\Gamma}$, $\bar{M}$ and $\bar{K}$ of the
surface BZ are labeled. Schematic evolution of bulk band energies as
a function of $x$ is shown. The L band inversion transition occurs
at $x \approx 0.04$, where a 3D gapless Dirac point is realized, and
the composition we study here (for which $x = 0.1$) is indicated by
the green arrow. A more detailed phase diagram based on our
experiments is shown in Fig.\ref{fig:BiSb_Fig3}c. [Adapted from D. Hsieh $et$ $al.$, \textit{Nature} \textbf{452}, 970 (2008) (Completed and Submitted in \textbf{2007}) \cite{10}].}
\end{figure*}

\section{Separation of insulating bulk from metallic surface states using incident photon energy modulated ARPES}

Three-dimensional topological order is predicted to occur in semiconductors with an inverted band gap, therefore 3D topological insulators are often searched for in systems where a band gap inversion is known to take place as a function of some control parameter. The experimental signature of being in the vicinity of a bulk band inversion is that the bulk band dispersion should be described by the massive Dirac equation rather than Schroedinger equation, since the system must be described by the massless Dirac equation exactly at the bulk band inversion point.

\begin{figure*}
\includegraphics[scale=0.65,clip=true, viewport=0.0in 0in 10.0in 7.4in]{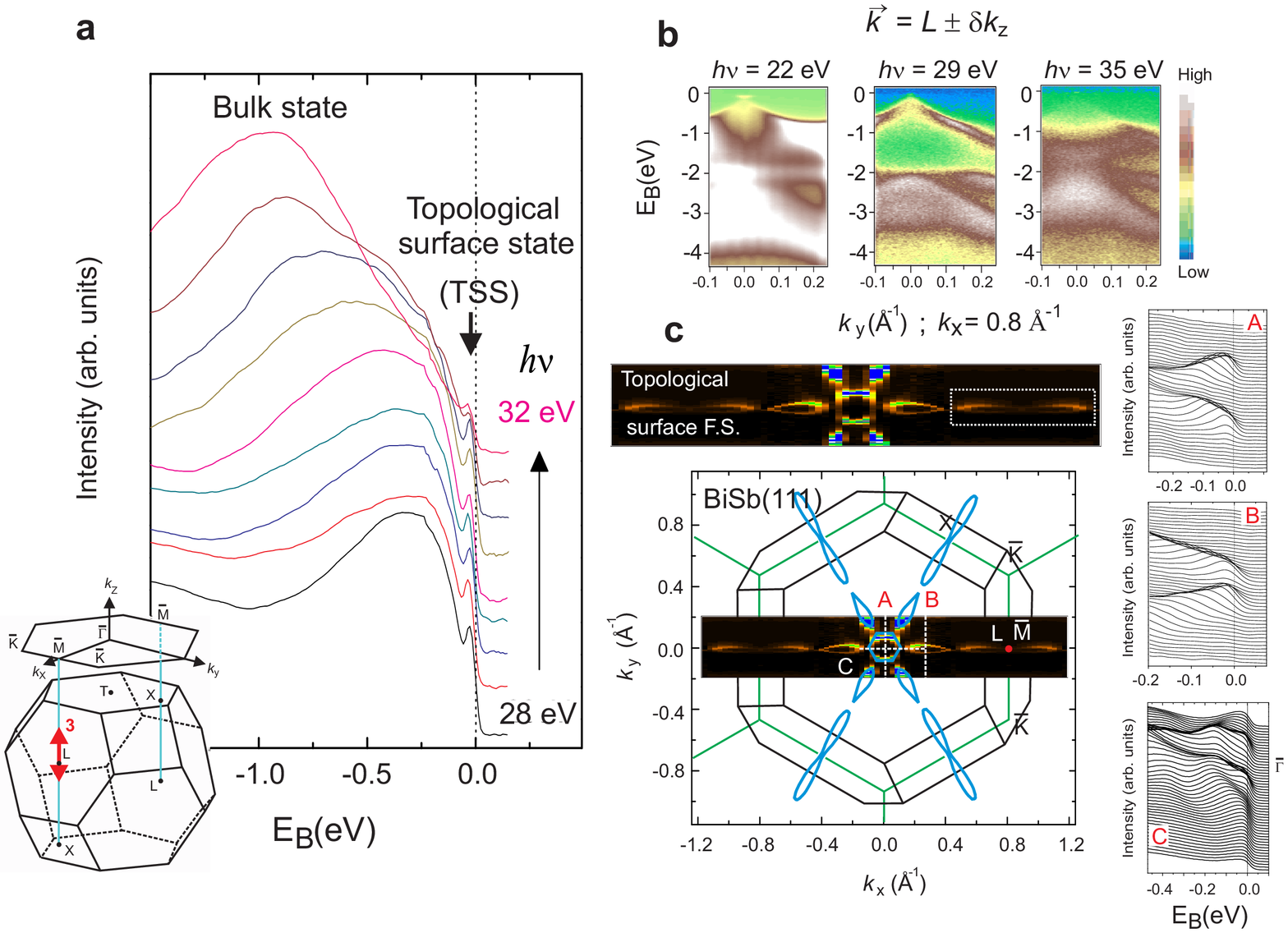}
\caption{\label{fig:BiSb_Fig2} \textbf{The first 3D topological insulator (\textbf{2007}): Topological Surface States and electronic band dispersion along the $\mathbf{k_z}$-direction in momentum space.} Surface states are experimentally
identified by studying their out-of-plane momentum dispersion
through the systematic variation of incident photon energy.
\textbf{a}, Energy distribution curves (EDCs) of
Bi$_{0.9}$Sb$_{0.1}$ with electrons at the Fermi level ($E_F$)
maintained at a fixed in-plane momentum of ($k_x$=0.8 \AA$^{-1}$,
$k_y$=0.0 \AA$^{-1}$) are obtained as a function of incident photon
energy to identify states that exhibit no dispersion perpendicular
to the (111)-plane along the direction shown by the double-headed
arrow labeled ``3" in the inset (see SM). Selected EDC data sets with photon energies of 28 eV to
32 eV in steps of 0.5 eV are shown for clarity. The non-energy
dispersive ($k_z$ independent) peaks near $E_F$ are the topological surface
states (TSS). \textbf{b}, ARPES intensity maps along cuts parallel to
$k_y$ taken with electrons at $E_F$ fixed at $k_x$ = 0.8 \AA$^{-1}$
with respective photon energies of $h \nu$ = 22 eV, 29 eV and 35 eV
(for a conversion map from photon energy to $k_z$ see SM). The faint $\Lambda$-shaped band at $h \nu$ = 22 eV and
$h \nu$ = 35 eV shows some overlap with the bulk valence band at L
($h \nu$ = 29 eV), suggesting that it is a resonant surface state
degenerate with the bulk state in some limited k-range near $E_F$.
The flat band of intensity centered about $-$2 eV in the $h \nu$ =
22 eV scan originates from Bi 5d core level emission from second
order light. \textbf{c}, Projection of the bulk BZ (black lines)
onto the (111) surface BZ (green lines). Overlay (enlarged in inset)
shows the high resolution Fermi surface (FS) of the metallic SS
mode, which was obtained by integrating the ARPES intensity (taken
with $h \nu$ = 20 eV) from $-$15 meV to 10 meV relative to $E_F$.
The six tear-drop shaped lobes of the surface FS close to
$\bar{\Gamma}$ (center of BZ) show some intensity variation between
them that is due to the relative orientation between the axes of the
lobes and the axis of the detector slit. The six-fold symmetry was
however confirmed by rotating the sample in the $k_x-k_y$ plane.
EDCs corresponding to the cuts A, B and C are also shown; these
confirm the gapless character of the surface states in bulk
insulating Bi$_{0.9}$Sb$_{0.1}$. [Adapted from D. Hsieh $et$ $al.$, \textit{Nature} \textbf{452}, 970 (2008) (Completed and Submitted in \textbf{2007})\cite{10}].}
\end{figure*}

The early theoretical treatments \cite{11,Murakami} focused on the
strongly spin-orbit coupled, band-inverted Bi$_{1-x}$Sb$_x$ series
as a possible realization of 3D topological order for the following reason. Bismuth is a semimetal with strong spin-orbit interactions. Its band
structure is believed to feature an indirect negative gap between
the valence band maximum at the T point of the bulk Brillouin zone
(BZ) and the conduction band minima at three equivalent L points
\cite{Lenoir,Liu} (here we generally refer to these as a single
point, L). The valence and conduction bands at L are derived from
antisymmetric (L$_a$) and symmetric (L$_s$) $p$-type orbitals,
respectively, and the effective low-energy Hamiltonian at this point
is described by the (3+1)-dimensional relativistic Dirac equation
\cite{Wolff, Fukuyama, Buot}. The resulting dispersion relation,
$E(\vec{k}) = \pm \sqrt{ {(\vec{v} \cdot \vec{k})}^2 + \Delta^2}
\approx \vec{v} \cdot \vec{k}$, is highly linear owing to the
combination of an unusually large band velocity $\vec{v}$ and a
small gap $\Delta$ (such that $\lvert \Delta / \lvert \vec{v} \rvert
\rvert \approx 5 \times 10^{-3} $\AA$^{-1}$) and has been used to
explain various peculiar properties of bismuth \cite{Wolff,
Fukuyama, Buot}. Substituting bismuth with antimony is believed to
change the critical energies of the band structure as follows (see
Fig.\ref{fig:BiSb_Fig1}). At an Sb concentration of $x \approx 4\%$, the gap $\Delta$
between L$_a$ and L$_s$ closes and a massless three-dimensional (3D)
Dirac point is realized. As $x$ is further increased this gap
re-opens with inverted symmetry ordering, which leads to a change in
sign of $\Delta$ at each of the three equivalent L points in the BZ.
For concentrations greater than $x \approx 7\%$ there is no overlap
between the valence band at T and the conduction band at L, and the
material becomes an inverted-band insulator. Once the band at T
drops below the valence band at L, at $x \approx 8\%$, the system
evolves into a direct-gap insulator whose low energy physics is
dominated by the spin-orbit coupled Dirac particles at L
\cite{11,Lenoir}.

High-momentum-resolution angle-resolved photoemission spectroscopy
performed with varying incident photon energy (IPEM-ARPES) allows
for measurement of electronic band dispersion along various momentum
space ($\vec{k}$-space) trajectories in the 3D bulk BZ. ARPES
spectra taken along two orthogonal cuts through the L point of the
bulk BZ of Bi$_{0.9}$Sb$_{0.1}$ are shown in Figs 1a and c. A
$\Lambda$-shaped dispersion whose tip lies less than 50 meV below
the Fermi energy ($E_F$) can be seen along both directions.
Additional features originating from surface states that do not
disperse with incident photon energy are also seen. Owing to the
finite intensity between the bulk and surface states, the exact
binding energy ($E_B$) where the tip of the $\Lambda$-shaped band
dispersion lies is unresolved. The linearity of the bulk
$\Lambda$-shaped bands is observed by locating the peak positions at
higher $E_B$ in the momentum distribution curves (MDCs), and the
energy at which these peaks merge is obtained by extrapolating
linear fits to the MDCs. Therefore 50 meV represents a lower bound
on the energy gap $\Delta$ between L$_a$ and L$_s$. The magnitude of
the extracted band velocities along the $k_x$ and $k_y$ directions
are $7.9 \pm 0.5 \times 10^4$ ms$^{-1}$ and $10.0 \pm 0.5 \times
10^5$ ms$^{-1}$, respectively, which are similar to the tight
binding values $7.6 \times 10^4$ ms$^{-1}$ and $9.1 \times 10^5$
ms$^{-1}$ calculated for the L$_a$ band of bismuth \cite{Liu}. Our
data are consistent with the extremely small effective mass of
$0.002m_e$ (where $m_e$ is the electron mass) observed in
magneto-reflection measurements on samples with $x = 11\%$
\cite{Hebel}. The Dirac point in graphene, co-incidentally, has a
band velocity ($|v_F| \approx 10^6$ ms$^{-1}$) \cite{Zhang}
comparable to what we observe for Bi$_{0.9}$Sb$_{0.1}$, but its
spin-orbit coupling is several orders of magnitude weaker, and the only known method of inducing a gap
in the Dirac spectrum of graphene is by coupling to an external
chemical substrate \cite{Zhou}. The Bi$_{1-x}$Sb$_x$ series thus
provides a rare opportunity to study relativistic Dirac Hamiltonian
physics in a 3D condensed matter system where the intrinsic (rest)
mass gap can be easily tuned.

\begin{figure*}
\includegraphics[scale=0.65,clip=true, viewport=0.0in 0in 10.5in 6.7in]{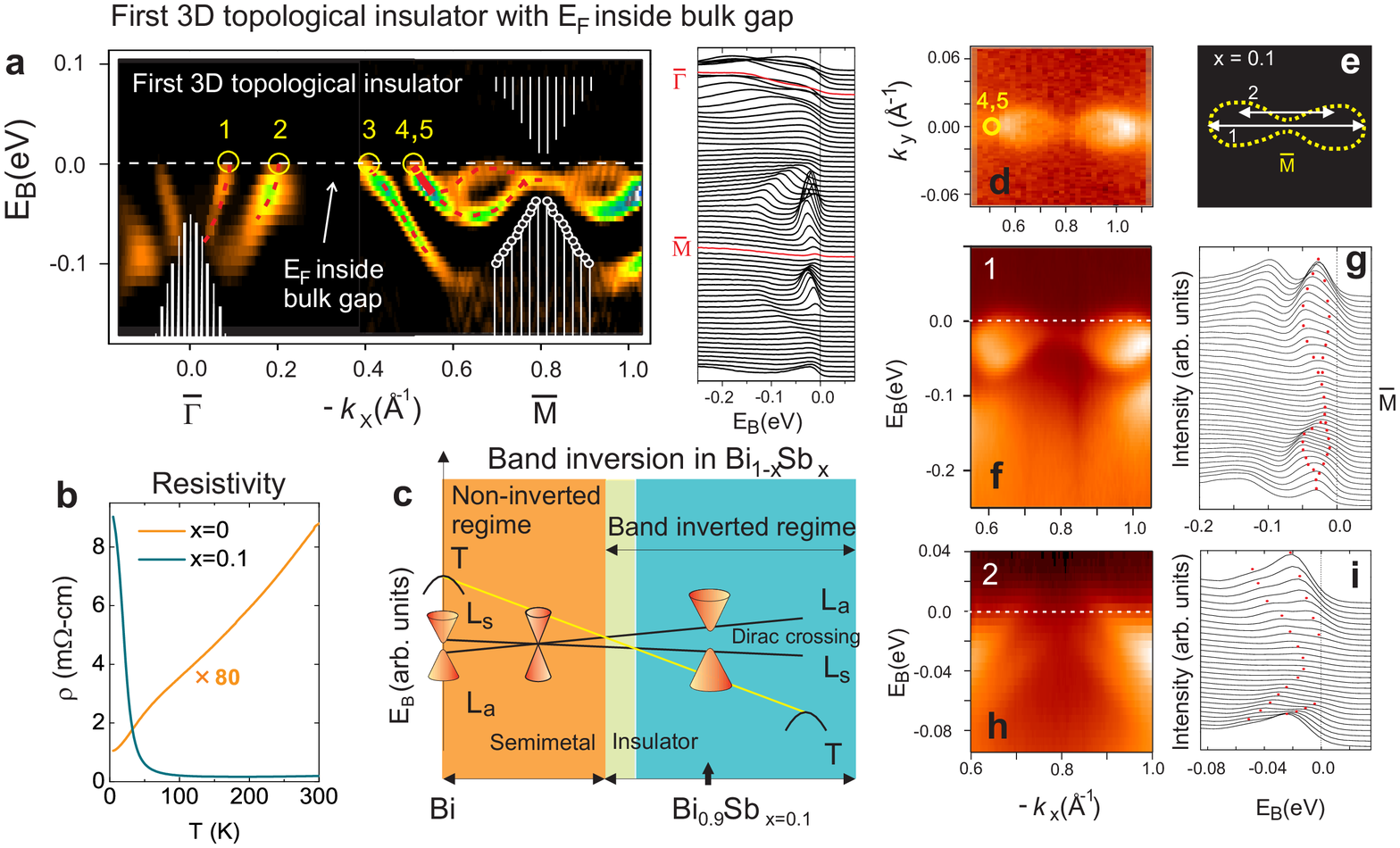}
\caption{\label{fig:BiSb_Fig3} \textbf{The first 3D topological insulator (\textbf{2007}): The topological gapless
surface states in bulk insulating (observed via the bulk band dispersion gap) Bi$_{0.9}$Sb$_{0.1}$.} \textbf{a},
The surface band dispersion second derivative image (SDI) of
Bi$_{0.9}$Sb$_{0.1}$ along $\bar{\Gamma} - \bar{M}$. The shaded
white area shows the projection of the bulk bands based on ARPES
data, as well as a rigid shift of the tight binding bands to sketch
the unoccupied bands above the Fermi level. To maintain high
momentum resolution, data were collected in two segments of momentum
space, then the intensities were normalized using background level
above the Fermi level. A non-intrinsic flat band of intensity near
$E_F$ generated by the SDI analysis was rejected to isolate the
intrinsic dispersion. The Fermi crossings of the surface state are
denoted by yellow circles, with the band near $-k_x \approx 0.5$
\AA$^{-1}$ counted twice owing to double degeneracy. The red lines
are guides to the eye. An in-plane rotation of the sample by
$60^{\circ}$ produced the same surface state dispersion. The EDCs
along $\bar{\Gamma} - \bar{M}$ are shown to the right. There are a
total of five crossings from $\bar{\Gamma} - \bar{M}$ which
indicates that these surface states are topologically non-trivial.
The number of surface state crossings in a material (with an odd
number of Dirac points) is related to the topological $Z_2$
invariant (see text). \textbf{b}, The resistivity curves of Bi and
Bi$_{0.9}$Sb$_{0.1}$ reflect the contrasting transport behaviours.
The presented resistivity curve for pure bismuth has been multiplied
by a factor of 80 for clarity. \textbf{c}, Schematic variation of
bulk band energies of Bi$_{1-x}$Sb$_x$ as a function of $x$ (based
on band calculations and on \cite{11, Lenoir}).
Bi$_{0.9}$Sb$_{0.1}$ is a direct gap bulk Dirac point insulator well
inside the inverted-band regime, and its surface forms a
``topological metal'' - the 2D analogue of the 1D edge states in
quantum spin Hall systems. \textbf{d}, ARPES intensity integrated
within $\pm 10$ meV of $E_F$ originating solely from the surface
state crossings. The image was plotted by stacking along the
negative $k_x$ direction a series of scans taken parallel to the
$k_y$ direction. \textbf{e}, Outline of Bi$_{0.9}$Sb$_{0.1}$ surface
state ARPES intensity near $E_F$ measured in \textbf{d}. White lines
show scan directions ``1'' and ``2''. \textbf{f}, Surface band
dispersion along direction ``1'' taken with $h \nu$ = 28 eV and the
corresponding EDCs (\textbf{g}). The surface Kramers degenerate
point, critical in determining the topological $Z_2$ class of a band
insulator, is clearly seen at $\bar{M}$, approximately $15 \pm 5$
meV below $E_F$. (We note that the scans are taken along the
negative $k_x$ direction, away from the bulk L point.) \textbf{h},
Surface band dispersion along direction ``2'' taken with $h \nu$
 = 28 eV and the corresponding EDCs (\textbf{i}). This scan no longer
passes through the $\bar{M}$-point, and the observation of two well
separated bands indicates the absence of Kramers degeneracy as
expected, which cross-checks the result in (\textbf{a}). [Adapted from D. Hsieh $et$ $al.$, \textit{Nature} \textbf{452}, 970 (2008) (Completed and Submitted in \textbf{2007})\cite{10}].}
\end{figure*}

Studying the band dispersion perpendicular to the sample surface
provides a way to differentiate bulk states from surface states in a
3D material. To visualize the near-$E_F$ dispersion along the 3D L-X
cut (X is a point that is displaced from L by a $k_z$ distance of
3$\pi/c$, where $c$ is the lattice constant), in Fig.2a we plot
energy distribution curves (EDCs), taken such that electrons at
$E_F$ have fixed in-plane momentum $(k_x, k_y)$ = (L$_x$, L$_y$) =
(0.8 \AA$^{-1}$, 0.0 \AA$^{-1}$), as a function of photon energy
($h\nu$). There are three prominent features in the EDCs: a
non-dispersing, $k_z$ independent, peak centered just below $E_F$ at
about $-$0.02 eV; a broad non-dispersing hump centered near $-$0.3
eV; and a strongly dispersing hump that coincides with the latter
near $h\nu$ = 29 eV. To understand which bands these features
originate from, we show ARPES intensity maps along an in-plane cut
$\bar{K} \bar{M} \bar{K}$ (parallel to the $k_y$ direction) taken
using $h\nu$ values of 22 eV, 29 eV and 35 eV, which correspond to
approximate $k_z$ values of L$_z -$ 0.3 \AA$^{-1}$, L$_z$, and L$_z$
+ 0.3 \AA$^{-1}$ respectively (Fig.2b). At $h\nu$ = 29 eV, the low
energy ARPES spectral weight reveals a clear $\Lambda$-shaped band
close to $E_F$. As the photon energy is either increased or
decreased from 29 eV, this intensity shifts to higher binding
energies as the spectral weight evolves from the $\Lambda$-shaped
into a $\cup$-shaped band. Therefore the dispersive peak in Fig.2a
comes from the bulk valence band, and for $h\nu$ = 29 eV the high
symmetry point L = (0.8, 0, 2.9) appears in the third bulk BZ. In
the maps of Fig.2b with respective $h\nu$ values of 22 eV and 35 eV,
overall weak features near $E_F$ that vary in intensity remain even
as the bulk valence band moves far below $E_F$. The survival of
these weak features over a large photon energy range (17 to 55 eV)
supports their surface origin. The non-dispersing feature centered
near $-0.3$ eV in Fig.2a comes from the higher binding energy
(valence band) part of the full spectrum of surface states, and the
weak non-dispersing peak at $-0.02$ eV reflects the low energy part
of the surface states that cross $E_F$ away from the $\bar{M}$ point
and forms the surface Fermi surface (Fig.2c).

\begin{figure*}
\includegraphics[scale=0.58,clip=true, viewport=0.0in 0.5in 11.3in 6.5in]{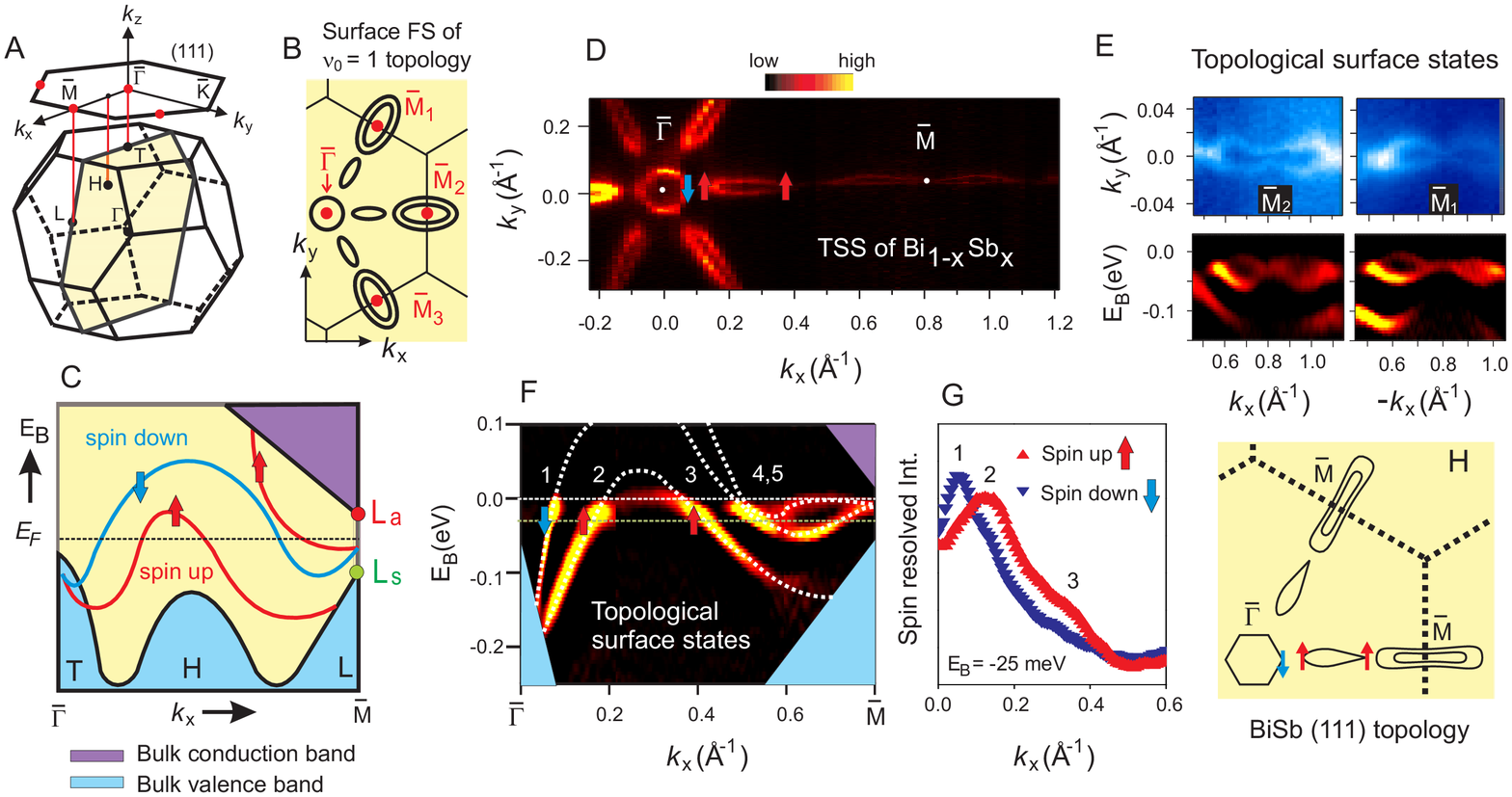}
\caption{\label{Sb_Fig1} \textbf{Spin texture of a topological insulator encodes Z$_2$ topological order of the bulk} (A) Schematic
sketches of the bulk Brillouin zone (BZ) and (111) surface BZ of the
Bi$_{1-x}$Sb$_x$ crystal series. The high symmetry points
(L,H,T,$\Gamma$,$\bar{\Gamma}$,\={M},\={K}) are identified. (B)
Schematic of Fermi surface pockets formed by the surface states (SS)
of a topological insulator that carries a Berry's phase. (C) Partner
switching band structure topology: Schematic of spin-polarized SS
dispersion and connectivity between $\bar{\Gamma}$ and \={M}
required to realize the FS pockets shown in panel-(B). $L_a$ and
$L_s$ label bulk states at $L$ that are antisymmetric and symmetric
respectively under a parity transformation (see text). (D)
Spin-integrated ARPES intensity map of the SS of
Bi$_{0.91}$Sb$_{0.09}$ at $E_F$. Arrows point in the measured
direction of the spin. (E) High resolution ARPES intensity map of
the SS at $E_F$ that enclose the \={M}$_1$ and \={M}$_2$ points.
Corresponding band dispersion (second derivative images) are shown
below. The left right asymmetry of the band dispersions are due to
the slight offset of the alignment from the
$\bar{\Gamma}$-\={M}$_1$(\={M}$_2$) direction. (F) Surface band
dispersion image along the $\bar{\Gamma}$-\={M} direction showing
five Fermi level crossings. The intensity of bands 4,5 is scaled up
for clarity (the dashed white lines are guides to the eye). The
schematic projection of the bulk valence and conduction bands are
shown in shaded blue and purple areas. (G) Spin-resolved momentum
distribution curves presented at $E_B$ = $-$25 meV showing single
spin degeneracy of bands at 1, 2 and 3. Spin up and down correspond
to spin pointing along the +$\hat{y}$ and -$\hat{y}$ direction
respectively. (H) Schematic of the spin-polarized surface FS
observed in our experiments. It is consistent with a $\nu_0$ = 1
topology (compare (B) and (H)). [Adapted from D. Hsieh $et$ $al.$, \textit{Science} \textbf{323}, 919 (2009) (Completed and Submitted in 2008)\cite{Science}].}
\end{figure*}

\section{Winding number count: Counting of surface Fermi surfaces enclosing Kramers points to identify topologically non-trivial surface spin-textured states}

Having established the existence of an energy gap in the bulk state
of Bi$_{0.9}$Sb$_{0.1}$ (Figs 1 and 2) and observed linearly
dispersive bulk bands uniquely consistent with strong spin-orbit
coupling model calculations \cite{Wolff, Fukuyama, Buot, Liu} (see
SM for full comparison with theoretical
calculation), we now discuss the topological character of its
surface states, which are found to be gapless (Fig.2c). In general,
the states at the surface of spin-orbit coupled compounds are
allowed to be spin split owing to the loss of space inversion
symmetry $[E(k,\uparrow) = E(-k,\uparrow)]$. However, as required by
Kramers' theorem, this splitting must go to zero at the four time
reversal invariant momenta (TRIM) in the 2D surface BZ. As discussed
in \cite{11, Fu:STI2}, along a path connecting two TRIM in the
same BZ, the Fermi energy inside the bulk gap will intersect these
singly degenerate surface states either an even or odd number of
times. When there are an even number of surface state crossings, the
surface states are topologically trivial because weak disorder (as
may arise through alloying) or correlations can remove \emph{pairs}
of such crossings by pushing the surface bands entirely above or
below $E_F$. When there are an odd number of crossings, however, at
least one surface state must remain gapless, which makes it
non-trivial \cite{11, Murakami, Fu:STI2}. The existence of such
topologically non-trivial surface states can be theoretically
predicted on the basis of the \emph{bulk} band structure only, using
the $Z_2$ invariant that is related to the quantum Hall Chern number
\cite{14}. Materials with band structures with $Z_2 = +1$
are ordinary Bloch band insulators that are topologically equivalent
to the filled shell atomic insulator, and are predicted to exhibit
an even number (including zero) of surface state crossings.
Materials with bulk band structures with $Z_2 = -1$ on the other
hand, which are expected to exist in rare systems with strong
spin-orbit coupling acting as an internal quantizing magnetic field
on the electron system \cite{Haldane(P-anomaly)}, and inverted bands
at an odd number of high symmetry points in their bulk 3D BZs, are
predicted to exhibit an odd number of surface state crossings,
precluding their adiabatic continuation to the atomic insulator
\cite{11, Murakami, Fu:STI2, 15,
8, 7}. Such ``topological quantum Hall metals''
\cite{Fu:STI2, 15} cannot be realized in a purely 2D
electron gas system such as the one realized at the interface of
GaAs/GaAlAs systems.

\begin{figure}
\includegraphics[scale=0.5,clip=true, viewport=0.0in 1.7in 6.8in 7.8in]{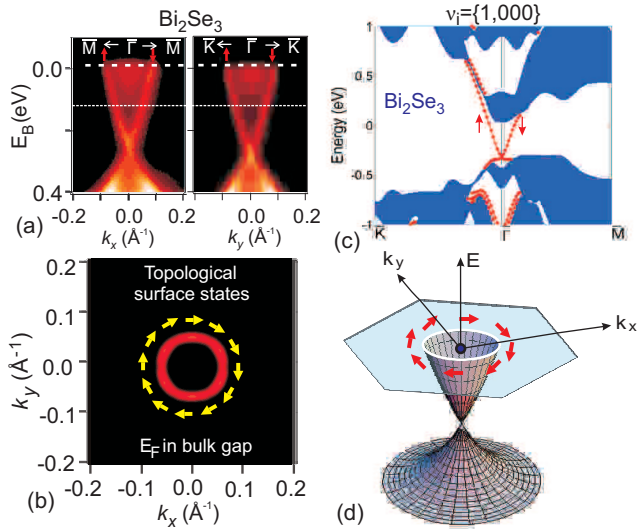}
\caption{\label{RMP_Fig12} \textbf{Spin-momentum locked helical fermions reveal topological-order of the bulk:} Spin-momentum locked helical
surface Dirac fermions are hallmark signatures of topological
insulators. (a) ARPES data for Bi$_2$Se$_3$ reveals surface electronic
states with a single spin-polarized Dirac cone. The
surface Fermi surface (b) exhibits a chiral left-handed spin
texture. Data is from a sample doped to a Fermi level [thin line in (a)] that is in the bulk gap. (c) Surface electronic structure of Bi$_2$Se$_3$ computed
in the local density approximation. The shaded regions describe
bulk states, and the red lines are surface states. (d)
Schematic of the spin polarized surface state dispersion in
the Bi$_2$X$_3$ topological insulators. [Adapted from Y. Xia $et$ $al$., \textit{Nature Phys.} \textbf{5}, 398 (2009).(Completed and Submitted in 2008) \cite{Xia}].}
\end{figure}

In our experimental case, namely the (111) surface of
Bi$_{0.9}$Sb$_{0.1}$, the four TRIM are located at $\bar{\Gamma}$
and three $\bar{M}$ points that are rotated by $60^{\circ}$ relative
to one another. Owing to the three-fold crystal symmetry (A7 bulk
structure) and the observed mirror symmetry of the surface Fermi
surface across $k_x = 0$ (Fig.2), these three $\bar{M}$ points are
equivalent (and we henceforth refer to them as a single point,
$\bar{M}$). The mirror symmetry $[E(k_y) = E(-k_y)]$ is also
expected from time reversal invariance exhibited by the system. The
complete details of the surface state dispersion observed in our
experiments along a path connecting $\bar{\Gamma}$ and $\bar{M}$ are
shown in Fig.3a; finding this information is made possible by our
experimental separation of surface states from bulk states. As for
bismuth (Bi), two surface bands emerge from the bulk band continuum
near $\bar{\Gamma}$ to form a central electron pocket and an
adjacent hole lobe \cite{Ast:Bi1, Hochst,Hofmann}. It has been
established that these two bands result from the spin-splitting of a
surface state and are thus singly degenerate \cite{Hirahara,
Hofmann}. On the other hand, the surface band that crosses $E_F$ at
$-k_x \approx 0.5$ \AA$^{-1}$, and forms the narrow electron pocket
around $\bar{M}$, is clearly doubly degenerate, as far as we can
determine within our experimental resolution. This is indicated by
its splitting below $E_F$ between $-k_x \approx 0.55$ \AA$^{-1}$ and
$\bar{M}$, as well as the fact that this splitting goes to zero at
$\bar{M}$ in accordance with Kramers theorem. In semimetallic single
crystal bismuth, only a single surface band is observed to form the
electron pocket around $\bar{M}$ \cite{Hengsberger, Ast:Bi2}.
Moreover, this surface state overlaps, hence becomes degenerate
with, the bulk conduction band at L (L projects to the surface
$\bar{M}$ point) owing to the semimetallic character of Bi (Fig.3b).
In Bi$_{0.9}$Sb$_{0.1}$ on the other hand, the states near $\bar{M}$
fall completely inside the bulk energy gap preserving their purely
surface character at $\bar{M}$ (Fig.3a). The surface Kramers doublet
point can thus be defined in the bulk insulator (unlike in Bi
\cite{Hirahara,Ast:Bi1, Hochst, Hofmann, Hengsberger, Ast:Bi2}) and
is experimentally located in Bi$_{0.9}$Sb$_{0.1}$ samples to lie
approximately 15 $\pm$ 5 meV below $E_F$ at $\vec{k} = \bar{M}$
(Fig.3a). For the precise location of this Kramers point, it is
important to demonstrate that our alignment is strictly along the
$\bar{\Gamma} - \bar{M}$ line. To do so, we contrast high resolution
ARPES measurements taken along the $\bar{\Gamma} - \bar{M}$ line
with those that are slightly offset from it (Fig.3e). Figs 3f-i show
that with $k_y$ offset from the Kramers point at $\bar{M}$ by less
than 0.02 \AA$^{-1}$, the degeneracy is lifted and only one band
crosses $E_F$ to form part of the bow-shaped electron distribution
(Fig.3d). Our finding of five surface state crossings (an odd rather
than an even number) between $\bar{\Gamma}$ and $\bar{M}$ (Fig.3a),
confirmed by our observation of the Kramers degenerate point at the
TRIM, indicates that these gapless surface states are topologically
non-trivial. This corroborates our bulk electronic structure result
that Bi$_{0.9}$Sb$_{0.1}$ is in the insulating band-inverted ($Z_2 =
-1$) regime (Fig.3c), which contains an odd number of bulk (gapped)
Dirac points, and is topologically analogous to an integer quantum
spin Hall insulator.

\begin{figure*}
\includegraphics[scale=0.6,clip=true, viewport=0.0in 0.7in 11.0in 5.4in]{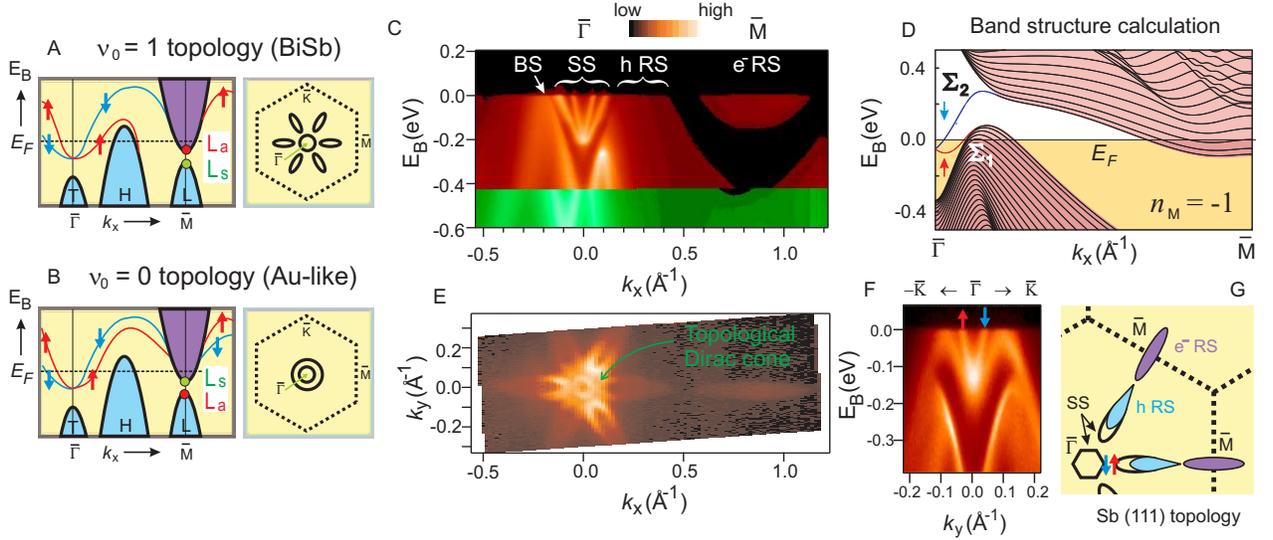}
\caption{\label{Sb_Fig2} \textbf{Topological character of parent compound revealed on the (111) surface states.} Schematic of the bulk band structure (shaded
areas) and surface band structure (red and blue lines) of Sb near
$E_F$ for a (A) topologically non-trivial and (B) topological
trivial (gold-like) case, together with their corresponding surface
Fermi surfaces are shown. (C) Spin-integrated ARPES spectrum of
Sb(111) along the $\bar{\Gamma}$-\={M} direction. The surface states
are denoted by SS, bulk states by BS, and the hole-like resonance
states and electron-like resonance states by h RS and e$^-$ RS
respectively. (D) Calculated surface state band structure of Sb(111)
based on the methods in \cite{20,Liu}. The continuum bulk energy bands are
represented with pink shaded regions, and the lines show the
discrete bands of a 100 layer slab. The red and blue single bands,
denoted $\Sigma_1$ and $\Sigma_2$, are the surface states bands with
spin polarization $\langle \vec{P} \rangle \propto +\hat{y}$ and
$\langle \vec{P} \rangle \propto -\hat{y}$ respectively. (E) ARPES
intensity map of Sb(111) at $E_F$ in the $k_x$-$k_y$ plane. The only
one FS encircling $\bar{\Gamma}$ seen in the data is formed by the
inner V-shaped SS band seen in panel-(C) and (F). The outer V-shaped
band bends back towards the bulk band best seen in data in
panel-(F). (F) ARPES spectrum of Sb(111) along the
$\bar{\Gamma}$-\={K} direction shows that the outer V-shaped SS band
merges with the bulk band. (G) Schematic of the surface FS of
Sb(111) showing the pockets formed by the surface states (unfilled)
and the resonant states (blue and purple). The purely surface state
Fermi pocket encloses only one Kramers degenerate point
($\vec{k}_T$), namely, $\bar{\Gamma}$(=$\vec{k}_T$), therefore
consistent with the $\nu_0$ = 1 topological classification of Sb
which is different from Au (compare (B) and (G)). As discussed in
the text, the hRS and e$^-$RS count trivially. [Adapted from D. Hsieh $et$ $al.$, \textit{Science} \textbf{323}, 919 (2009) (Completed and Submitted in 2008) \cite{Science}].}
\end{figure*}

Our experimental results taken collectively strongly suggest that
Bi$_{0.9}$Sb$_{0.1}$ is quite distinct from graphene \cite{Zhang,
Novoselov} and represents a novel state of quantum matter: a
strongly spin-orbit coupled insulator with an odd number of Dirac
points with a negative $Z_2$ topological Hall phase, which realizes
the ``parity anomaly without Fermion doubling". Our work further
demonstrates a general methodology for possible future
investigations of \emph{novel topological orders} in exotic quantum
matter.

\section{Spin-resolving the surface states to identify the non-trivial topological phase and establish a 2D helical metal protected from backscattering}

\begin{figure}
\includegraphics[scale=0.35,clip=true, viewport=0.0in 0.0in 10.3in 8.65in]{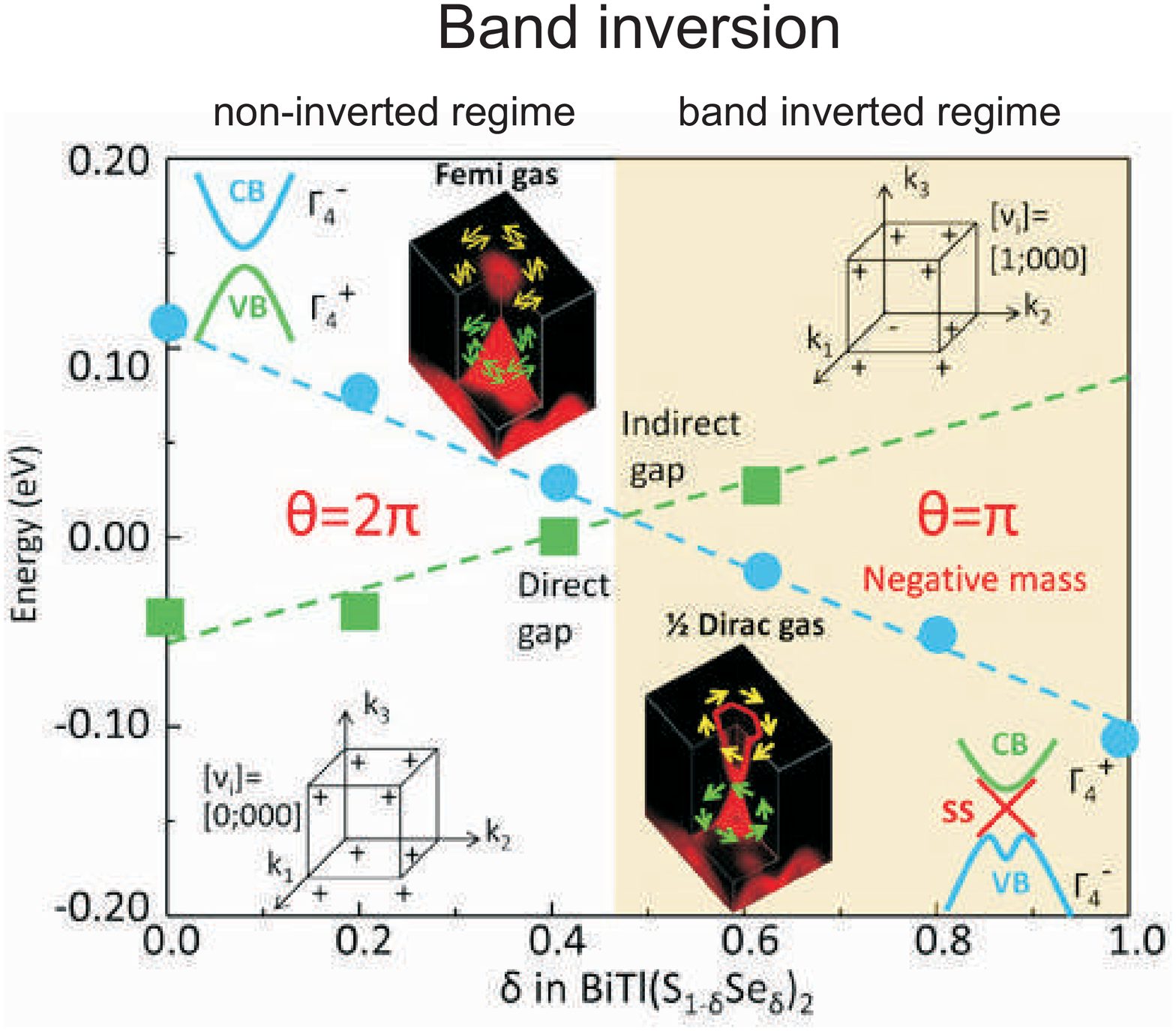}
\label{QPT}
\caption{\textbf{Observation of bulk band inversion across a topological quantum phase transition}. Energy levels of $\Gamma_4^-$. (blue circles) and $\Gamma_4^+$ (green squares) bands are obtained from ARPES measurements as a function of composition $\delta$. CB: conduction band; VB: valence
band. Parity eigenvalues (+ or –) of Bloch states are
shown. The topological invariants, $\nu_i$, obtained from the
parity eigenvalues are presented as [$\theta$/$\pi$ = $\nu_0$;$\nu_1$,$\nu_2$,$\nu_3$] where $\theta$ =
$\pi \nu_0$ is the axion angle \cite{12} and $\nu_0$ is the strong invariant. [Adapted from S.-Y. Xu $et$ $al.$, \textit{Science} \textbf{332} 560 (2011). \cite{Xu}]}
\end{figure}

Strong topological materials are distinguished from ordinary
materials such as gold by a topological quantum number, $\nu_0$ = 1
or 0 respectively \cite{14,15}. For Bi$_{1-x}$Sb$_x$, theory has shown
that $\nu_0$ is determined solely by the character of the bulk
electronic wave functions at the $L$ point in the three-dimensional
(3D) Brillouin zone (BZ). When the lowest energy conduction band
state is composed of an antisymmetric combination of atomic $p$-type
orbitals ($L_a$) and the highest energy valence band state is
composed of a symmetric combination ($L_s$), then $\nu_0$ = 1, and
vice versa for $\nu_0$ = 0 \cite{11}. Although the bonding nature
(parity) of the states at $L$ is not revealed in a measurement of
the bulk band structure, the value of $\nu_0$ can be determined from
the spin-textures of the surface bands that form when the bulk is
terminated. In particular, a $\nu_0$ = 1 topology requires the
terminated surface to have a Fermi surface (FS) that supports a
non-zero Berry's phase (odd as opposed to even multiple of $\pi$),
which is not realizable in an ordinary spin-orbit material.

\begin{figure*}
\includegraphics[scale=0.62,clip=true, viewport=0.0in 0in 11.0in 6.0in]{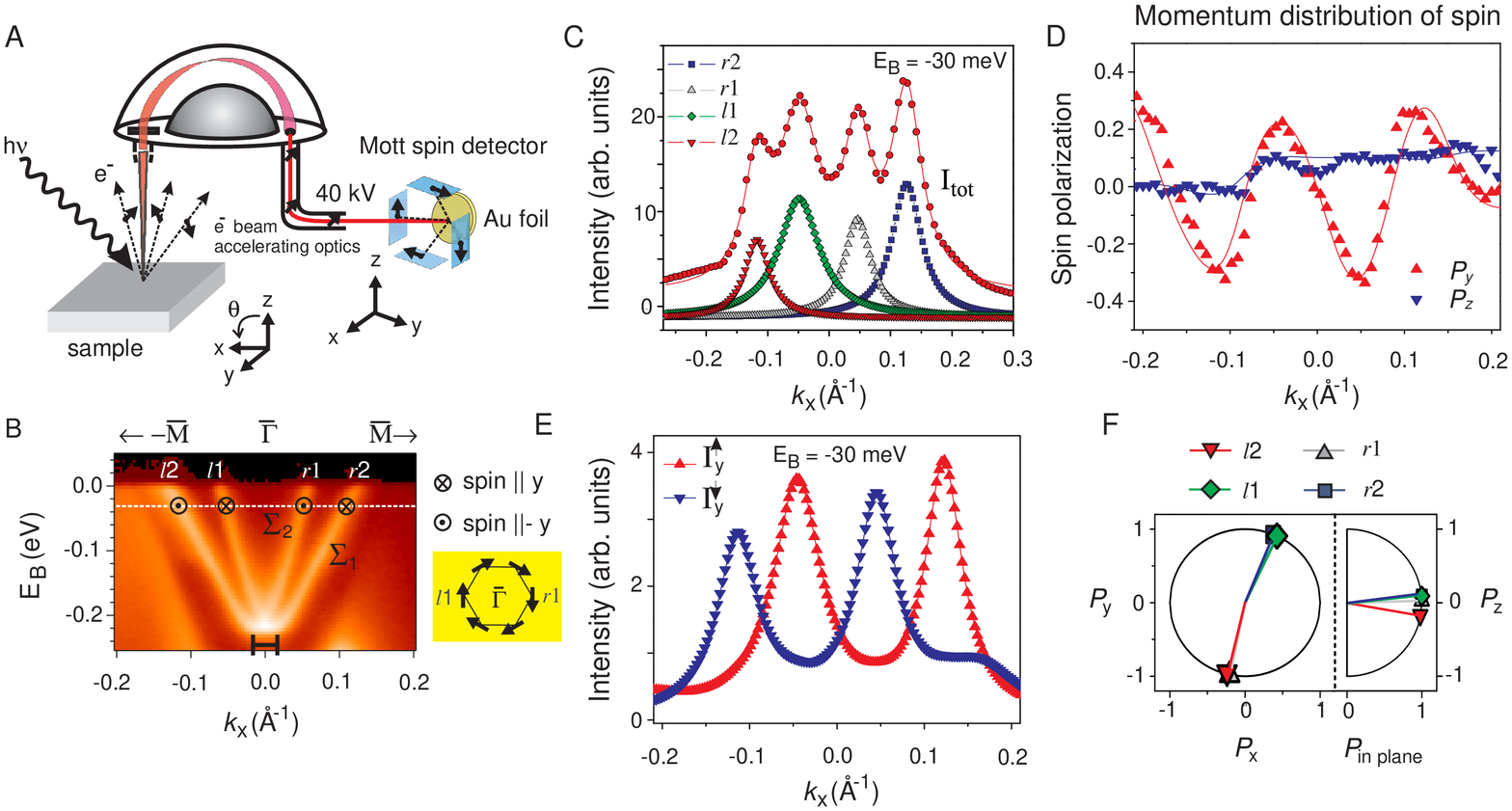}
\caption{\label{Sb_Fig3} \textbf{Spin-texture of topological surface states and topological chirality.} (A) Experimental geometry of the spin-resolved ARPES
study. At normal emission ($\theta$ = 0$^{\circ}$), the sensitive
$y'$-axis of the Mott detector is rotated by 45$^{\circ}$ from the
sample $\bar{\Gamma}$ to $-$\={M} ($\parallel -\hat{x}$) direction,
and the sensitive $z'$-axis of the Mott detector is parallel to the
sample normal ($\parallel \hat{z}$). (B) Spin-integrated ARPES
spectrum of Sb(111) along the $-$\={M}-$\bar{\Gamma}$-\={M}
direction. The momentum splitting between the band minima is
indicated by the black bar and is approximately 0.03 \AA$^{-1}$. A
schematic of the spin chirality of the central FS based on the
spin-resolved ARPES results is shown on the right. (C) Momentum
distribution curve of the spin averaged spectrum at $E_B$ = $-$30
meV (shown in (B) by white line), together with the Lorentzian peaks
of the fit. (D) Measured spin polarization curves (symbols) for the
detector $y'$ and $z'$ components together with the fitted lines
using the two-step fitting routine \cite{26}. (E) Spin-resolved spectra
for the sample $y$ component based on the fitted spin polarization
curves shown in (D). Up (down) triangles represent a spin direction
along the +(-)$\hat{y}$ direction. (F) The in-plane and out-of-plane
spin polarization components in the sample coordinate frame obtained
from the spin polarization fit. Overall spin-resolved data and the
fact that the surface band that forms the central electron pocket
has $\langle \vec{P} \rangle \propto -\hat{y}$ along the +$k_x$
direction, as in (E), suggest a left-handed chirality (schematic in
(B) and see text for details). [Adapted from D. Hsieh $et$ $al.$, \textit{Science} \textbf{323}, 919 (2009) \cite{Science}].}
\end{figure*}

In a general inversion symmetric spin-orbit insulator, the bulk
states are spin degenerate because of a combination of space
inversion symmetry $[E(\vec{k},\uparrow) = E(-\vec{k},\uparrow)]$
and time reversal symmetry $[E(\vec{k},\uparrow) =
E(-\vec{k},\downarrow)]$. Because space inversion symmetry is broken
at the terminated surface, the spin degeneracy of surface bands can
be lifted by the spin-orbit interaction [19-21]. However, according
to Kramers theorem [16], they must remain spin degenerate at four
special time reversal invariant momenta ($\vec{k}_T$ =
$\bar{\Gamma}$, \={M}) in the surface BZ [11], which for the (111)
surface of Bi$_{1-x}$Sb$_x$ are located at $\bar{\Gamma}$ and three
equivalent \={M} points [see Fig.\ref{Sb_Fig1}(A)].

\begin{figure*}
\includegraphics[scale=0.7,clip=true, viewport=-0.0in 0in 11.0in 5.5in]{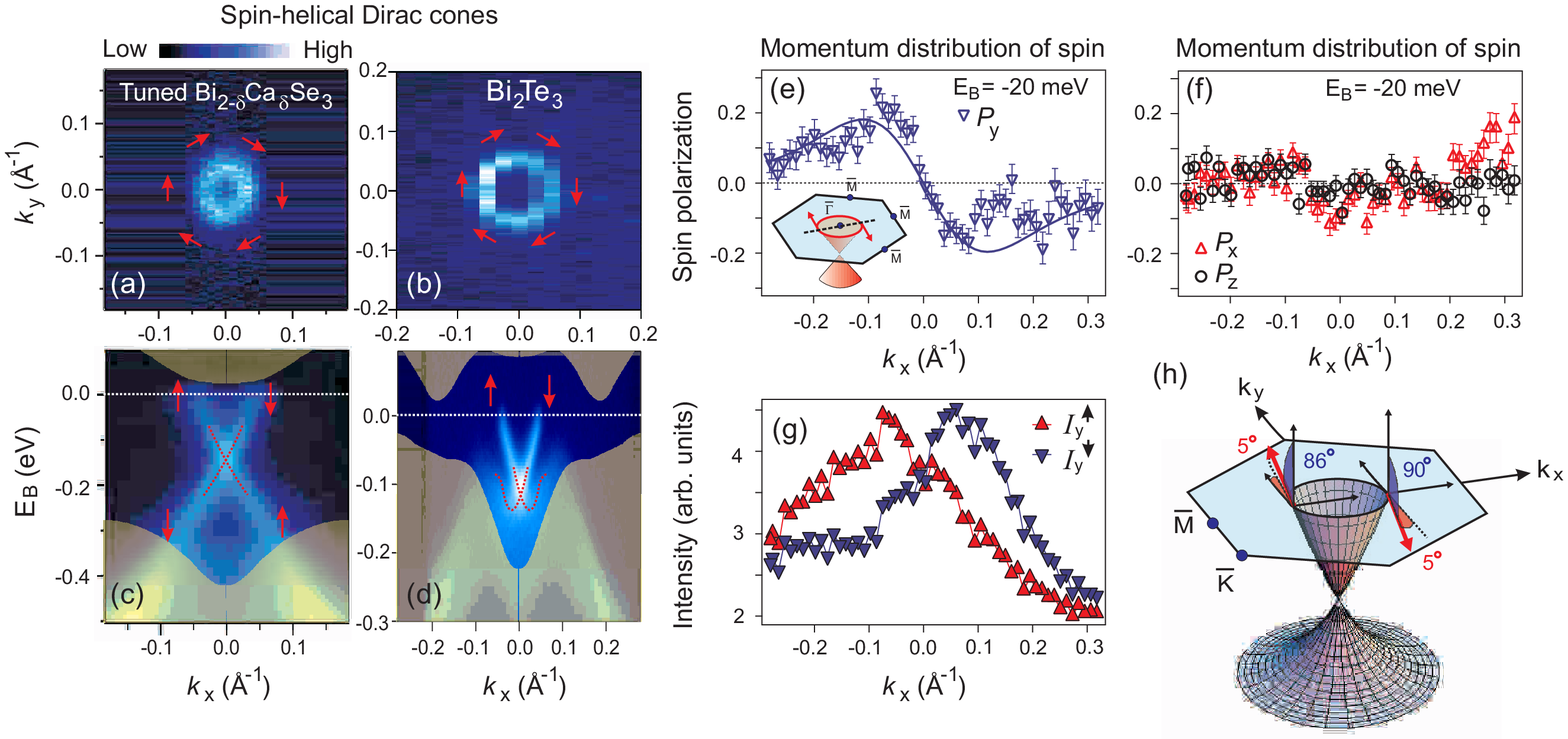}
\caption{\label{fig:Fig1} \textbf{First detection of Topological-Order: spin-momentum locking of spin-helical Dirac electrons in Bi$_2$Se$_3$ and Bi$_2$Te$_3$ using spin-resolved ARPES.} (a) ARPES intensity map at $E_F$ of the (111)
surface of tuned Bi$_{2-\delta}$Ca$_{\delta}$Se$_3$ (see text) and
(b) the (111) surface of Bi$_2$Te$_3$. Red arrows denote the
direction of spin around the Fermi surface. (c) ARPES dispersion of
tuned Bi$_{2-\delta}$Ca$_{\delta}$Se$_3$ and (d) Bi$_2$Te$_3$ along
the $k_x$ cut. The dotted red lines are guides to the eye. The
shaded regions in (c) and (d) are our calculated projections of the
bulk bands of pure Bi$_2$Se$_3$ and Bi$_2$Te$_3$ onto the (111)
surface respectively. (e) Measured $y$ component of
spin-polarization along the $\bar{\Gamma}$-\={M} direction at $E_B$
= -20 meV, which only cuts through the surface states. Inset shows a
schematic of the cut direction. (f) Measured $x$ (red triangles) and
$z$ (black circles) components of spin-polarization along the
$\bar{\Gamma}$-\={M} direction at $E_B$ = -20 meV. Error bars in (e)
and (f) denote the standard deviation of $P_{x,y,z}$, where typical
detector counts reach $5\times10^5$; Solid lines are numerical fits
\cite{21}. (g) Spin-resolved spectra obtained from the $y$ component
spin polarization data. The non-Lorentzian lineshape of the
$I_y^{\uparrow}$ and $I_y^{\downarrow}$ curves and their non-exact
merger at large $|k_{x}|$ is due to the time evolution of the
surface band dispersion, which is the dominant source of statistical
uncertainty. a.u., arbitrary units. (h) Fitted values of the spin
polarization vector \textbf{P} = ($S_x$,$S_y$,$S_z$) are
(sin(90$^{\circ}$)cos(-95$^{\circ}$),
sin(90$^{\circ}$)sin(-95$^{\circ}$), cos(90$^{\circ}$)) for
electrons with +$k_x$ and (sin(86$^{\circ}$)cos(85$^{\circ}$),
sin(86$^{\circ}$)sin(85$^{\circ}$), cos(86$^{\circ}$)) for electrons
with -$k_x$, which demonstrates the topological helicity of the
spin-Dirac cone. The angular uncertainties are of order
$\pm$10$^{\circ}$ and the magnitude uncertainty is of order
$\pm$0.15. [Adapted from D. Hsieh $et$ $al.$, \textit{Nature} \textbf{460}, 1101 (2009). \cite{Nature_2009}].}
\end{figure*}

Depending on whether $\nu_0$ equals 0 or 1, the Fermi surface
pockets formed by the surface bands will enclose the four
$\vec{k}_T$ an even or odd number of times respectively. If a Fermi
surface pocket does not enclose $\vec{k}_T$ (= $\bar{\Gamma}$,
\={M}), it is irrelevant for the topology \cite{11,20}. Because the wave
function of a single electron spin acquires a geometric phase factor
of $\pi$ \cite{16,17} as it evolves by 360$^{\circ}$ in momentum space along
a Fermi contour enclosing a $\vec{k}_T$, an odd number of Fermi
pockets enclosing $\vec{k}_T$ in total implies a $\pi$ geometrical
(Berry's) phase \cite{11}. In order to realize a $\pi$ Berry's phase the
surface bands must be spin-polarized and exhibit a partner switching
\cite{11} dispersion behavior between a pair of $\vec{k}_T$. This means
that any pair of spin-polarized surface bands that are degenerate at
$\bar{\Gamma}$ must not re-connect at \={M}, or must separately
connect to the bulk valence and conduction band in between
$\bar{\Gamma}$ and \={M}. The partner switching behavior is realized
in Fig. \ref{Sb_Fig1}(C) because the spin down band connects to and is
degenerate with different spin up bands at $\bar{\Gamma}$ and \={M}.
The partner switching behavior is realized in Fig. \ref{Sb_Fig2}(A) because the
spin up and spin down bands emerging from $\bar{\Gamma}$ separately
merge into the bulk valence and conduction bands respectively
between $\bar{\Gamma}$ and \={M}.

\begin{figure*}
\includegraphics[scale=0.55,clip=true, viewport=0.0in 0.0in 10.0in 8.5in]{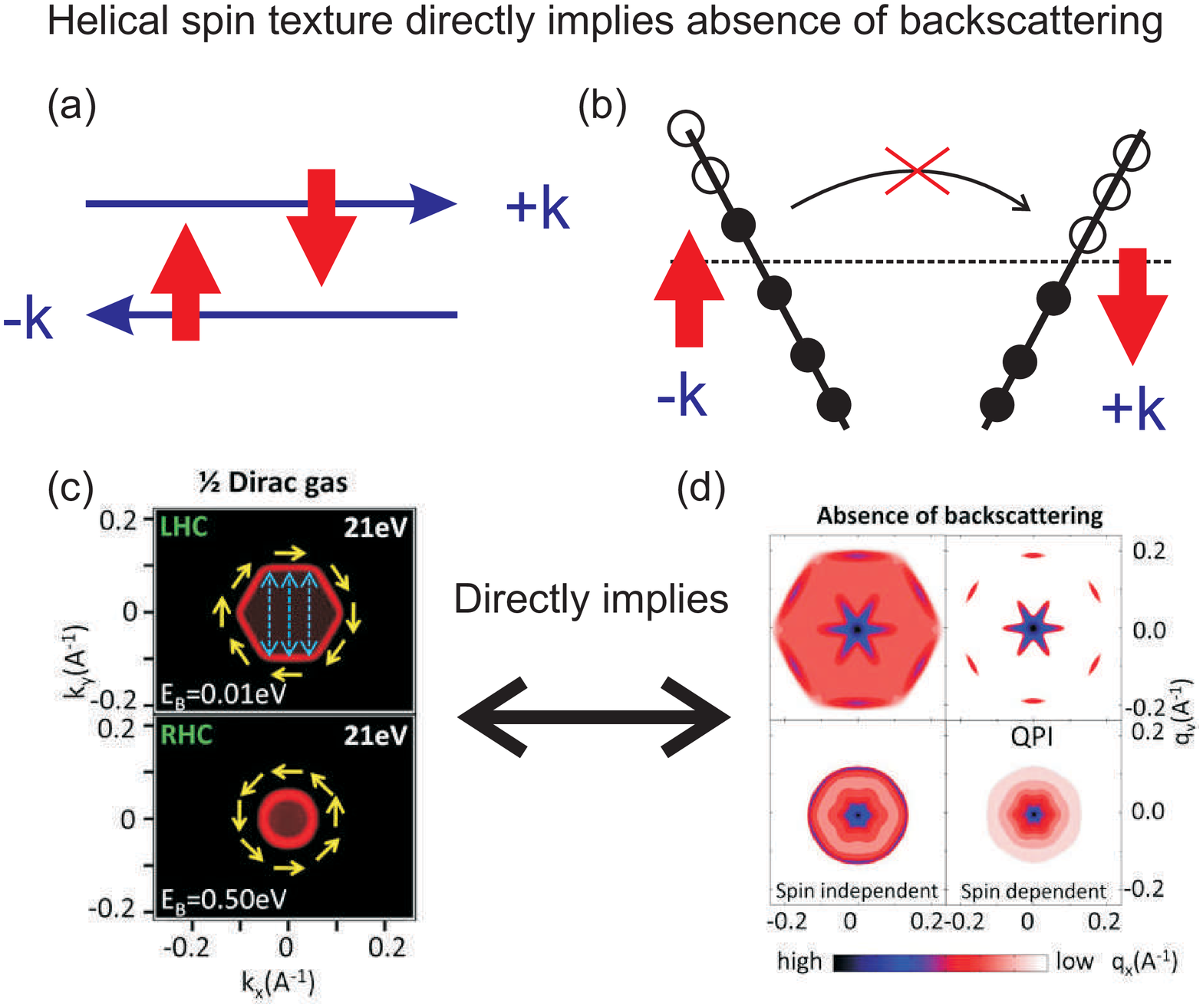}
\caption{\label{backscatter} \textbf{Helical spin texture naturally leads to absence of elastic backscattering for surface transport: No "U" turn on a 3D topological insulator surface.} (a) Our measurement of a helical spin texture in both Bi$_{1-x}$Sb$_x$ and in Bi$_2$Se$_3$ directly shows that there is (b) an absence of backscattering. (c) ARPES measured FSs are shown with spin
directions based on polarization measurements. L(R)HC
stands for left(right)-handed chirality. (d) Spin independent and spin dependent scattering
profiles on FSs in (c) relevant for surface quasi-particle
transport are shown which is sampled by the quasi-particle
interference (QPI) modes. [Adapted from S.-Y. Xu $et$ $al.$, \textit{Science} \textbf{332} 560 (2011). \cite{Xu}]}
\end{figure*}

We first investigate the spin properties of the topological
insulator phase \cite{Science}. Spin-integrated ARPES \cite{19} intensity maps of the
(111) surface states of insulating Bi$_{1-x}$Sb$_x$ taken at the
Fermi level ($E_F$) [Figs \ref{Sb_Fig1}(D)\&(E)] show that a hexagonal FS
encloses $\bar{\Gamma}$, while dumbbell shaped FS pockets that are
much weaker in intensity enclose \={M}. By examining the surface
band dispersion below the Fermi level [Fig.\ref{Sb_Fig1}(F)] it is clear that
the central hexagonal FS is formed by a single band (Fermi crossing
1) whereas the dumbbell shaped FSs are formed by the merger of two
bands (Fermi crossings 4 and 5) \cite{10}.

\begin{figure}
\includegraphics[scale=1,clip=true, viewport=0.0in 0.0in 5.3in 3.5in]{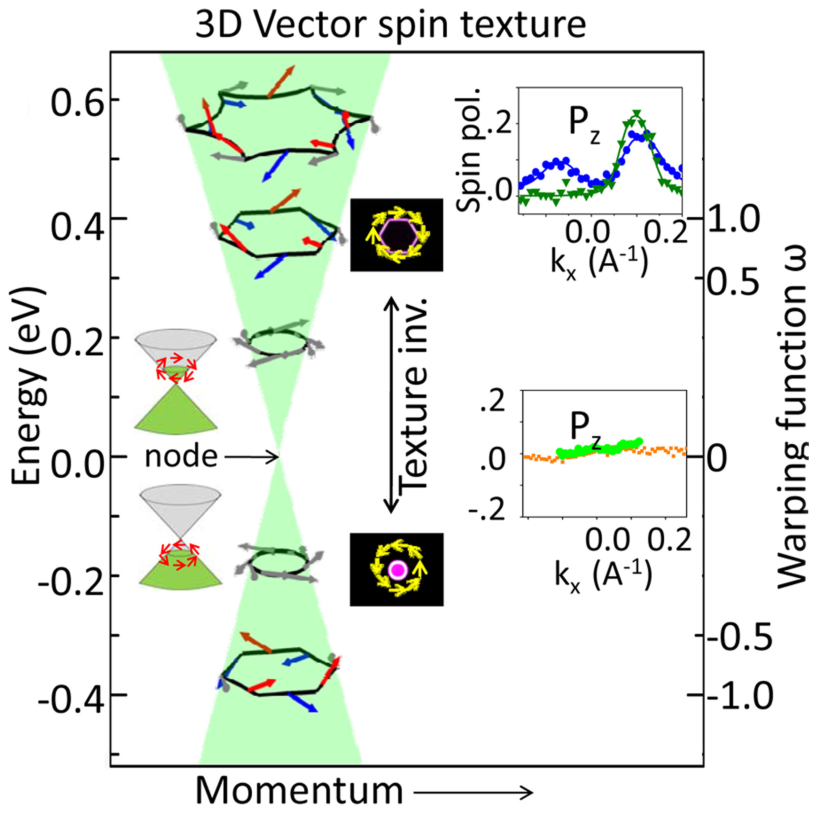}
\caption{\label{texture} \textbf{Spin texture evolution of topological surface bands as a function of energy away from the Dirac node} (left axis) and geometrical warping factor $\omega$ (right axis). The warping factor is defined as $\omega = \frac{k_F(\bar{\Gamma}-\bar{M})-k_F(\bar{\Gamma}-\bar{K})}{k_F(\bar{\Gamma}-\bar{M})+k_F(\bar{\Gamma}-\bar{K})}\times\frac{2+\sqrt{3}}{2-\sqrt{3}}$ where $\omega$ = 0, $\omega$ = 1, and $\omega$ $>$ 1 implies circular, hexagonal and snowflake-shaped FSs respectively. The sign of $\omega$ indicates texture chirality for LHC (+) or RHC (-). Insets: out-of-plane 3D spin-polarization measurements at corresponding FSs. [Adapted from S.-Y. Xu $et$ $al.$, \textit{Science} \textbf{332} 560 (2011). \cite{Xu}]}
\end{figure}

This band dispersion resembles the partner switching dispersion
behavior characteristic of topological insulators. To check this
scenario and determine the topological index $\nu_0$, we have
carried out spin-resolved photoemission spectroscopy. Fig.\ref{Sb_Fig1}(G) shows
a spin-resolved momentum distribution curve taken along the
$\bar{\Gamma}$-\={M} direction at a binding energy $E_B$ = $-$25 meV
[Fig.\ref{Sb_Fig1}(G)]. The data reveal a clear difference between the spin-up
and spin-down intensities of bands 1, 2 and 3, and show that bands 1
and 2 have opposite spin whereas bands 2 and 3 have the same spin
(detailed analysis discussed later in text). The former observation
confirms that bands 1 and 2 form a spin-orbit split pair, and the
latter observation suggests that bands 2 and 3 (as opposed to bands
1 and 3) are connected above the Fermi level and form one band. This
is further confirmed by directly imaging the bands through raising
the chemical potential via doping (SM). Irrelevance of bands 2 and 3 to the topology is
consistent with the fact that the Fermi surface pocket they form
does not enclose any $\vec{k}_T$. Because of a dramatic intrinsic
weakening of signal intensity near crossings 4 and 5, and the small
energy and momentum splitting of bands 4 and 5 lying at the
resolution limit of modern spin-resolved ARPES spectrometers, no
conclusive spin information about these two bands can be drawn from
the methods employed in obtaining the data sets in Figs \ref{Sb_Fig1}(G)\&(H).
However, whether bands 4 and 5 are both singly or doubly degenerate
does not change the fact that an odd number of spin-polarized FSs
enclose the $\vec{k}_T$, which provides evidence that
Bi$_{1-x}$Sb$_x$ has $\nu_0$ = 1 and that its surface supports a
non-trivial Berry's phase. This directly implies an absence of backscattering in electronic transport along the surface (Fig.\ref{backscatter}).

Shortly after the discovery of Bi$_{1-x}$Sb$_x$, physicists sought to find a simpler version of a 3D topological insulator consisting of a single surface state instead of five. This is because the surface structure of Bi$_{1-x}$Sb$_x$ was rather complicated
and the band gap was rather small. This motivated
a search for topological insulators with a larger band gap
and simpler surface spectrum. A second generation of
3D topological insulator materials, especially
Bi$_2$Se$_3$, offers the potential for topologically protected
behavior in ordinary crystals at room temperature
and zero magnetic field. Starting in 2008, work by the Princeton
group used spin-ARPES and first-principles calculations to
study the surface band structure of Bi$_2$Se$_3$ and observe
the characteristic signature of a topological insulator in
the form of a single Dirac cone that is spin-polarized (Figs \ref{RMP_Fig12} and \ref{fig:Fig1}) such that it also carries a non-trivial Berry's phase \cite{Xia,Nature_2009}. Concurrent theoretical work by \cite{Zhang_nphys} used electronic structure methods to show
that Bi$_2$Se$_3$ is just one of several new large band-gap
topological insulators. These other materials were soon after also identified using this ARPES technique we describe \cite{Chen,Hsieh_PRL}.

\begin{figure}
\includegraphics[scale=0.32,clip=true, viewport=0.0in 0in 11.0in 5.0in]{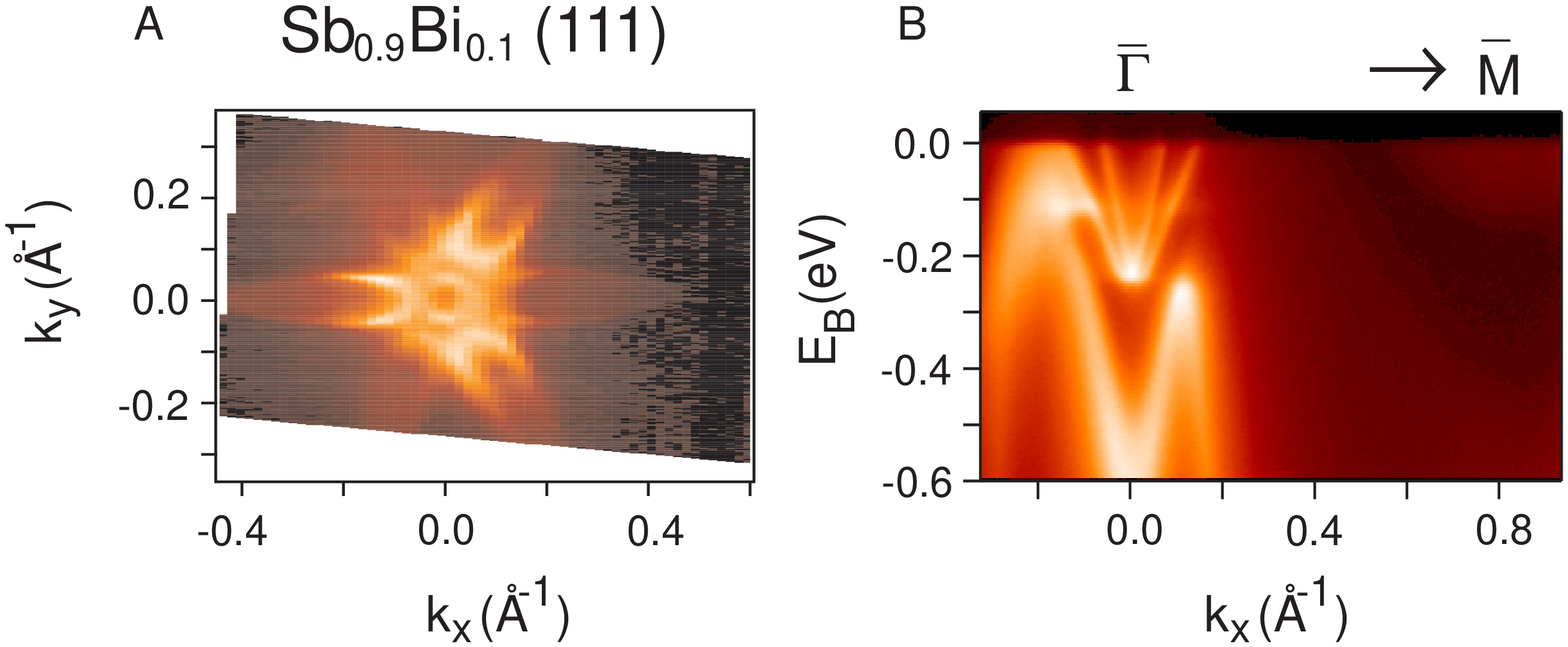}
\caption{\label{fig:Sb_FigS5} \textbf{Robustness against disorder} Spin split surface states survive alloying
disorder in Sb$_{0.9}$Bi$_{0.1}$. (\textbf{A}) ARPES intensity map
at $E_F$ of single crystal Sb$_{0.9}$Bi$_{0.1}$(111) in the
$k_x$-$k_y$ plane taken using 20 eV photons. (\textbf{B}) ARPES
intensity map of Sb$_{0.9}$Bi$_{0.1}$(111) along the
$\bar{\Gamma}$-\={M} direction taken with $h\nu$ = 22 eV photons.
The band dispersion is not symmetric about $\bar{\Gamma}$ because of
the three-fold rotational symmetry of the bulk states about the
$\langle111\rangle$ axis. [Adapted from D. Hsieh $et$ $al.$, \textit{Science} \textbf{323}, 919 (2009) \cite{Science}].}
\end{figure}

\begin{figure*}
\includegraphics[scale=0.84,clip=true, viewport=0.0in 0.0in 7.0in 5.8in]{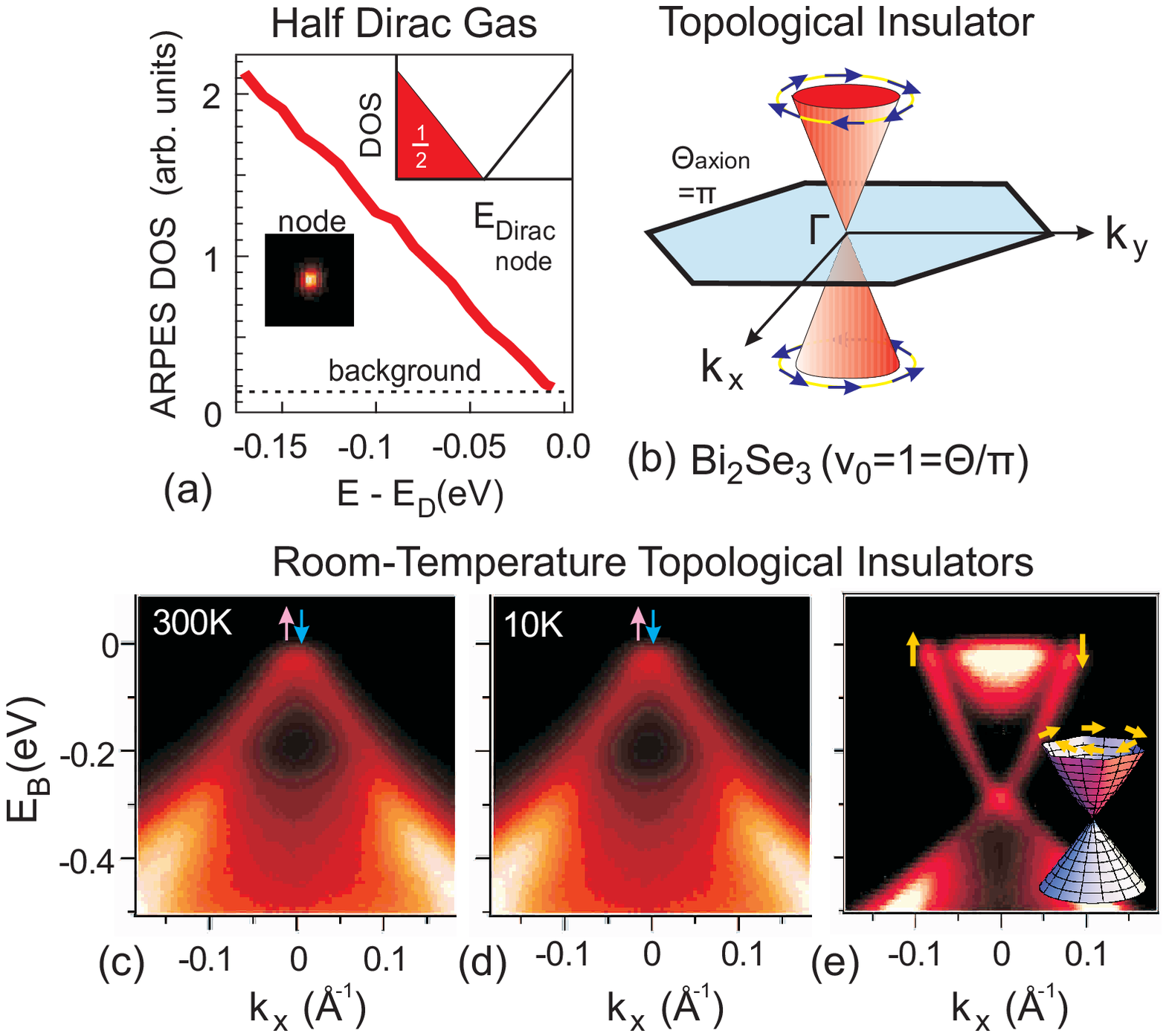}
\caption{\label{RMP_Fig13} \textbf{Observation of room temperature (300K) topological order without applied magnetic field in Bi$_2$Se$_3$:} (a)
Crystal momentum integrated ARPES data near Fermi level
exhibit linear fall-off of density of states, which, combined
with the spin-resolved nature of the states, suggest that a half
Fermi gas is realized on the topological surfaces. (b) Spin texture
map based on spin-ARPES data suggest that the spin-chirality changes sign across the Dirac point. (c) The Dirac
node remains well defined up a temperature of 300 K suggesting
the stability of topological effects up to the room temperature.
(d) The Dirac cone measured at a temperature of 10 K. (e) Full Dirac cone. [Adapted from D. Hsieh $et$ $al.$, \textit{Nature} \textbf{460}, 1101 (2009). \cite{Nature_2009}].}
\end{figure*}

The Bi$_2$Se$_3$ surface state is found from spin-ARPES and theory to be a nearly idealized single Dirac
cone as seen from the experimental data in Figs.\ref{RMP_Fig12} and \ref{RMP_Fig13}. An added advantage is that Bi$_2$Se$_3$ is stoichiometric (i.e., a pure
compound rather than an alloy such as Bi$_{1-x}$Sb$_x$) and
hence can be prepared, in principle, at higher purity.
While the topological insulator phase is predicted to be
quite robust to disorder, many experimental probes of
the phase, including ARPES of the surface band structure,
are clearer in high-purity samples. Finally and perhaps
most important for applications, Bi$_2$Se$_3$ has a large
band gap of around 0.3 eV (3600 K). This indicates that in its
high-purity form Bi$_2$Se$_3$ can exhibit topological insulator
behavior at room temperature and greatly increases
the potential for applications, which we discuss in greater depth later in the review.

\section{Identifying the origin of 3D topological order via a bulk band gap inversion transition}

We investigated the quantum origin of topological order in this
class of materials. It has been theoretically speculated that the
novel topological order originates from the parities of the
electrons in pure Sb and not Bi \cite{11,Lenoir}. It was also noted \cite{20} that
the origin of the topological effects can only be tested by
measuring the spin-texture of the Sb surface, which has not been
measured. Based on quantum oscillation and magneto-optical studies,
the bulk band structure of Sb is known to evolve from that of
insulating Bi$_{1-x}$Sb$_x$ through the hole-like band at H rising
above $E_F$ and the electron-like band at $L$ sinking below $E_F$ \cite{Lenoir}. The relative energy ordering of the $L_a$ and $L_s$ states in
Sb again determines whether the surface state pair emerging from
$\bar{\Gamma}$ switches partners [Fig.\ref{Sb_Fig2}(A)] or not [Fig.\ref{Sb_Fig2}(B)]
between $\bar{\Gamma}$ and \={M}, and in turn determines whether
they support a non-zero Berry's phase.

In a conventional spin-orbit metal such as gold, a free-electron
like surface state is split into two parabolic spin-polarized
sub-bands that are shifted in $\vec{k}$-space relative to each other \cite{18}. Two concentric spin-polarized Fermi surfaces are created, one
having an opposite sense of in-plane spin rotation from the other,
that enclose $\bar{\Gamma}$. Such a Fermi surface arrangement, like
the schematic shown in figure \ref{Sb_Fig2}(B), does not support a non-zero
Berry's phase because the $\vec{k}_T$ are enclosed an even number of
times (2 for most known materials).

However, for Sb, this is not the case. Figure \ref{Sb_Fig2}(C) shows a
spin-integrated ARPES intensity map of Sb(111) from $\bar{\Gamma}$
to \={M}. By performing a systematic incident photon energy
dependence study of such spectra, previously unavailable with He
lamp sources \cite{24}, it is possible to identify two V-shaped surface
states (SS) centered at $\bar{\Gamma}$, a bulk state located near
$k_x$ = $-$0.25 \AA$^{-1}$ and resonance states centered about $k_x$
= 0.25 \AA$^{-1}$ and \={M} that are hybrid states formed by surface
and bulk states \cite{19} (SM). An examination of the ARPES
intensity map of the Sb(111) surface and resonance states at $E_F$
[Fig.\ref{Sb_Fig2}(E)] reveals that the central surface FS enclosing
$\bar{\Gamma}$ is formed by the inner V-shaped SS only. The outer
V-shaped SS on the other hand forms part of a tear-drop shaped FS
that does \textit{not} enclose $\bar{\Gamma}$, unlike the case in
gold. This tear-drop shaped FS is formed partly by the outer
V-shaped SS and partly by the hole-like resonance state. The
electron-like resonance state FS enclosing \={M} does not affect the
determination of $\nu_0$ because it must be doubly spin degenerate
(SM). Such a FS geometry [Fig.\ref{Sb_Fig2}(G)] suggests that the
V-shaped SS pair may undergo a partner switching behavior expected
in Fig.\ref{Sb_Fig2}(A). This behavior is most clearly seen in a cut taken along
the $\bar{\Gamma}$-\={K} direction since the top of the bulk valence
band is well below $E_F$ [Fig.\ref{Sb_Fig2}(F)] showing only the inner V-shaped
SS crossing $E_F$ while the outer V-shaped SS bends back towards the
bulk valence band near $k_x$ = 0.1 \AA$^{-1}$ before reaching $E_F$.
The additional support for this band dispersion behavior comes from
tight binding surface calculations on Sb [Fig.\ref{Sb_Fig2}(D)], which closely
match with experimental data below $E_F$. Our observation of a
single surface band forming a FS enclosing $\bar{\Gamma}$ suggests
that pure Sb is likely described by $\nu_0$ = 1, and that its
surface may support a Berry's phase.

Confirmation of a surface $\pi$ Berry's phase rests critically on a
measurement of the relative spin orientations (up or down) of the SS
bands near $\bar{\Gamma}$ so that the partner switching is indeed
realized, which cannot be done without spin resolution. Spin
resolution was achieved using a Mott polarimeter that measures two
orthogonal spin components of a photoemitted electron \cite{27,28}. These
two components are along the $y'$ and $z'$ directions of the Mott
coordinate frame, which lie predominantly in and out of the sample
(111) plane respectively. Each of these two directions represents a
normal to a scattering plane defined by the photoelectron incidence
direction on a gold foil and two electron detectors mounted on
either side (left and right) [Fig.\ref{Sb_Fig3}(A)]. Strong spin-orbit coupling
of atomic gold is known to create an asymmetry in the scattering of
a photoelectron off the gold foil that depends on its spin component
normal to the scattering plane \cite{28}. This leads to an asymmetry
between the left intensity ($I^L_{y',z'}$) and right intensity
($I^R_{y',z'}$) given by $A_{y',z'} =
(I^L_{y',z'}-I^R_{y',z'})/(I^L_{y',z'}+I^R_{y',z'})$, which is
related to the spin polarization $P_{y',z'} = (1/S_{eff})\times
A_{y',z'}$ through the Sherman function $S_{eff}$ = 0.085 \cite{27,28}.
Spin-resolved momentum distribution curve data sets of the SS bands
along the $-$\={M}-$\bar{\Gamma}$-\={M} cut at $E_B$ = $-$30 meV
[Fig.\ref{Sb_Fig3}(B)] are shown for maximal intensity. Figure \ref{Sb_Fig3}(D) displays
both $y'$ and $z'$ polarization components along this cut, showing
clear evidence that the bands are spin polarized, with spins
pointing largely in the (111) plane. In order to estimate the full
3D spin polarization vectors from a two component measurement (which
is not required to prove the partner switching or the Berry's
phase), we fit a model polarization curve to our data following the
recent demonstration in Ref-\cite{26}, which takes the polarization
directions associated with each momentum distribution curve peak
[Fig.\ref{Sb_Fig3}(C)] as input parameters, with the constraint that each
polarization vector has length one (in angular momentum units of
$\hbar$/2). Our fitted polarization vectors are displayed in the
sample ($x,y,z$) coordinate frame [Fig.\ref{Sb_Fig3}(F)], from which we derive
the spin-resolved momentum distribution curves for the spin
components parallel ($I_y^{\uparrow}$) and anti-parallel
($I_y^{\downarrow}$) to the $y$ direction (SM) as shown
in figure \ref{Sb_Fig3}(E). There is a clear difference in $I_y^{\uparrow}$ and
$I_y^{\downarrow}$ at each of the four momentum distribution curve
peaks indicating that the surface state bands are spin polarized
[Fig.\ref{Sb_Fig3}(E)], which is possible to conclude even without a full 3D
fitting. Each of the pairs $l2/l1$ and $r1/r2$ have opposite spin,
consistent with the behavior of a spin split pair, and the spin
polarization of these bands are reversed on either side of
$\bar{\Gamma}$ in accordance with the system being time reversal
symmetric $[E(\vec{k},\uparrow) = E(-\vec{k},\downarrow)]$
[Fig.\ref{Sb_Fig3}(F)]. The measured spin texture of the Sb(111) surface states
(Fig.\ref{Sb_Fig3}), together with the connectivity of the surface bands
(Fig.\ref{Sb_Fig2}), uniquely determines its belonging to the $\nu_0$ = 1 class.
Therefore the surface of Sb carries a non-zero ($\pi$) Berry's phase
via the inner V-shaped band and pure Sb can be regarded as the
parent metal of the Bi$_{1-x}$Sb$_x$ topological insulator class, in
other words, the topological order originates from the Sb wave
functions. A recent work \cite{Xu} has demonstrated a topological quantum phase transition as a function of chemical composition from a non-inverted to an inverted semiconductor as a clear example of the origin of topological order (Fig.\ref{QPT}).

\begin{figure}
\includegraphics[scale=0.27,clip=true, viewport=0.0in 0.0in 12.5in 11.4in]{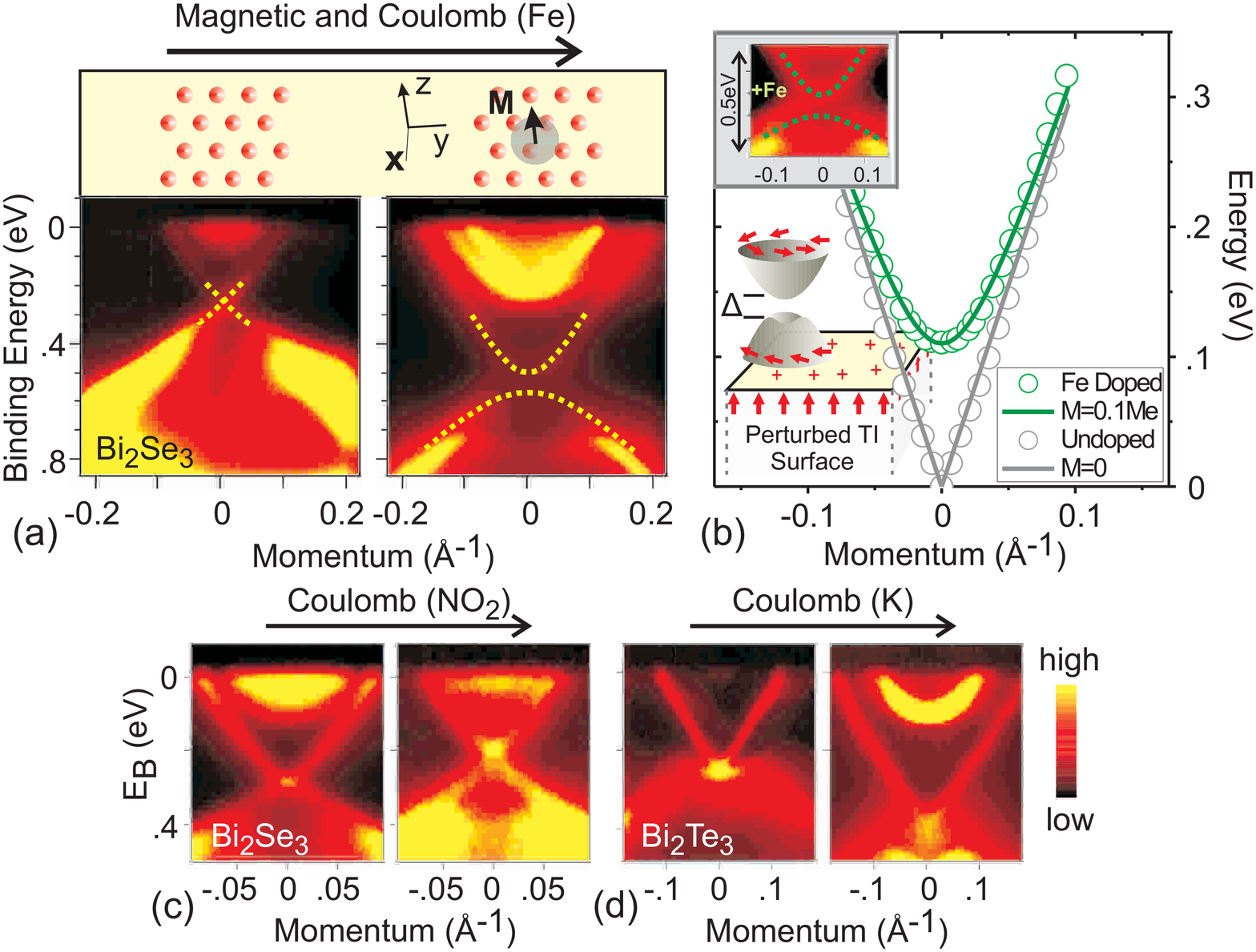}
\caption{\label{RMP_Fig15} \textbf{Protection by time reversal symmetry:} Topological surface states are robust in the presence of strong nonmagnetic disorder but open a gap in the presence of time reversal symmetry breaking magnetic impurities and disorder. (a) Magnetic impurity such as Fe on the surface of Bi$_2$Se$_3$ opens a gap at the Dirac point. The magnitude of the gap is set by the interaction of Fe ions with the Se surface and the time reversal symmetry breaking disorder potential introduced on the surface. (b) A comparison of surface band dispersion with and without Fe doping. (c,d) Non-magnetic disorder created via molecular absorbent NO$_2$ or alkali atom adsorption (K or Na) on the surface leaves the Dirac node intact in both Bi$_2$Se$_3$ and Bi$_2$Te$_3$. [Adapted from L. Wray $et$ $al$., \textit{Nature Phys.} \textbf{7}, 32 (2011). \cite{Wray}]}
\end{figure}

Our spin polarized measurement methods (Fig.\ref{Sb_Fig1} and \ref{Sb_Fig3}) uncover a new
type of topological quantum number $n_M$ which provides information
about the chirality properties. Topological band theory suggests
that the bulk electronic states in the mirror ($k_y$ = 0) plane can
be classified in terms of a number $n_M$ (=$\pm$1) that describes
the handedness (either left or right handed) or chirality of the
surface spins which can be directly measured or seen in
spin-resolved experiments \cite{20}. We now determine the value of $n_M$
from our data. From figure \ref{Sb_Fig1}, it is seen that a single (one) surface
band, which switches partners at \={M}, connects the bulk valence
and conduction bands, so $|n_M|$ = 1 (SM). The sign of
$n_M$ is related to the direction of the spin polarization $\langle
\vec{P} \rangle$ of this band \cite{20}, which is constrained by mirror
symmetry to point along $\pm\hat{y}$. Since the central
electron-like FS enclosing $\bar{\Gamma}$ intersects six mirror
invariant points [see Fig.\ref{Sb_Fig3}(B)], the sign of $n_M$ distinguishes two
distinct types of handedness for this spin polarized FS. Figures
\ref{Sb_Fig1}(F) and \ref{Sb_Fig3} show that for both Bi$_{1-x}$Sb$_x$ and Sb, the surface
band that forms this electron pocket has $\langle \vec{P} \rangle
\propto -\hat{y}$ along the $k_x$ direction, suggesting a
left-handed rotation sense for the spins around this central FS thus
$n_M$ = $-$1. Therefore, both insulating Bi$_{1-x}$Sb$_x$ and pure
Sb possess equivalent chirality properties $-$ a definite spin
rotation sense (left-handed chirality, see Fig.\ref{Sb_Fig3}(B)) and a
topological Berry's phase. Recently a chirality transition across the surface Dirac point of a 3D topological insulator has also been observed using spin-ARPES \cite{Xu} (Fig.\ref{texture}).

These spin-resolved experimental measurements reveal an intimate and
straightforward connection between the topological numbers ($\nu_0$,
$n_M$) and the physical observables. The $\nu_0$ determines whether
the surface electrons support a non-trivial Berry's phase, and if
they do, the $n_M$ determines the spin handedness of the Fermi
surface that manifests this Berry's phase. The 2D Berry's phase is a
critical signature of topological order and is not realizable in
isolated 2D electron systems, nor on the surfaces of conventional
spin-orbit or exchange coupled magnetic materials. A non-zero
Berry's phase is known, theoretically, to protect an electron system
against the almost universal weak-localization behavior in their low
temperature transport \cite{11,13} and is expected to form the key
element for fault-tolerant computation schemes \cite{13,29}, because the
Berry's phase is a geometrical agent or mechanism for protection
against quantum decoherence \cite{30}. Its remarkable realization on the
Bi$_{1-x}$Sb$_x$ surface represents an unprecedented example of a 2D
$\pi$ Berry's phase, and opens the possibility for building
realistic prototype systems to test quantum computing modules. In
general, our results demonstrate that spin-ARPES is a powerful probe
of 3D topological order, which opens up a new search front for topological materials for novel spin-devices
and fault-tolerant quantum computing.

\begin{figure*}
\includegraphics[scale=.75,clip=true, viewport=-0.0in 0in 6.8in 6.8in]{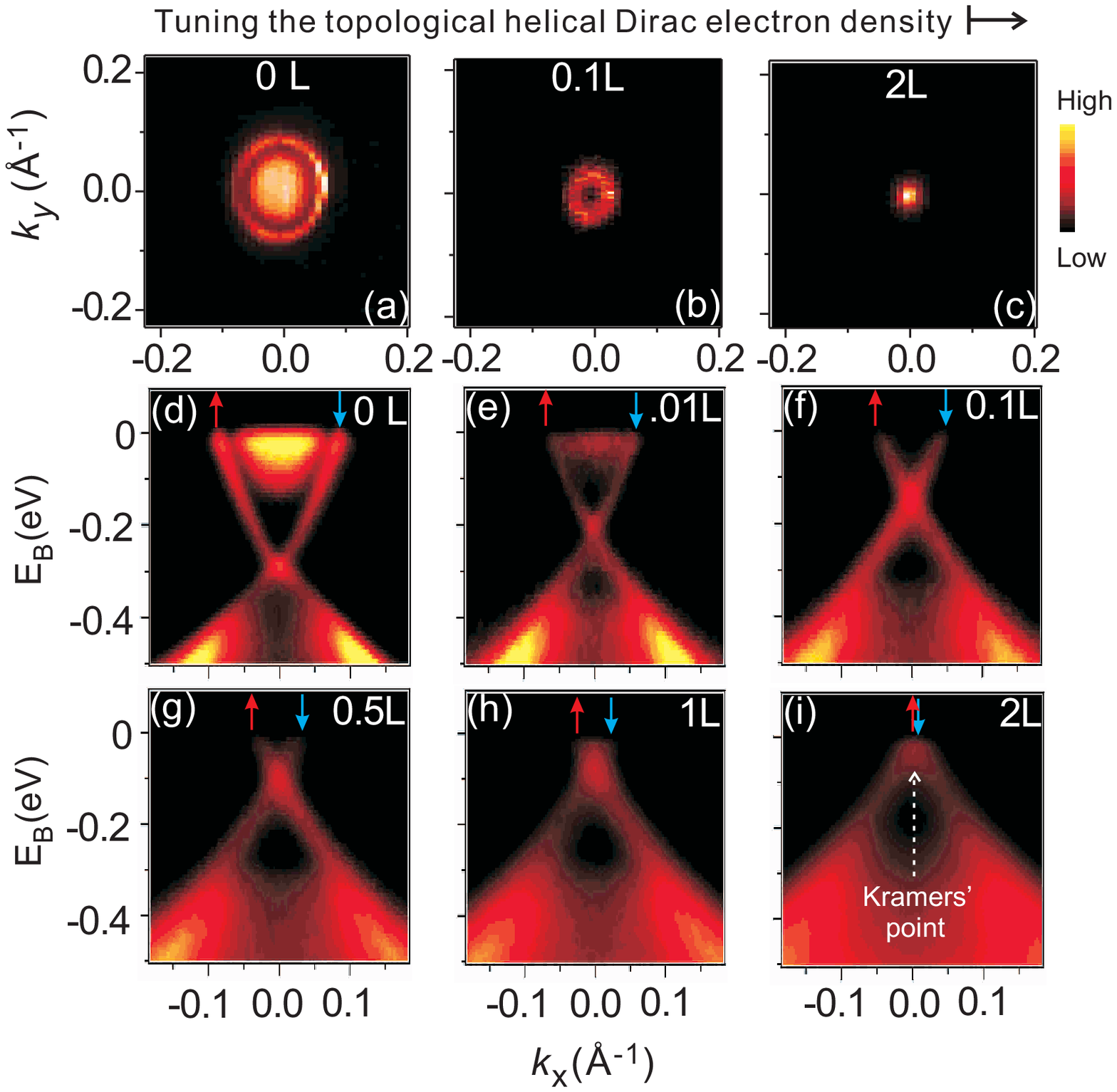}
\caption{\label{fig:Fig3} \textbf{Surface Gating : Tuning the density of helical Dirac electrons to the spin-degenerate Kramers point and topological transport regime.}
(a) A high resolution ARPES mapping of the surface Fermi surface
(FS) near $\bar{\Gamma}$ of Bi$_{2-\delta}$Ca$_{\delta}$Se$_3$
(111). The diffuse intensity within the ring originates from the
bulk-surface resonance state \cite{15}. (b) The FS after 0.1
Langmuir (L) of NO$_2$ is dosed, showing that the resonance state is
removed. (c) The FS after a 2 L dosage, which achieves the Dirac
charge neutrality point. (d) High resolution ARPES surface band
dispersions through after an NO$_2$ dosage of 0 L, (e) 0.01 L, (f)
0.1 L, (g) 0.5 L, (h) 1 L and (i) 2 L. The arrows denote the spin
polarization of the bands. We note that due to an increasing level
of surface disorder with NO$_2$ adsorption, the measured spectra
become progressively more diffuse and the total photoemission
intensity from the buried Bi$_{2-\delta}$Ca$_{\delta}$Se$_3$ surface
is gradually reduced. [Adapted from D. Hsieh $et$ $al.$, \textit{Nature} \textbf{460}, 1101 (2009). \cite{Nature_2009}].}
\end{figure*}

\section{Topological protection and tunability of the surface states of a 3D topological insulator}

The predicted topological protection of the surface states of Sb
implies that their metallicity cannot be destroyed by weak time
reversal symmetric perturbations. In order to test the robustness of
the measured gapless surface states of Sb, we introduce such a
perturbation by randomly substituting Bi into the Sb crystal matrix. Another motivation for performing such an experiment
is that the formalism developed by Fu and Kane \cite{11} to calculate the
$Z_2$ topological invariants relies on inversion symmetry being
present in the bulk crystal, which they assumed to hold true even in
the random alloy Bi$_{1-x}$Sb$_x$. However, this formalism is simply
a device for simplifying the calculation and the non-trivial
$\nu_0=1$ topological class of Bi$_{1-x}$Sb$_x$ is predicted to hold
true even in the absence of inversion symmetry in the bulk crystal
\cite{11}. Therefore introducing light Bi substitutional disorder into
the Sb matrix is also a method to examine the effects of alloying
disorder and possible breakdown of bulk inversion symmetry on the
surface states of Sb(111). We have performed spin-integrated ARPES
measurements on single crystals of the random alloy
Sb$_{0.9}$Bi$_{0.1}$. Figure~\ref{fig:Sb_FigS5} shows that both the
surface band dispersion along $\bar{\Gamma}$-\={M} as well as the
surface state Fermi surface retain the same form as that observed in
Sb(111), and therefore the `topological metal' surface state of
Sb(111) fully survives the alloy disorder. Since Bi alloying is seen
to only affect the band structure of Sb weakly, it is reasonable to
assume that the topological order is preserved between Sb and
Bi$_{0.91}$Sb$_{0.09}$
as we observed.

In a simpler fashion compared to Bi$_{1-x}$Sb$_x$, the topological insulator behavior in Bi$_2$Se$_3$
is associated with a single band inversion at the surface Brillouin zone center. Owing to its larger bandgap compared with Bi$_{1-x}$Sb$_x$, ARPES has shown that its topological properties are preserved at room temperature \cite{Nature_2009}. Two defining properties of topological insulators —
spin-momentum locking of surface states and $\pi$ Berry
phase — can be clearly demonstrated in the Bi$_2$Se$_3$ series.
The surface states are expected to be protected by time-reversal symmetry
symmetry, which implies that the surface Dirac node
should be robust in the presence of nonmagnetic disorder
but open a gap in the presence of time-reversal symmetry breaking perturbations.
Magnetic impurities such as Fe or Mn on the
surface of Bi$_2$Se$_3$ open a gap at the Dirac point [Figs.
\ref{RMP_Fig15}(a) and \ref{RMP_Fig15}(b)] \cite{Hor,Wray}. The magnitude of the gap is likely set by the interaction
of Fe ions with the Se surface and the time-reversal symmetry breaking
disorder potential introduced on the surface. Nonmagnetic
disorder created via molecular absorbent NO$_2$
or alkali atom adsorption (K or Na) on the surface
leaves the Dirac node intact [Figs. 15(c) and 15(d)] in
both Bi$_2$Se$_3$ and Bi$_2$Te$_3$ \cite{Nature_2009,Xia_arxiv}. These
results are consistent with the fact that the topological surface states are protected by time-reversal symmetry.

Many of the interesting theoretical proposals that utilize
topological insulator surfaces require the chemical
potential to lie at or near the surface Dirac point. This is
similar to the case in graphene, where the chemistry of
carbon atoms naturally locates the Fermi level at the
Dirac point. This makes its density of carriers highly
tunable by an applied electrical field and enables applications
of graphene to both basic science and microelectronics.
The surface Fermi level of a topological insulator
depends on the detailed electrostatics of the surface
and is not necessarily at the Dirac point. Moreover, for
naturally grown Bi$_2$Se$_3$ the bulk Fermi energy is not
even in the gap. The observed n-type behavior is believed
to be caused Se vacancies. By appropriate chemical
modifications, however, the Fermi energy of both the
bulk and the surface can be controlled. This allowed \cite{Nature_2009} to reach the sweet
spot in which the surface Fermi energy is tuned to the
Dirac point (Fig.\ref{fig:Fig3}). This was achieved by doping bulk
with a small concentration of Ca, which compensates the
Se vacancies, to place the Fermi level within the bulk
band gap. The surface was the hole doped by exposing the
surface to NO$_2$ gas to place the Fermi level at the Dirac
point. These results collectively show how ARPES can be used to study the topological protection and tunability properties of the 2D surface of a 3D topological insulator.

\section{Future directions: Identifying Majorana Platforms and Topological Superconductors}

\begin{figure*}
\includegraphics[scale=0.3,clip=true, viewport=0.0in 0.0in 21.5in 13.5in]{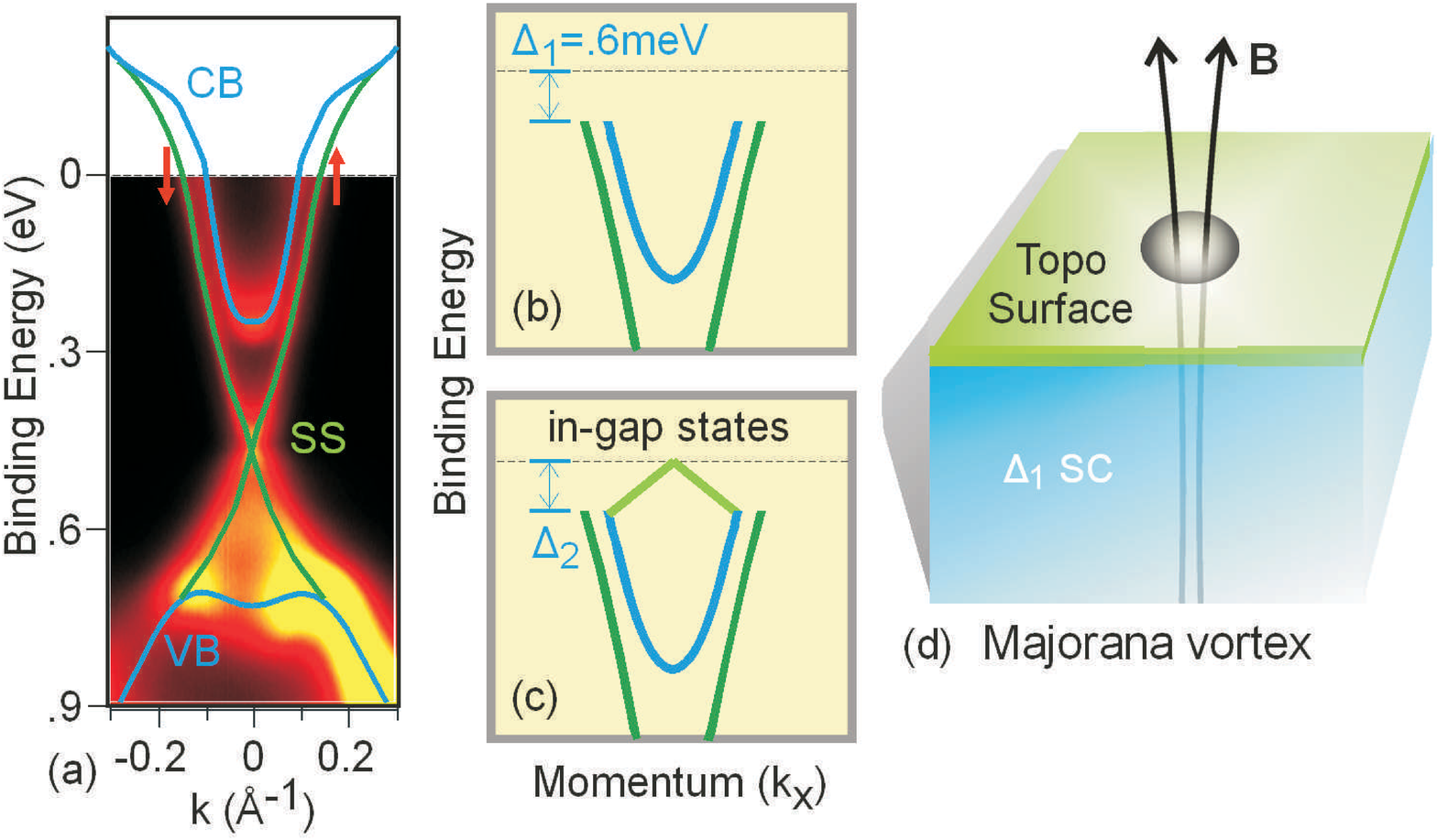}
\caption{\label{Figure13} \textbf{A Majorana platform.} (a) Topologically protected surface states cross the Fermi level before merging with the
bulk valence and conduction bands in a lightly doped topological insulator. (b) If the superconducting wavefunction has even
parity, the surface states will be gapped by the proximity effect, and vortices on the crystal surface will host braidable Majorana
fermions. (c) If superconducting parity is odd, the material will be a so-called "topological superconductor", and new states
will appear below T$_c$ to span the bulk superconducting gap. (d) Majorana fermion surface vortices are found at the end of
bulk vortex lines and could be manipulated for quantum computation if superconducting pairing is even. [Adapted from L. Wray $et$ $al$., \textit{Nature Phys.} \textbf{7}, 32 (2011). \cite{Wray}]}
\end{figure*}

Recent measurements \cite{Wray} show that surface
instabilities cause the spin-helical topological insulator
band structure of Bi$_2$Se$_3$ to remain well defined and non-
degenerate with bulk electronic states at the Fermi level
of optimally doped superconducting Cu$_{0.12}$Bi$_2$Se$_3$, and
that this is also likely to be the case for superconduct-
ing variants of p-type Bi$_2$Te$_3$. These surface states pro-
vide a highly unusual physical setting in which super-
conductivity cannot take a conventional form, and is expected to realize one of two novel states that have not
been identified elsewhere in nature. If superconducting
pairing has even parity, as is nearly universal among the
known superconducting materials, the surface electrons
will achieve a 2D non-Abelian superconductor state with
non-commutative Majorana fermion vortices that can potentially be manipulated to store quantum information.
Surface vortices will be found at the end of bulk vortex
lines as drawn in Fig.\ref{Figure13}. If superconducting pairing
is odd, the resulting state is a novel state of matter known
as a ``topological superconductor" with Bogoliubov surface quasi-particles present below the superconducting critical temperature of 3.8 K. As drawn in Fig.\ref{Figure13}(c), these low temperature surface states would be gapless,
likely making it impossible to adiabatically manipulate
surface vortices for quantum computation. The unique
physics and applications of the topological superconductor state are distinct from any known material system,
and will be an exciting vista for theoretical and experimental exploration if they are achieved for the first time
in Cu$_x$Bi$_2$Se$_3$.

\textbf{Acknowledgement}
The authors acknowledge collaborations with A. Bansil, R. J. Cava, J.H. Dil, A. V. Fedorov, Liang Fu, Y. S. Hor, S. Jia, H. Lin, F. Meier, N. P. Ong, J. Osterwalder, L. Patthey, D. Qian and A. Yazdani for collaborations and U.S. DOE Grants No. DE-FG-02-05ER46200, No. AC03-76SF00098, and No. DE-FG02-07ER46352. M.Z.H. acknowledges visiting-scientist support from Lawrence Berkeley National Laboratory and additional support from the A.P. Sloan Foundation.

\vspace{3cm}

\section{Supplementary materials}

\setcounter{figure}{0}

\subsection{Growth method for high-quality single crystals}

\begin{figure*}
\renewcommand{\thefigure}{S\arabic{figure}}
\includegraphics[scale=0.65,clip=true, viewport=0.0in 0in 8.0in 6.3in]{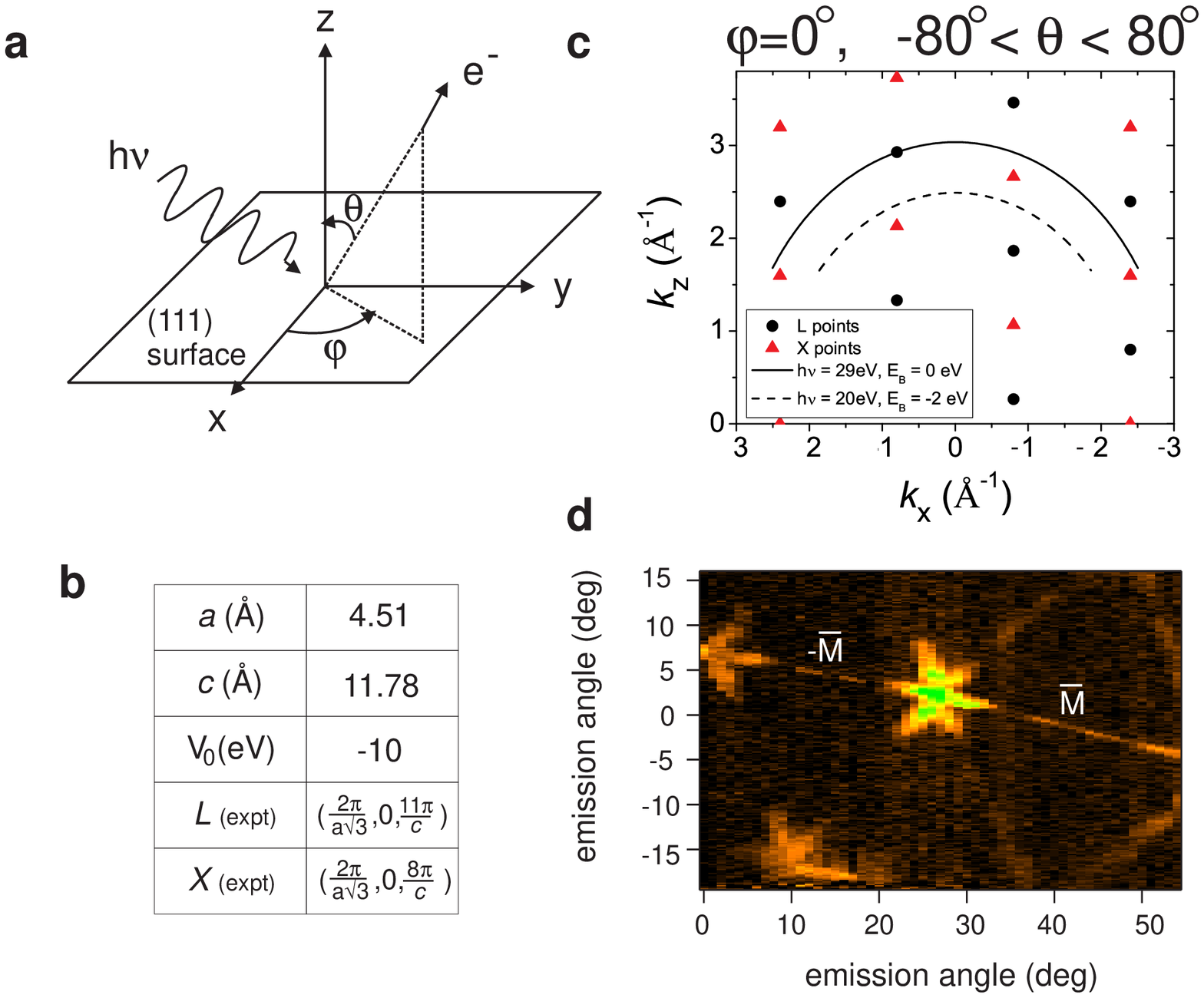}
\caption{\label{fig:BiSb_FigS1} \textbf{Method of locating high symmetry bulk reciprocal
lattice points of Bi$_{0.9}$Sb$_{0.1}$ using incident photon energy
modulated ARPES.} \textbf{a,} Geometry of an ARPES experiment.
\textbf{b,} Key parameters relevant to the calculation of the
positions of the high symmetry points in the 3D BZ. The lattice
constants refer to the rhombohedral A7 lattice structure.
\textbf{c,} Location of $L$ and $X$ points of the bulk BZ in the
$k_x$-$k_z$ plane together with the constant energy contours that
can be accessed by changing the angle $\theta$. \textbf{d,} Near
$E_F$ intensity map ($h\nu$ = 55 eV) of the Fermi surface formed by
the surface states covering an entire surface BZ, used to help
locate various in-plane momenta, in units of the photoelectron
emission angle along two orthogonal spatial directions. The electron
pockets near \={M} in Fig.\ref{fig:BiSb_Fig2}c (main text) appear as lines in Fig.\ref{fig:BiSb_FigS1}d
due to relaxed k-resolution in order to cover a large k-space in a
single shot. [Adapted from D. Hsieh $et$ $al.$, \textit{Nature} \textbf{452}, 970 (2008) \cite{10}].}
\end{figure*}

\begin{figure*}
\renewcommand{\thefigure}{S\arabic{figure}}
\includegraphics[scale=0.55,clip=true, viewport=0.0in 0in 10.0in 8.5in]{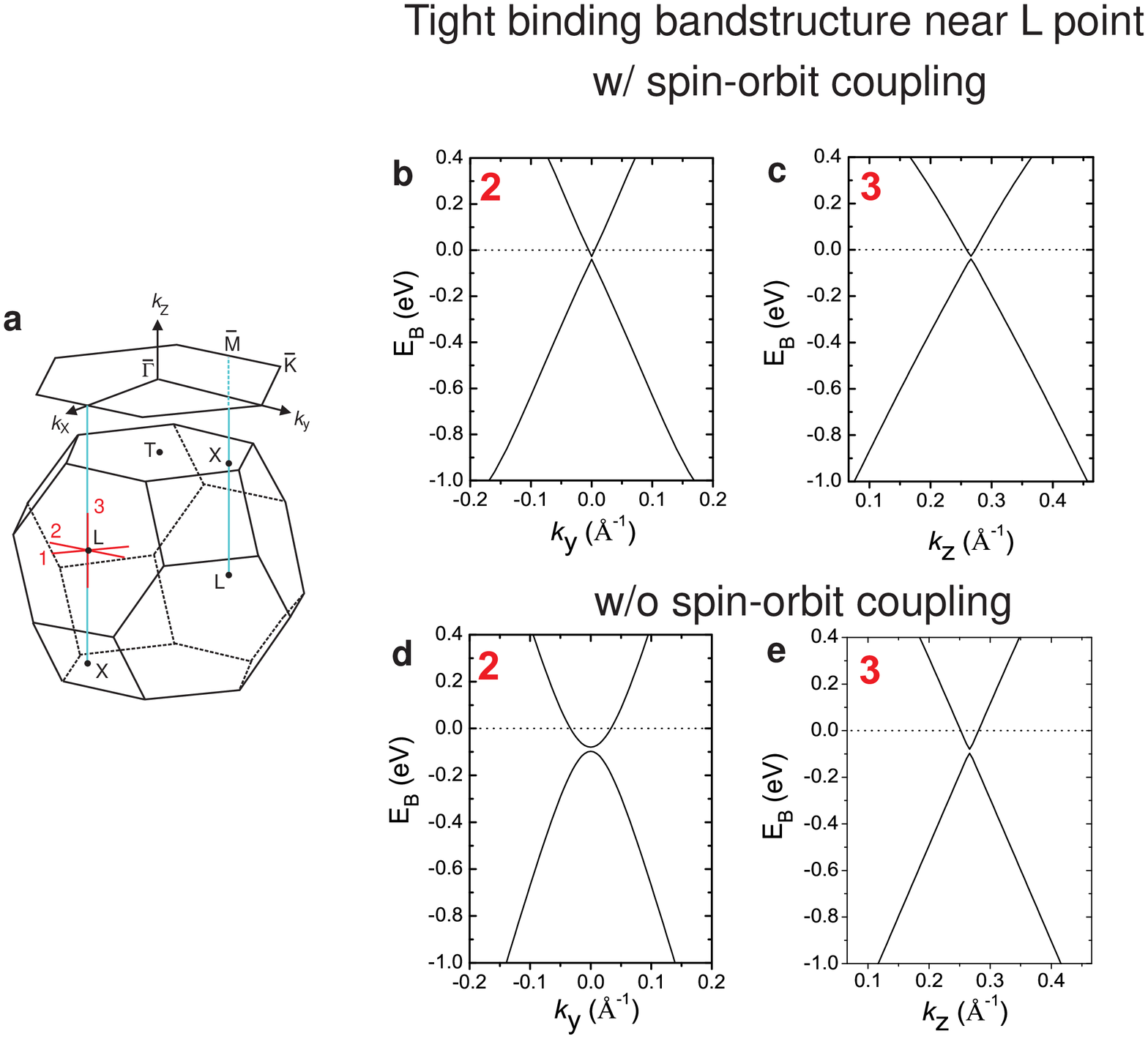}
\caption{\label{fig:BiSb_FigS2} \textbf{Spin-orbit coupling has a profound effect on the
band structure of bismuth near $L$ point.} \textbf{a,} Schematic of
the bulk 3D BZ and the projected BZ of the (111) surface.
\textbf{b,c,} Calculated tight binding band structure of bismuth
including a spin-orbit coupling strength of 1.5 eV along two
orthogonal cuts through the $L$ point in the 1$^{st}$ bulk BZ.
\textbf{d,e,} Tight binding band structure along the same two
directions as \textbf{b} and \textbf{c} calculated without
spin-orbit coupling. The inter-band gap of 13.7 meV is barely
visible on the scale of \textbf{b} and \textbf{c}. [Adapted from D. Hsieh $et$ $al.$, \textit{Nature} \textbf{452}, 970 (2008) \cite{10}].}
\end{figure*}

The undoped and doped Bi$_{1-x}$Sb$_x$ single-crystal samples ($0\leq x \leq0.17$) used for ARPES experiments were each cleaved from a boule grown from a stoichiometric mixture of high-purity elements. The boule was
cooled from 650 $^{\circ}$C to 270 $^{\circ}$C over a period of five
days and was annealed for seven days at 270 $^{\circ}$C. The samples
naturally cleaved along the (111) plane, which resulted in shiny
flat silver surfaces. X-ray diffraction measurements were used to
check that the samples were single phase, and confirmed that the
Bi$_{0.9}$Sb$_{0.1}$ single crystals presented in this paper have a
rhombohedral A7 crystal structure (point group $R\bar{3}m$), with
room-temperature (T=300K) lattice parameters $a$ = 4.51 \AA\ and $c$
= 11.78 \AA\ indexed using a rhombohedral unit cell. The X-ray
diffraction patterns of the cleaved crystals exhibit only the (333),
(666), and (999) peaks, showing that the cleaved surface is oriented
along the trigonal (111) axis. Room-temperature data were recorded
on a Bruker D8 diffractometer using Cu K$\alpha$  radiation
($\lambda$ = 1.54 \AA) and a diffracted-beam monochromator. The
in-plane crystal orientation was determined by Laue X-ray
diffraction. During the angle-resolved photoemission spectroscopy
(ARPES) measurements a fine alignment was achieved by carefully
studying the band dispersions and Fermi surface symmetry as an
internal check for crystal orientation.

\subsection{Resistivity characterization}

Temperature-dependent resistivity measurements were carried out on
single-crystal samples in a Quantum Design PPMS-9 instrument, using
a standard four-probe technique on approximately 4 $\times$ 1
$\times$1-mm$^3$, rectangular samples with the current in the basal
plane, which was perpendicular to the trigonal axis. The four
contacts were made by using room-temperature silver paste. The data
for samples with concentrations ranging from $x$ = 0 to $x$ = 0.17
showed a systematic change from semimetallic to insulating-like
behaviour with increasing $x$, in agreement with previously
published works \cite{Fukuyama}, which was used as a further check of the
antimony concentrations. Conventional magnetic and transport
measurements \cite{7,Lenoir,Kopelovic} such as these cannot separately measure the contributions of the surface and bulk states to the total signal.
ARPES, on the other hand, is a momentum-selective technique \cite{19},
which allows for a separation of 2D (surface) from 3D (bulk)
dispersive energy bands. This capability is especially important for
Bi$_{1-x}$Sb$_x$ because the Dirac point lies at a single point in
the 3D Brillouin zone, unlike for 2D graphene, where the Dirac
points can be studied at any arbitrary perpendicular momentum along
a line \cite{Novoselov,Bostwick}.

\subsection{ARPES}

Spin-integrated angle-resolved photoemission spectroscopy (ARPES)
measurements were performed with 14 to 30 eV photons on beam line
5-4 at the Stanford Synchrotron Radiation Laboratory, and with 28 to
32 eV photons on beam line 12 at the Advanced Light Source, both
endstations being equipped with a Scienta hemispherical electron
analyzer (see VG Scienta manufacturer website for instrument
specifications). Spin-resolved ARPES measurements were performed at
the SIS beam line at the Swiss Light Source using the COPHEE
spectrometer (\cite{31} p.15) with a single 40 kV classical Mott detector
and photon energies of 20 and 22 eV. The typical energy and momentum
resolution was 15 meV and 1.5\% of the surface Brillouin zone (BZ)
respectively at beam line 5-4, 9 meV and 1\% of the surface BZ
respectively at beam line 12, and 80 meV and 3\% of the surface BZ
respectively at SIS using a pass energy of 3 eV. Cleaving these samples \textit{in situ} between 10 K and 55 K at chamber pressures less than 5
$\times10^{-11}$ torr resulted in shiny flat surfaces, characterized
\textit{in situ} by low energy electron diffraction (LEED) to be
clean and well ordered with the same symmetry as the bulk. This is consistent with photoelectron
diffraction measurements that show no substantial structural relaxation of the Sb(111) surface \cite{32}.

\subsection{Systematic methods for separating bulk from surface
electronic states}

ARPES is a photon-in, electron-out technique \cite{19}. Photoelectrons
ejected from a sample by a monochromatic beam of radiation are
collected by a detector capable of measuring its kinetic energy
$E_{kin}$. By varying the detector angles, $\theta$ and $\varphi$,
relative to the sample surface normal, the momentum of the
photoelectrons, $\textbf{K}$, can also be determined (as illustrated
in \ref{fig:BiSb_FigS1}). By employing the commonly used
free-electron final state approximation, we can fully convert from
the measured kinetic energy and momentum values of the photoelectron
to the binding energy, $E_B$, and Bloch momentum values $\textbf{k}$
of its initial state inside the crystal, via

\begin{align}
|E_B| &= h\nu - W - E_{kin}\notag\\
k_x &= K_x = \frac{1}{\hbar}\sqrt{2 m_e
E_{kin}}\textnormal{sin}\theta\notag\\
k_z &=
\frac{1}{\hbar}\sqrt{2m_e(E_{kin}\textnormal{cos}^2\theta-V_0)}\notag
\end{align}
where we have set $\varphi$ = 0, $W$ is the work function, $m_e$ is
the electron mass and $V_0$ is an experimentally determined
parameter, which is approximately $-$10 eV for bismuth \cite{Jezequel,Ast:Bi3D}.
Features in the ARPES spectra originating from bulk initial states
(dispersive along the $k_z$-direction) were distinguished from those
originating from surface initial states (non-dispersive along the
$k_z$-direction) by studying their dependence on incident photon
energy, $h\nu$, and converting this to dependence on $k_z$ via the
displayed equations. ARPES data were collected at beamlines 12.0.1
and 10.0.1 of the Advanced Light Source at the Lawrence Berkeley
National Laboratory, as well as at the PGM beamline of the
Synchrotron Radiation Center in Wisconsin, with incident photon
energies ranging from 17 eV to 55 eV, energy resolutions ranging
from 9 meV to 40 meV and momentum resolution better than 1.5\% of
the surface Brillouin zone, using Scienta electron analysers. The
combination of high spatial resolution and high crystalline quality
enabled us to probe only the highly ordered and cleanest regions of
our samples. Single-crystal Bi$_{1-x}$Sb$_x$ samples were cleaved in
situ at a temperature of 15 K and chamber pressures less than 8
$\times$ 10$^{-11}$ torr, and high surface quality was checked
throughout the measurement process by monitoring the EDC linewidths
of the surface state. To measure the near-$E_F$ dispersion of an
electronic band along a direction normal to the sample surface, such
as the direction from $X (2\pi/\sqrt{3}a, 0, 8\pi/c)$ to $L
(2\pi/\sqrt{3}a, 0, 11\pi/c)$ shown in Fig. \ref{fig:BiSb_Fig2}a, EDCs were taken at
several incident photon energies. The kinetic energy of the
photoelectron at $E_F$ is different for each value of $h\nu$, so the
angle was first adjusted and then held fixed for each $h\nu$ so as
to keep $k_x$ constant at $2\pi/\sqrt{3}a$ = 0.8 \AA$^{-1}$ for
electrons emitted near $E_F$. To ensure that the in-plane momentum
remained constant at \={M} (the $L$-$X$ line projects onto \={M})
for each EDC, a complete near-$E_F$ intensity map was generated for
each photon energy to precisely locate the \={M}-point (see Fig.\ref{fig:BiSb_FigS1}d). We note that because the bulk crystal has
only three-fold rotational symmetry about the $k_z$-axis, the
reciprocal lattice does not have mirror symmetry about the $k_x$ = 0
plane. Therefore, scans taken at +$\theta$ and -$\theta$  for the
same photon energy probe different points in the bulk 3D Brillouin
zone; this is responsible for the absence of the bulk
$\Lambda$-shaped band in Fig. \ref{fig:BiSb_Fig3}f.

\subsection{Confirming the bulk nature of electronic bands by
comparison with theoretical calculations}

\begin{figure*}
\renewcommand{\thefigure}{S\arabic{figure}}
\includegraphics[scale=0.75,clip=true, viewport=-0.15in 0in 11.0in 6.0in]{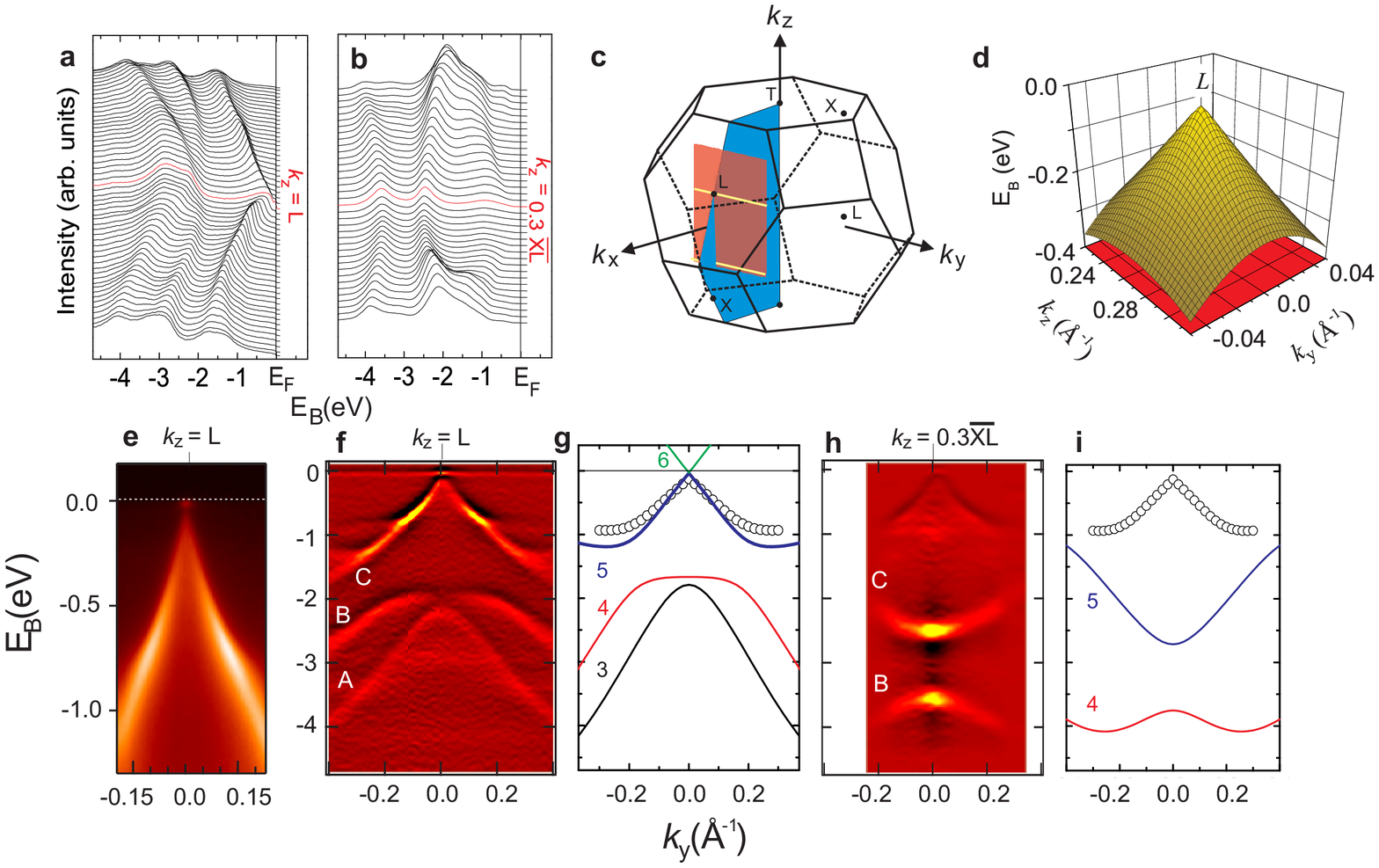}
\caption{\label{fig:BiSb_FigS3} \textbf{Identification of the bulk band features of
Bi$_{0.9}$Sb$_{0.1}$. Experimental band-structure determined by
ARPES is compared to bulk tight-binding calculations of bismuth to
further identify the deeper lying bulk bands and their symmetry
origins.} \textbf{a,} Energy distribution curves (EDCs) along a
k-space cut given by the upper yellow line shown in schematic
\textbf{c} which goes through the bulk $L$ point in the 3$^{rd}$ BZ
($h\nu$ = 29 eV). The corresponding ARPES intensity in the vicinity
of $L$ is shown in \textbf{e}. \textbf{b,} EDCs along the lower
yellow line of \textbf{c} which goes through the point a fraction
0.3 of the k-distance from $X$ to $L$ ($h\nu$ = 20 eV), showing a
dramatic change of the deeper lying band dispersions. (This cut was
taken at a $k_x$ value equal in magnitude but opposite in sign to
that in \textbf{a} as described in the SM text). \textbf{f,h,} The
ARPES second derivative images (SDI) of the raw data shown in
\textbf{a} and \textbf{b} to reveal the band dispersions. The flat
band of intensity at $E_F$ is an artifact of taking SDI.
\textbf{g,i,} Tight binding band calculations of bismuth including
spin-orbit coupling, using Liu and Allen model \cite{Liu}, along the
corresponding experimental cut directions shown in \textbf{f} and
\textbf{h}. The bands (colored solid lines) labelled 3 to 6 are
derived from the symmetries associated with the 6$p$-orbitals and
their dispersion is thus strongly influenced by spin-orbit coupling.
The inter-band gap between bands 5 and 6 is barely visible on the
scale of Fig. \ref{fig:BiSb_FigS2}g. The circled curves mark the surface state
dispersion, which is present at all measured photon energies (no
$k_z$ dispersion). There is a close match of the bulk band
dispersion between the data and calculations, confirming the
presence of strong spin-orbit coupling. \textbf{d,} Tight binding
valence band (5) dispersion of bismuth in the $k_y$-$k_z$ momentum
plane showing linearity along both directions. The close match
between data and calculation along $k_y$ suggests that the
dispersion near $E_F$ along $k_z$ is also linear. [Adapted from D. Hsieh $et$ $al.$, \textit{Nature} \textbf{452}, 970 (2008) \cite{10}].}
\end{figure*}

In an ARPES experiment (Fig.\ref{fig:BiSb_FigS1}a), three dimensional (3D) dispersive bulk electronic states can be identified as those that disperse with incident photon energy, whereas surface states do not. As an
additional check that we have indeed correctly identified the bulk
bands of Bi$_{0.9}$Sb$_{0.1}$ in Figs \ref{fig:BiSb_Fig1} and \ref{fig:BiSb_Fig2}, we also measured the
dispersion of the deeper lying bands well below the Fermi level
($E_F$) and compared them to tight binding theoretical calculations
of the bulk bands of pure bismuth following the model of Liu and
Allen (1995) \cite{Liu}. A tight-binding approach is known to be valid
since Bi$_{0.9}$Sb$_{0.1}$ is not a strongly correlated electron
system. As Bi$_{0.9}$Sb$_{0.1}$ is a random alloy (Sb does not form
a superlattice \cite{Lenoir}) with a relatively small Sb concentration
($\sim$0.2 Sb atoms per rhombohedral unit cell), the deeper lying
band structure of Bi$_{0.9}$Sb$_{0.1}$ is expected to follow that of
pure Bi because the deeper lying (localized wave function) bands of
Bi$_{0.9}$Sb$_{0.1}$ are not greatly affected by the substitutional
disorder, and no additional back folded bands are expected to arise.
Since these deeper lying bands are predicted to change dramatically
with $k_z$, they help us to finely determine the experimentally
probed $k_z$ values. Fig.\ref{fig:BiSb_FigS2}f shows the ARPES second derivative image
(SDI) of a cut parallel to \={K}\={M}\={K} that passes through the
$L$ point of the 3D Brillouin zone (BZ), and Fig.\ref{fig:BiSb_FigS2}h shows a
parallel cut that passes through the 0.3 $XL$ point (Fig.\ref{fig:BiSb_FigS2}c). The
locations of these two cuts in the 3D bulk BZ were calculated from
the ARPES kinematic relations, from which
we can construct the constant energy contours shown in Fig. \ref{fig:BiSb_FigS1}c. By
adjusting $\theta$ such that the in-plane momentum $k_x$ is fixed at
approximately 0.8 \AA$^{-1}$ (the surface \={M} point), at a photon
energy $h\nu$ =29 eV, electrons at the Fermi energy ($E_B$=0 eV)
have a $k_z$ that corresponds to the $L$ point in the 3$^{rd}$ bulk
BZ. By adjusting $\theta$ such that the in-plane momentum $k_x$ is
fixed at approximately -0.8 \AA$^{-1}$, at a photon energy $h\nu$ =
20 eV, electrons at a binding energy of -2 eV have a $k_z$ near 0.3
$XL$.

\begin{figure*}
\renewcommand{\thefigure}{S\arabic{figure}}
\includegraphics[scale=0.6,clip=true, viewport=0.0in 0in 10.0in 8.0in]{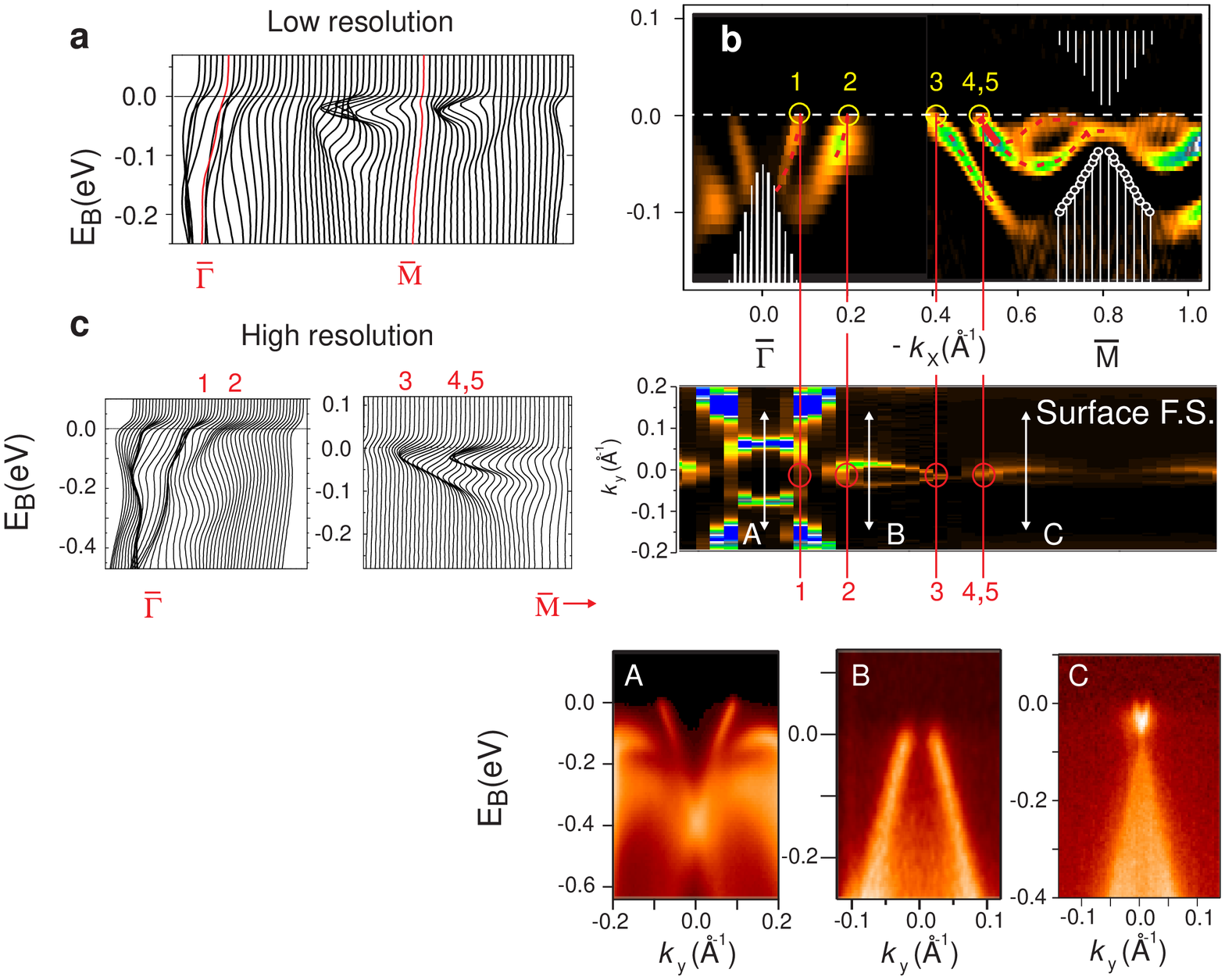}
\caption{\label{fig:BiSb_FigS4} \textbf{The Kramers' point, the gapless nature and topology
of surface states in insulating Bi$_{0.9}$Sb$_{0.1}$ is revealed
through high spatial and k-resolution ARPES.} \textbf{a,} Energy
distribution curves (EDCs) of a low resolution ARPES scan along the
surface $\bar{\Gamma}$-\={M} cut of Bi$_{0.9}$Sb$_{0.1}$.
\textbf{b,} The surface band dispersion second derivative image
along $\bar{\Gamma}$-\={M} obtained by piecing together four high
resolution ARPES scans. See main text Fig.\ref{fig:BiSb_Fig3} for explanation of other
features. \textbf{c,} EDCs of high resolution ARPES scans in the
vicinity of surface Fermi crossings 1 and 2 and crossings 3, 4 and 5
(left panels). These crossings form the surface Fermi surface shown
in the upper right panel of \textbf{c} (see also main text Fig.\ref{fig:BiSb_Fig2}).
High resolution ARPES scans along cut directions A, B and C are
further evidence for a surface Fermi surface. [Adapted from D. Hsieh $et$ $al.$, \textit{Nature} \textbf{452}, 970 (2008) \cite{10}].}
\end{figure*}

There is a clear $k_z$ dependence of the dispersion of measured
bands A, B and C, pointing to their bulk nature. The bulk origin of
bands A, B and C is confirmed by their good agreement with tight
binding calculations (bands 3, 4 and 5 in Figs \ref{fig:BiSb_FigS2}g and i), which
include a strong spin-orbit coupling constant of 1.5 eV derived from
bismuth \cite{Liu}. Band 3 drops below -5 eV at the 0.3 $XL$ point. The
slight differences between the experimentally measured band energies
and the calculated band energies at $k_y$ = 0 \AA$^{-1}$ shown in
Fig.\ref{fig:BiSb_FigS2}f-i are due to the fact that the ARPES data were collected in
a single shot, taken in constant $\theta$ mode. This means that
electrons detected at different binding energies will have slightly
different values of $k_z$, whereas the
presented tight binding calculations show all bands at a single
$k_z$. We checked that the magnitude of these band energy
differences is indeed accounted for by this explanation. Even though
the $L_a$ and $L_s$ bands in Bi$_{0.9}$Sb$_{0.1}$ are inverted
relative to those of pure semimetallic Bi, calculations show that
near $E_F$, apart from an insulating gap, they are ``mirror" bands
in terms of k dispersion (see bands 5 and 6 in Fig.\ref{fig:BiSb_FigS2}g). Such a
close match to calculations, which also predict a linear dispersion
along the $k_z$ cut near $E_F$ (Fig.\ref{fig:BiSb_FigS2}d), provides strong support
that the dispersion of band C, near $E_F$, is in fact linear along
$k_z$. Focusing on the $\Lambda$-shaped valence band at $L$, the
EDCs (Fig.\ref{fig:BiSb_FigS2}a) show a single peak out to $k_y \approx \pm0.15$
\AA$^{-1}$ demonstrating that it is composed of a single band
feature. Outside this range however, an additional feature develops
on the low binding energy side of the main peak in the EDCs, which
shows up as two well separated bands in the SDI image (Fig.\ref{fig:BiSb_Fig2}f) and
signals a splitting of the band into bulk representative and surface
representative components (Fig.\ref{fig:BiSb_FigS2}a,f). Unlike the main peak that
disperses strongly with incident photon energy, this shoulder-like
feature is present and retains the same $\Lambda$-shaped dispersion
near this k-region (open circles in Figs \ref{fig:BiSb_FigS2}g and i) for all photon
energies used, supporting its 2D surface character. This behaviour
is quite unlike bulk band C, which attains the $\Lambda$-shaped
dispersion only near 29 eV (see main text Fig. \ref{fig:BiSb_Fig2}b).

\subsection{Spin-orbit coupling is responsible for the unique Dirac-like
dispersion behaviour of the bulk bands near $E_F$}

According to theoretical models, a strongly spin-orbit coupled bulk
band structure is necessary for topological surface states to
exist \cite{Fu:STI2,11,15,Roy,Murakami}. Therefore it is important to show that our
experimentally measured bulk band structure of Bi$_{0.9}$Sb$_{0.1}$
can only be accounted for by calculations that explicitly include a
large spin-orbit coupling term. As shown in the previous section,
the measured bulk band dispersion of Bi$_{0.9}$Sb$_{0.1}$ generally
follows the calculated bulk bands of pure Bi from a tight binding
model. The dispersion of the bulk valence and conduction bands of
pure bismuth near $E_F$ at the $L$ point from such a tight binding
calculation \cite{Liu} with a spin-orbit coupling constant of 1.5 eV are
shown in Fig. \ref{fig:BiSb_FigS3}b and c, which show a high degree of linearity. The
high degree of linearity can be understood from a combination of the
large Fermi velocity ($v_F$ $\approx$ 6 eV \AA\ along $k_y$) and
small inter-band (below $E_F$) gap $\Delta$ = 13.7 meV (Fig. \ref{fig:BiSb_FigS3}).
This calculated inter-band gap of Bi (13.7 meV) is smaller than our
measured lower limit of 50 meV (main text Fig. \ref{fig:BiSb_Fig1}a) for the
insulating gap of Bi$_{0.9}$Sb$_{0.1}$. To illustrate the importance
of spin-orbit coupling in determining the band structure near $L$,
we show the dispersion along $k_y$ and $k_z$ calculated without
spin-orbit coupling (Fig. \ref{fig:BiSb_FigS3}d and e). While the dispersion along
$k_z$ is not drastically altered by neglecting the spin-orbit
coupling, the dispersion along $k_y$ changes from being linear to
highly parabolic. This is further evidence that our measured Dirac
point can be accounted for only by including spin-orbit coupling.
\textit{A strong spin-orbit coupling constant acts as an internal
quantizing magnetic field for the electron system \cite{Haldane(P-anomaly)} which can give
rise to a quantum spin Hall effect without any externally applied
magnetic field \cite{Bernevig(QSHE),Sheng(QSHE),8,Sheng}. Therefore, the existence or the
spontaneous emergence of the surface or boundary states does not
require an external magnetic field.}

\subsection{Matching the surface state Fermi crossings and the topology
of the surface Fermi surface in bulk insulating
B\lowercase{i}$_{0.9}$S\lowercase{b}$_{0.1}$}

In order to count the number of singly degenerate surface state
Fermi crossings \cite{14,21,Kim} along the $\bar{\Gamma}$-\={M} cut of
the surface BZ, high photon energy ARPES scans, which allow mapping
of the entire k range from $\bar{\Gamma}$-\={M} to fall within the
detector window at the expense of lower instrument resolution, were
taken to preliminarily identify the k-space locations of the Fermi
crossings (Fig. \ref{fig:BiSb_FigS4}a). Having determined where these surface state
Fermi crossings lie in k-space, we performed several high resolution
ARPES scans, each covering a successive small k interval in the
detector window, in order to construct a high resolution band
mapping of the surface states from $\bar{\Gamma}$ to \={M}. The
second derivative image of the surface band dispersion shown in
Fig.\ref{fig:BiSb_FigS4}b was constructed by piecing together four such high
resolution scans. Fig.\ref{fig:BiSb_FigS4}c shows energy distribution curves of high
resolution ARPES scans in the vicinity of each surface Fermi
crossing, which together give rise to the surface Fermi surface
shown. No previous work \cite{14,Hochst,Hofmann,Hirahara,Hengsberger,Ast:Bi2} has reported the band
dispersion near the $L$-point (thus missing the Dirac bands) or
resolved the Kramers point near the \={M} point, which is crucial to
determine the topology of the surface states. For this reason there
is no basis for one-to-one comparison with previous work, since no
previous ARPES data exists in the analogous k-range. Note that
surface band dispersions along the cuts A, B and C are highly
linear. This is indirect evidence for the existence of the bulk
Dirac point since surface states are formed when the bulk state wave
functions are subjected to the boundary conditions at the cleaved
plane.

\subsection{Two-step fitting analysis procedure of
Spin-Resolved ARPES
measurements of insulating Bi$_{1-x}$Sb$_{x}$}

\begin{figure*}
\renewcommand{\thefigure}{S\arabic{figure}}
\includegraphics[scale=0.54,clip=true, viewport=0.0in 0in 11.5in 6.1in]{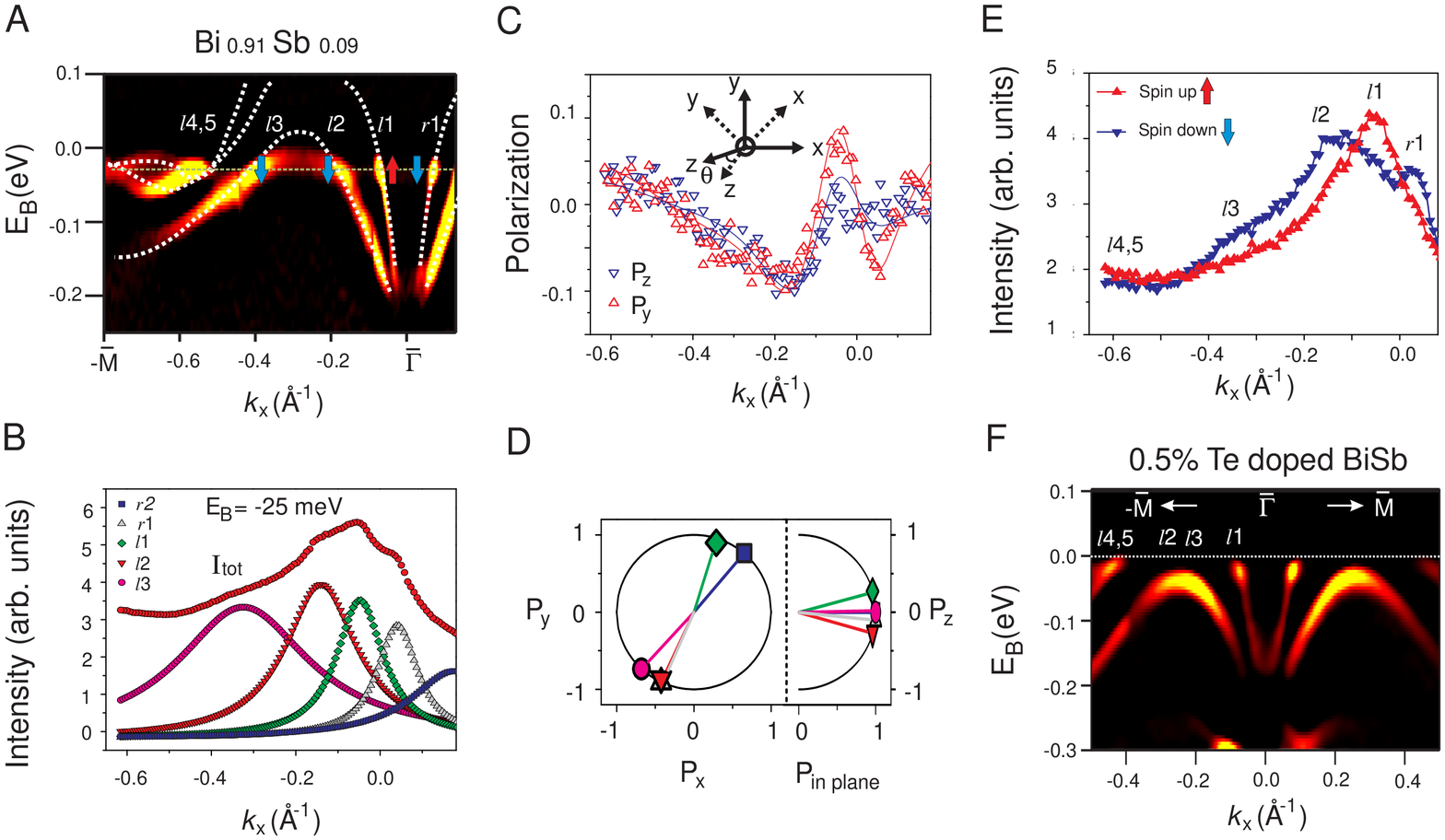}
\caption{\label{fig:Sb_FigS2} (\textbf{A}) The surface band dispersion ARPES
second derivative image (SDI) along the $\bar{\Gamma}$ to -\={M}
direction of bulk insulating Bi$_{0.91}$Sb$_{0.09}$. Dashed white
lines are guides to the eye. The intensity of bands $\textit{l}4,5$
is scaled up for clarity. (\textbf{B}) MDC of the spin averaged
spectrum at $E_B$ = -25 meV [green line below $E_F$ in (\textbf{A})]
using a photon energy $h\nu$ = 22 eV, together with the Lorentzian
peaks of the fit. (\textbf{C}) Measured spin polarization curves
(symbols) for the $y'$ and $z'$ (Mott coordinate) components
together with the fitted lines (SM). The relative
orientation of the sample (un-primed) to Mott (primed) coordinates
is shown in the inset. The polar angle $\theta$ is rotated during
the measurement to access different values of $k_x$. At normal
emission ($\theta = 0^{\circ}$), the $z'$ and $z$ axes are parallel
and the $y'$ axis is rotated from the $y$ axis by 45$^{\circ}$.
(\textbf{D}) The in-plane and out-of-plane spin polarization
components in the sample coordinate frame obtained from the spin
polarization fit. The symbols refer to those in (\textbf{B}). The
fitted parameters are consistent with 100\% polarized spins.
(\textbf{E}) Spin resolved spectra for the $y$ component based on
the fitted spin polarization curves shown in (\textbf{C}). Spin up
(down) refers to the spin direction being approximately parallel to
the +(-)$\hat{y}$ direction. (\textbf{F}) The surface band
dispersion SDI centered about $\bar{\Gamma}$ of
(Bi$_{0.925}$Sb$_{0.075})_{0.995}$Te$_{0.005}$. Electron doping
through Te reveals that bands $l2$ and $l3$ are connected above
$E_F$. [Adapted from D. Hsieh $et$ $al.$, \textit{Science} \textbf{323}, 919 (2009) \cite{Science}].}
\end{figure*}

Here we present details of the spin-resolved ARPES analysis on bulk
insulating Bi$_{0.91}$Sb$_{0.09}$ that show how we arrive at a
spin-resolved surface band dispersion such as that presented in
Figure \ref{Sb_Fig1}(G) in the main text. In the VUV incident photon energy
regime that we use, spin conserving photoemission processes (where
the electric field of light only acts on the orbital degree of
freedom of the electron inside a solid) dominate over spin
non-conserving processes (which arise from coupling to the magnetic
field of light) \cite{33}. Therefore we are confident that the
photo-emission process does not change the spin polarization of the
electrons. Figure~\ref{fig:Sb_FigS2}(B) shows a spin averaged
momentum distribution curve (MDC) along the $\bar{\Gamma}$ to -\={M}
direction taken at $E_B$ = -25 meV, indicated by the green line
shown in Figure~\ref{fig:Sb_FigS2}(A). This MDC was obtained by
summing the signal coming from both left and right electron
detectors in the Mott polarimeter (see diagram in Fig.\ref{Sb_Fig3}(A) of the
main text). Lorentzian lineshapes denoted $I^i$ and a non-polarized
background $B$ are fitted to this MDC, which are used as inputs to
the two-step fitting routine developed by Meier $et$ $al.$ \cite{26} in
the following way. To begin with, a spin polarization vector
$\vec{P}_{M}^i = (P_{x'}^i, P_{y'}^i, P_{z'}^i) =
(\cos\theta_i\cos\phi_i, \cos\theta_i\sin\phi_i, \sin\theta_i)$ is
assigned to each band, where $\theta_i$ and $\phi_i$ are referenced
to the primed Mott coordinate frame. Here it is necessary to assume
a spin magnitude of one because only two spin components are
measured by a single Mott detector. Such an assumption is likely
valid since even though the spin polarization is no longer a good
quantum number due to spin-orbit coupling, the bands near
$\bar{\Gamma}$ are expected to exhibit a high degree of spin
polarization since the spin-orbit coupling is smallest near
$\bar{\Gamma}$. Moreover, common strong spin-orbit coupled materials
such as gold have been experimentally shown to exhibit 100\% spin
polarized surface states \cite{18}. A spin-resolved spectrum is then
defined for each peak $i$ using $I_{\alpha}^{i;\uparrow,\downarrow}
= I^i(1 \pm P^i_{\alpha})/6$, where $\alpha = x', y', z'$, and + and
$-$ correspond to the spin direction being parallel ($\uparrow$) or
antiparallel ($\downarrow$) to $\alpha$. The full spin-resolved
spectrum is then given by $I^{\uparrow,\downarrow}_{\alpha} = \sum_i
I_{\alpha}^{i;\uparrow,\downarrow} + B/6$, where $B$ is the
unpolarized background, from which the spin polarization of each
spatial component can be obtained as $P_{\alpha} =
(I_{\alpha}^{\uparrow}-I_{\alpha}^{\downarrow})/(I_{\alpha}^{\uparrow}+I_{\alpha}^{\downarrow})$.
This latter expression is a function of $\theta_i$ and $\phi_i$ and
is used to fit to the experimental data.

The spin polarization data for the $y^{\prime}$ and $z^{\prime}$
components (i.e. $P_{y'}$ and $P_{z'}$) are obtained by taking the
difference between the intensities of the left-right (or top-bottom)
electron detectors over their sum, all divided by the Sherman
function, which is calibrated using the methods in (\cite{31}, p.36).
Typical electron counts on the detector reach $5 \times 10^5$, which
places an error bar of approximately $\pm$0.01 for each point on our
polarization curves. To account for unequal sensitivities between a
detector pair, we applied a small multiplicative factor to the
intensity from one detector to ensure that the unpolarized
background intensity yields zero polarization. Resultant curves are
shown in Figure~\ref{fig:Sb_FigS2}(C). The best fit parameters
($P_{x'}^i, P_{y'}^i, P_{z'}^i$), which are expressed in the sample
coordinates through an appropriate coordinate transformation [inset
of Fig.~\ref{fig:Sb_FigS2}(C)] are shown in
Figure~\ref{fig:Sb_FigS2}(D). Even though the measured polarization
only reaches a magnitude of around $\pm$0.1, this is similarly seen
in studies of Bi thin films \cite{21} and is due to the non-polarized
background and overlap of adjacent peaks with different spin
polarization. These effects are extremely sensitive to the sample
alignment due to the very narrow Fermi surface features. The fitted
parameters [Fig.~\ref{fig:Sb_FigS2}(D)] are consistent with spins
being nearly aligned along the $\pm\hat{y}$ direction, with bands
$l1$ and $r1$ having nearly opposite spin as required by time
reversal symmetry, and with these spins nearly parallel to those of
$l1$ and $r1$ respectively measured for Sb [main text Fig.\ref{Sb_Fig3}(F)]. The
small departures from ideality likely originate from the scan
direction not being exactly along $\bar{\Gamma}$-\={M}. Bands $l1$
and $l2$ display opposite spin, which indicates that they form a
Kramers pair split by spin-orbit coupling, and the fact that bands
$l2$ and $l3$ have the same spin suggests that they originate from
the same band.

To show that bands $l2$ and $l3$ connect above $E_F$ as we have
drawn in Figure~\ref{fig:Sb_FigS2}(A), and are thus expected to have
the same spin, we map the surface band dispersion of Te doped
Bi$_{1-x}$Sb$_x$ that is known to be an electron donor \cite{35}.
Figure~\ref{fig:Sb_FigS2}(F) shows that the hole band formed by
crossings 2 and 3 in insulating Bi$_{1-x}$Sb$_x$
[Fig.~\ref{fig:Sb_FigS2}(A)] has sunk completely below $E_F$ with
0.5\% Te doping, and is in fact the same band.

\subsection{Method of using incident photon energy modulated
ARPES to separate the bulk from surface electronic states of Sb}

\begin{figure}
\renewcommand{\thefigure}{S\arabic{figure}}
\includegraphics[scale=0.43,clip=true, viewport=0.0in 0in 8.8in 9.5in]{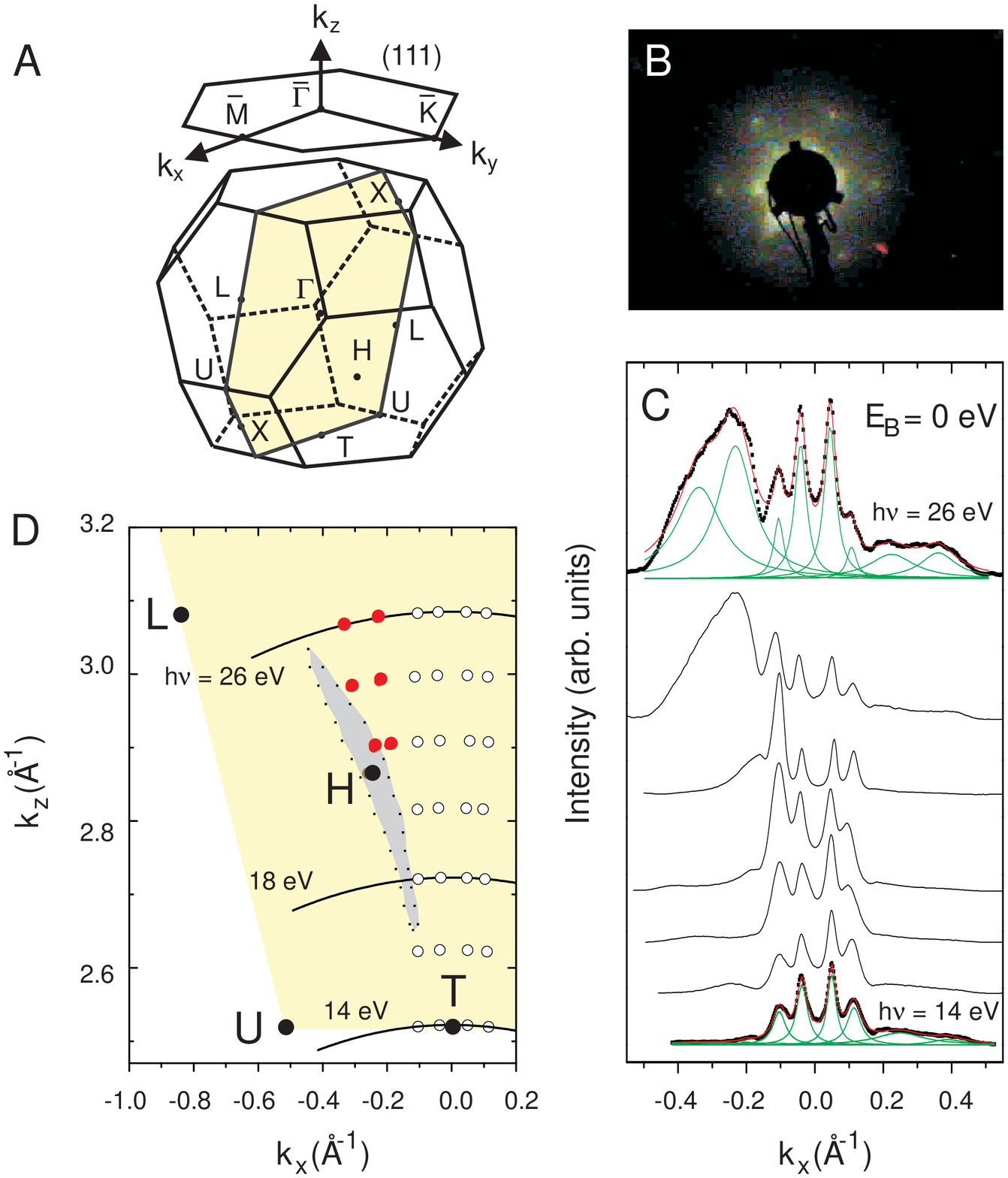}
\caption{\label{fig:Sb_FigS3} (\textbf{A}) Schematic of the bulk BZ of Sb and
its (111) surface BZ. The shaded region denotes the momentum plane
in which the following ARPES spectra were measured. (\textbf{B})
LEED image of the \textit{in situ} cleaved (111) surface exhibiting
a hexagonal symmetry. (\textbf{C}) Select MDCs at $E_F$ taken with
photon energies from 14 eV to 26 eV in steps of 2 eV, taken in the
$TXLU$ momentum plane. Peak positions in the MDCs were determined by
fitting to Lorentzians (green curves). (\textbf{D}) Experimental 3D
bulk Fermi surface near H (red circles) and 2D surface Fermi surface
near $\bar{\Gamma}$ (open circles) projected onto the $k_x$-$k_z$
plane, constructed from the peak positions found in (\textbf{C}).
The $k_z$ values are determined using calculated constant $h\nu$
contours (black curves) (SM). The shaded gray
region is the theoretical hole Fermi surface calculated in \cite{36}. [Adapted from D. Hsieh $et$ $al.$, \textit{Science} \textbf{323}, 919 (2009) \cite{Science}].}
\end{figure}

In this section we detail incident photon energy modulated ARPES
experiments on the low lying electronic states of single crystal
Sb(111), which we employ to isolate the surface from bulk-like
electronic bands over the entire BZ. Figure~\ref{fig:Sb_FigS3}(C)
shows momentum distributions curves (MDCs) of electrons emitted at
$E_F$ as a function of $k_x$ ($\parallel$ $\bar{\Gamma}$-\={M}) for
Sb(111). The out-of-plane component of the momentum $k_z$ was
calculated for different incident photon energies ($h\nu$) using the
free electron final state approximation with an experimentally
determined inner potential of 14.5 eV \cite{37,38}. There are four peaks
in the MDCs centered about $\bar{\Gamma}$ that show no dispersion
along $k_z$ and have narrow widths of $\Delta k_x \approx$ 0.03
\AA$^{-1}$. These are attributed to surface states and are similar
to those that appear in Sb(111) thin films \cite{37}. As $h\nu$ is
increased beyond 20 eV, a broad peak appears at $k_x \approx$ -0.2
\AA$^{-1}$, outside the $k$ range of the surface states near
$\bar{\Gamma}$, and eventually splits into two peaks. Such a strong
$k_z$ dispersion, together with a broadened linewidth ($\Delta k_x
\approx$ 0.12 \AA$^{-1}$), is indicative of bulk band behavior, and
indeed these MDC peaks trace out a Fermi surface
[Fig.~\ref{fig:Sb_FigS3}(D)] that is similar in shape to the hole
pocket calculated for bulk Sb near H \cite{36}. Therefore by choosing an
appropriate photon energy (e.g. $\leq$ 20 eV), the ARPES spectrum at
$E_F$ along $\bar{\Gamma}$-\={M} will have contributions from only
the surface states. The small bulk electron pocket centered at L is
not accessed using the photon energy range we employed [Fig.~\ref{fig:Sb_FigS3}(D)].

\begin{figure*}
\renewcommand{\thefigure}{S\arabic{figure}}
\includegraphics[scale=0.55,clip=true, viewport=0.0in 0in 11.5in 7.0in]{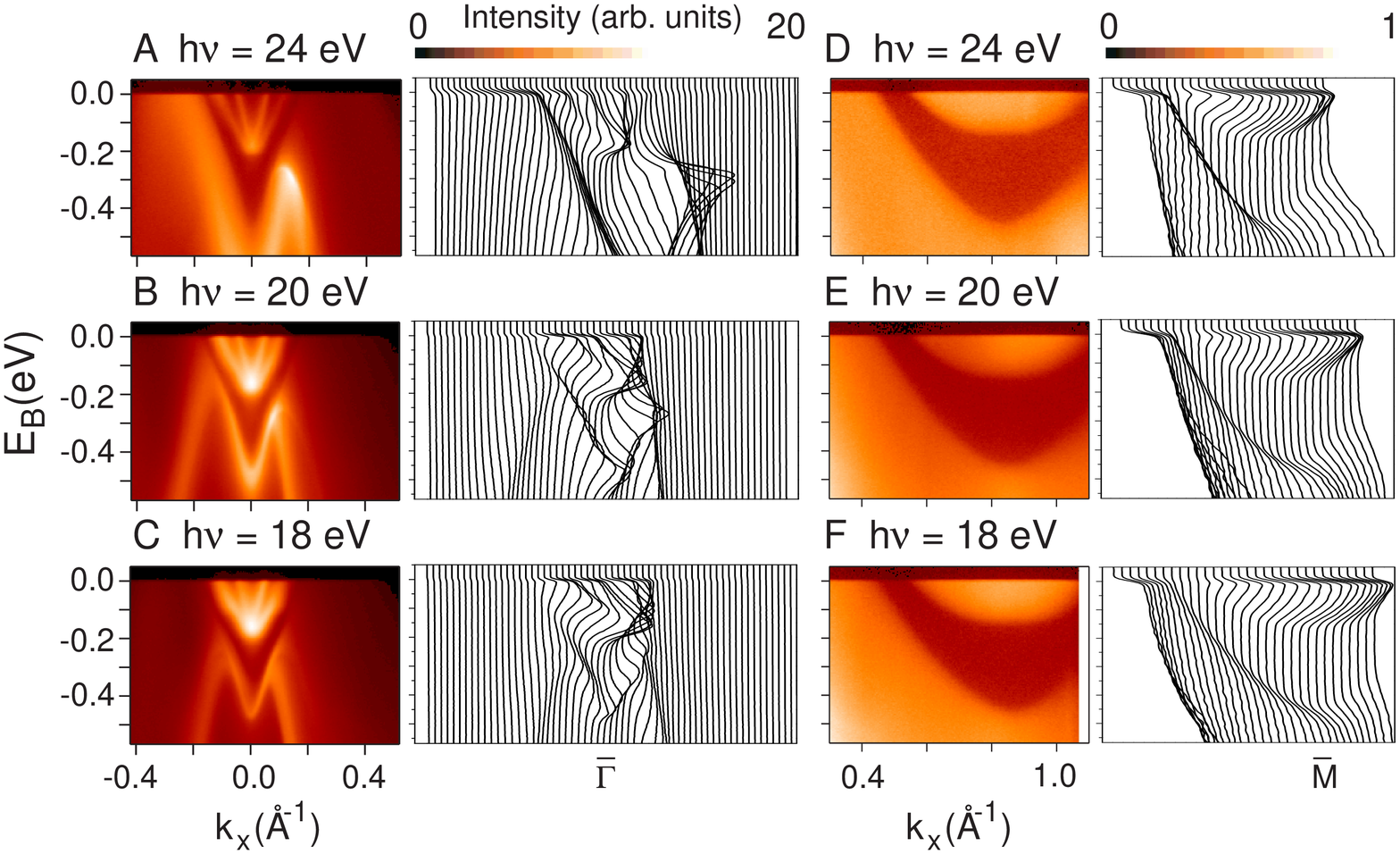}
\caption{\label{fig:Sb_FigS4} ARPES intensity maps of Sb(111) as a function of
$k_x$ near $\bar{\Gamma}$ (\textbf{A})-(\textbf{C}) and \={M}
(\textbf{D})-(\textbf{F}) and their corresponding energy
distribution curves, taken using $h\nu$ = 24 eV, 20 eV and 18 eV
photons. The intensity scale of (\textbf{D})-(\textbf{F}) is a
factor of about twenty smaller than that of
(\textbf{A})-(\textbf{C}) due to the intrinsic weakness of the ARPES
signal near \={M}. [Adapted from D. Hsieh $et$ $al.$, \textit{Science} \textbf{323}, 919 (2009) \cite{Science}].}
\end{figure*}

Now we describe the experimental procedure used to distinguish pure
surface states from resonant states on Sb(111) through their
spectral signatures. ARPES spectra along $\bar{\Gamma}$-\={M} taken
at three different photon energies are shown in
Fig.~\ref{fig:Sb_FigS4}. Near $\bar{\Gamma}$ there are two rather
linearly dispersive electron like bands that meet exactly at
$\bar{\Gamma}$ at a binding energy $E_B \sim$ -0.2 eV. This behavior
is consistent with a pair of spin-split surface bands that become
degenerate at the time reversal invariant momentum ($\vec{k}_T$)
$\bar{\Gamma}$ due to Kramers degeneracy. The surface origin of this
pair of bands is established by their lack of dependence on $h\nu$
[Fig.~\ref{fig:Sb_FigS4}(A)-(C)]. A strongly photon energy
dispersive hole like band is clearly seen on the negative $k_x$ side
of the surface Kramers pair, which crosses $E_F$ for $h\nu=$ 24 eV
and gives rise to the bulk hole Fermi surface near H
[Fig.~\ref{fig:Sb_FigS3}(D)]. For $h\nu\leq$ 20 eV, this band shows
clear back folding near $E_B \approx$ -0.2 eV indicating that it has
completely sunk below $E_F$. Further evidence for its bulk origin
comes from its close match to band calculations
[Fig.~\ref{fig:Sb_FigS3}(D)]. Interestingly, at photon energies such
as 18 eV where the bulk bands are far below $E_F$, there remains a
uniform envelope of weak spectral intensity near $E_F$ in the shape
of the bulk hole pocket seen with $h\nu$ = 24 eV photons, which is
symmetric about $\bar{\Gamma}$. This envelope does not change shape
with $h\nu$ suggesting that it is of surface origin. Due to its weak
intensity relative to states at higher binding energy, these
features cannot be easily seen in the energy distribution curves
(EDCs) in Fig.~\ref{fig:Sb_FigS4}(A)-(C), but can be clearly
observed in the MDCs shown in Fig.~\ref{fig:Sb_FigS3}(C) especially
on the positive $k_x$ side. Centered about the \={M} point, we also
observe a crescent shaped envelope of weak intensity that does not
disperse with $k_z$ [Fig.~\ref{fig:Sb_FigS4}(D)-(F)], pointing to
its surface origin. Unlike the sharp surface states near
$\bar{\Gamma}$, the peaks in the EDCs of the feature near \={M} are
much broader ($\Delta E \sim$80 meV) than the spectrometer
resolution (15 meV). The origin of this diffuse ARPES signal is not
due to surface structural disorder because if that were the case,
electrons at $\bar{\Gamma}$ should be even more severely scattered
from defects than those at \={M}. In fact, the occurrence of both
sharp and diffuse surface states originates from a $k$ dependent
coupling to the bulk. As seen in Fig.\ref{Sb_Fig2}(D) of the main text, the
spin-split Kramers pair near $\bar{\Gamma}$ lie completely within
the gap of the projected bulk bands near $E_F$ attesting to their
purely surface character. In contrast, the weak diffuse hole like
band centered near $k_x$ = 0.3 \AA$^{-1}$ and electron like band
centered near $k_x$ = 0.8 \AA$^{-1}$ lie completely within the
projected bulk valence and conduction bands respectively, and thus
their ARPES spectra exhibit the expected lifetime broadening due to
coupling with the underlying
bulk continuum \cite{39}.

\subsection{Method of counting spin Fermi surface $\vec{k}_T$ enclosures in pure Sb}

\begin{figure}
\renewcommand{\thefigure}{S\arabic{figure}}
\includegraphics[scale=0.32,clip=true, viewport=0.0in 0in 10.8in 6.8in]{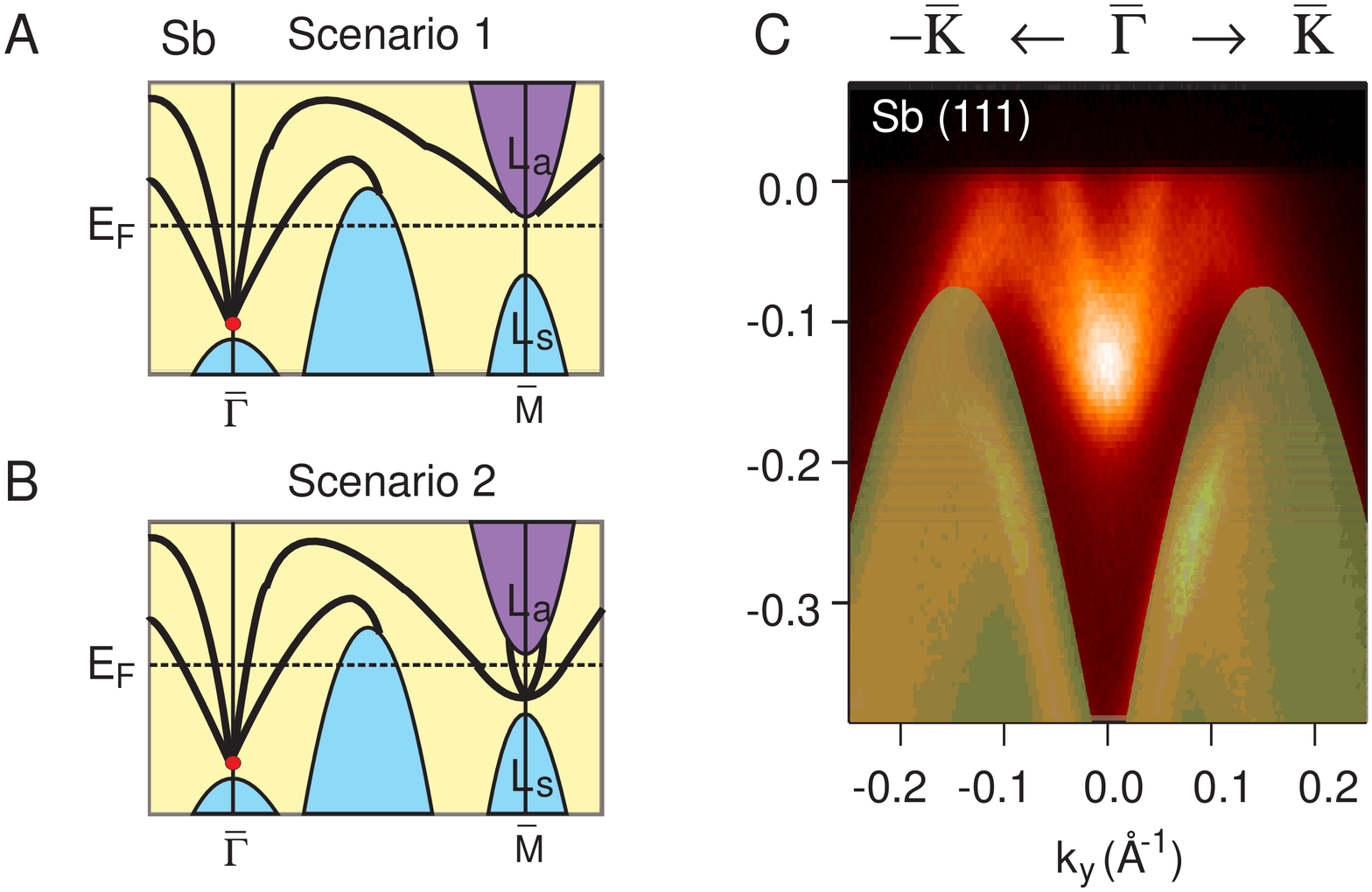}
\caption{\label{fig:Sb_FigS1} (\textbf{A}) Schematic of the surface band
structure of Sb(111) under a time reversal symmetric perturbation
that lifts the bulk conduction (L$_a$) band above the Fermi level
($E_F$). Here the surface bands near \={M} are also lifted completed
above $E_F$. (\textbf{B}) Alternatively the surface band near \={M}
can remain below $E_F$ in which case it must be doubly spin
degenerate at \={M}. (\textbf{C}) ARPES intensity plot of the
surface states along the -\={K}$-\bar{\Gamma}-$\={K} direction. The
shaded green regions denote the theoretical projection of the bulk
valence bands, calculated using the full potential linearized
augmented plane wave method using the local density approximation
including the spin-orbit interaction (method described in 40). Along
this direction, it is clear that the outer V-shaped surface band
that was observed along the -\={M}$-\bar{\Gamma}-$\={M} now merges
with the bulk valence band. [Adapted from D. Hsieh $et$ $al.$, \textit{Science} \textbf{323}, 919 (2009) \cite{Science}].}
\end{figure}

In this section we give a detailed explanation of why the surface
Fermi contours of Sb(111) that overlap with the projected bulk Fermi
surfaces can be neglected when determining the $\nu_0$ class of the
material. Although the Fermi surface formed by the surface resonance
near \={M} encloses the $\vec{k}_T$ \={M}, we will show that this
Fermi surface will only contribute an even number of enclosures and
thus not alter the overall evenness or oddness of $\vec{k}_T$
enclosures. Consider some time reversal symmetric perturbation that
lifts the bulk conduction L$_a$ band completely above $E_F$ so that
there is a direct excitation gap at L. Since this perturbation
preserves the energy ordering of the L$_a$ and L$_s$ states, it does
not change the $\nu_0$ class. At the same time, the weakly surface
bound electrons at \={M} can evolve in one of two ways. In one case,
this surface band can also be pushed up in energy by the
perturbation such that it remains completely inside the projected
bulk conduction band [Fig.~\ref{fig:Sb_FigS1}(A)]. In this case
there is no more density of states at $E_F$ around \={M}.
Alternatively the surface band can remain below $E_F$ so as to form
a pure surface state residing in the projected bulk gap. However by
Kramers theorem, this SS must be doubly spin degenerate at \={M} and
its FS must therefore enclose \={M} twice
[Fig.~\ref{fig:Sb_FigS1}(B)]. In determining $\nu_0$ for
semi-metallic Sb(111), one can therefore neglect all segments of the
FS that lie within the projected areas of the bulk FS [Fig.\ref{Sb_Fig2}(G) of
main text] because they can only contribute an even number of FS
enclosures, which does not change the modulo 2 sum of $\vec{k}_T$
enclosures.

In order to further experimentally confirm the topologically
non-trivial surface band dispersion shown in figures \ref{Sb_Fig2}(C) and (D) of
the main text, we show ARPES intensity maps of Sb(111) along the
-\={K}$-\bar{\Gamma}-$\={K} direction. Figure~\ref{fig:Sb_FigS1}(C)
shows that the inner V-shaped band that was observed along the
-\={M}$-\bar{\Gamma}-$\={M} direction retains its V-shape along the
-\={K}$-\bar{\Gamma}-$\={K} direction and continues to cross the
Fermi level, which is expected since it forms the central hexagonal
Fermi surface. On the other hand, the outer V-shaped band that was
observed along the -\={M}$-\bar{\Gamma}-$\={M} direction no longer
crosses the Fermi level along the -\={K}$-\bar{\Gamma}-$\={K}
direction, instead folding back below the Fermi level around $k_y$ =
0.1 \AA$^{-1}$ and merging with the bulk valence band
[Fig.~\ref{fig:Sb_FigS1}(C)]. This confirms that it is the
$\Sigma_{1(2)}$ band starting from $\bar{\Gamma}$ that connects to
the bulk valence (conduction) band, in agreement with the
calculations shown in figure \ref{Sb_Fig2}(D) of the main text.

\subsection{Physical interpretation of $n_M$ : the mirror Chern number and an analogy with the spin-Chern number}

\begin{figure*}
\renewcommand{\thefigure}{S\arabic{figure}}
\includegraphics[scale=0.55,clip=true, viewport=0.0in 0in 11.0in 7.3in]{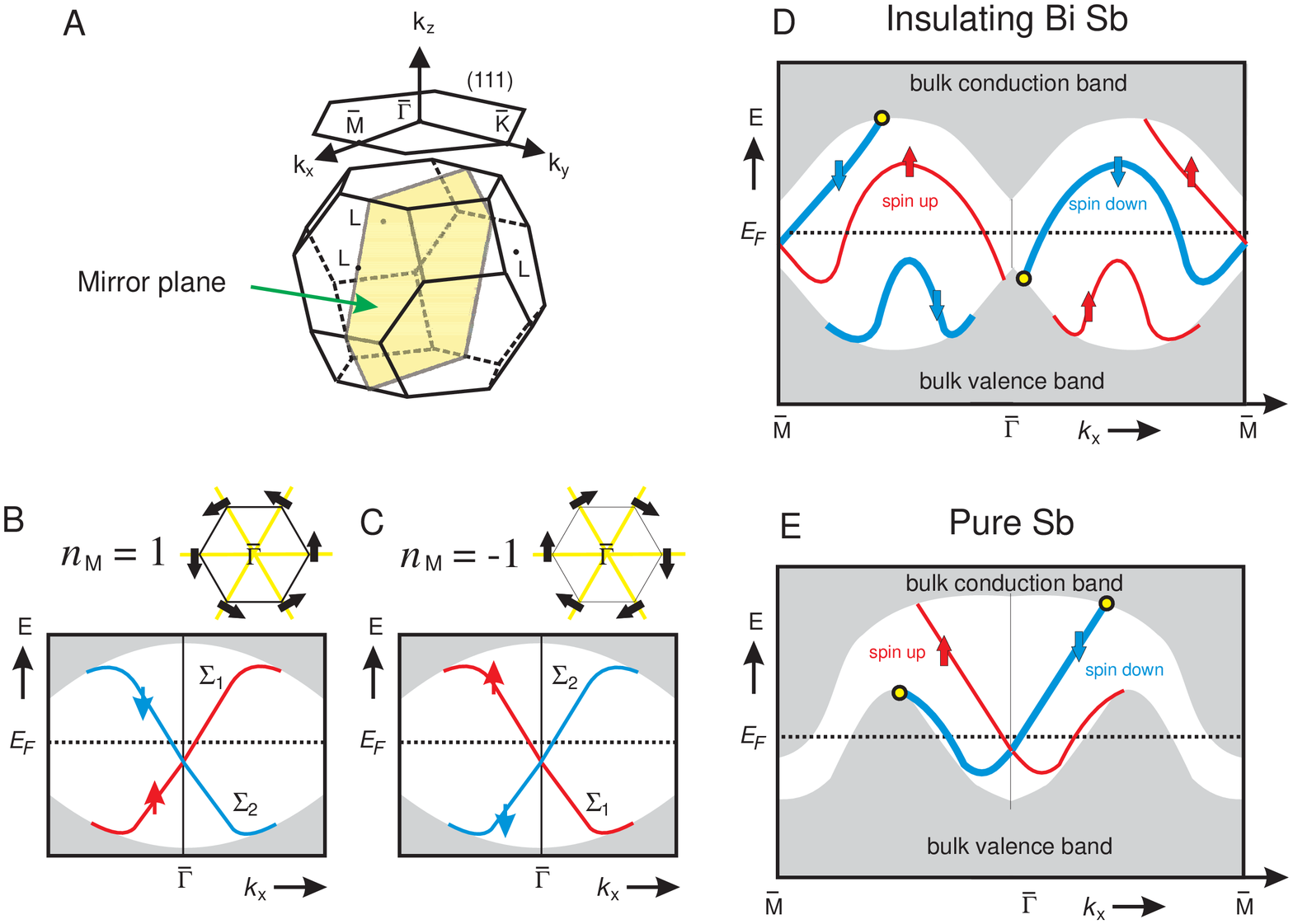}
\caption{\label{fig:Sb_FigS6} \textbf{Measurement of topological mirror Chern number: The k-space mirror symmetry and the topological spin states on the surface.} (\textbf{A}) 3D bulk Brillouin zone and the
mirror plane in reciprocal space. (\textbf{B}) Schematic spin
polarized surface state band structure for a mirror Chern number
($n_M$) of +1 and (\textbf{C}) -1. Spin up and down mean parallel and anti-parallel to $\hat{y}$ respectively. The upper (lower)
shaded gray region corresponds to the projected bulk conduction
(valence) band. The hexagons are schematic spin polarized surface Fermi surfaces for different $n_M$, with yellow lines denoting the mirror planes. (\textbf{D}) Schematic representation of surface state band structure of insulating Bi$_{1-x}$Sb$_x$ and (\textbf{E}) semi metallic Sb both showing a $n_M = -1$ topology. Yellow circles indicate where the spin down band (bold) connects the bulk valence and conduction bands. [Adapted from D. Hsieh $et$ $al.$, \textit{Science} \textbf{323}, 919 (2009) \cite{Science}].}
\end{figure*}

In this section we will describe how a mirror Chern number arises
from the crystal symmetry of Bi$_{1-x}$Sb$_x$. Electronic states in
the mirror plane ($k_y = 0$) [Fig.~\ref{fig:Sb_FigS6}(A)] are
eigenstates of the mirror operator $M(\hat y)$ with eigenvalues $\pm
i$. $M(\hat y)$ is closely related to, but not exactly the same as
the spin operator $S_y$. It may be written as $M(\hat y) = P
C_2(\hat y)$: the product of the parity operator
$P:(x,y,z)\rightarrow(-x,-y,-z)$ and a twofold rotation operator
$C_2(\hat y)$: $(x,y,z)\rightarrow(-x,y,-z)$. For a free spin, $P$
does not affect the pseudovector spin, and $C_2(\hat y)$ simply
rotates the spin. Thus, $M(\hat y)$ = exp$[- i\pi S_y /\hbar]$. For
spin eigenstates $S_y = \pm \hbar/2$, this gives $M(\hat y) = \mp
i$. In the crystal with spin-orbit interaction on the other hand,
$S_y$ is no longer a good quantum number, but $M(\hat y)$ still is.
The energy bands near the Fermi energy in Bi$_{1-x}$Sb$_x$ are
derived from states with even orbital mirror symmetry and satisfy
$M(\hat y) \propto - i$ sign$(\langle S_y\rangle)$, as detailed in
\cite{20} and summarized below.

Unlike the bulk states which are doubly spin degenerate, the surface
state spin degeneracy is lifted due to the loss of crystal inversion
symmetry at the surface, giving rise to the typical Dirac like
dispersion relations near time reversal invariant momenta
[Fig.~\ref{fig:Sb_FigS6}(B)\&(C)]. For surface states in the mirror
plane $k_y = 0$ with $M(\hat y) = \pm i$, the spin split dispersion
near $k_x=0$ has the form $E = \pm \hbar v k_x$.  Assuming no other
band crossings occur, the sign of the velocity $v$ is determined by
the topological mirror Chern number ($n_M$) describing the bulk band
structure. When $n_M = 1$, the situation in
figure~\ref{fig:Sb_FigS6}(B) is realized where it is the spin up
($\langle S_{y}\rangle
\parallel \hat{y}$) band that connects the bulk valence to
conduction band going in the positive $k_x$ direction (i.e. the spin
up band has a velocity in the positive $x$ direction). For $n_M =
-1$ the opposite holds true [Fig.~\ref{fig:Sb_FigS6}(C)]. These two
possibilities also lead to two distinct chiralities of the central
Fermi surface as shown in figures~\ref{fig:Sb_FigS6}(B)\&(C). From
our spin-resolved ARPES data on both insulating Bi$_{1-x}$Sb$_{x}$
and pure Sb, we find that the surface polarized band dispersions are
consistent with $n_M = -1$ [Figs~\ref{fig:Sb_FigS6}(D)\&(E)],
suggesting that their bulk electron wavefunctions exhibit the
anomalous value $n_M = -1$ predicted in (42), which is not
realizable in free electron systems with full rotational symmetry.

There is an intimate physical connection between a 2D quantum spin
Hall insulator and the 2D k-space mirror plane of a 3D topological
insulator. In the former case, the occupied energy bands for each
spin eigenvalue will be associated with an ordinary Chern integer
$n_{\uparrow,\downarrow}$, from which a non-zero spin-Chern number
can be defined $n_s = (n_{\uparrow}-n_{\downarrow})/2$. In the
latter case, it is the mirror eigenvalue of the occupied energy
bands that have associated with them Chern integers $n_{+i,-i}$,
from which a non-zero mirror Chern number can be defined $n_M =
(n_{+i}-n_{-i})/2$.

\end{document}